\documentclass[aps,onecolumn,notitlepage,nofootinbib]{revtex4-2}
\usepackage{epsfig,amssymb,amsmath,mathrsfs,amsfonts,amsthm,graphicx,float,color}
\usepackage[colorlinks=true]{hyperref}
\usepackage[active]{srcltx}
\usepackage{subfigure}
\usepackage{enumitem}
 \def\map#1{\mathcal #1}
\def\<{\langle}\def\>{\rangle}
\def\bb{\langle\!\langle}\def\kk{\rangle\!\rangle}
\def\Tr{\operatorname{Tr}}\def\:{\hbox{\bf
    :}}

\def\R{\mathbb R}
\def\N{\mathbb N}
\def\C{\mathbb C}

\def\Span{\mathsf{Span}}
\def\Supp{\mathsf{Supp}}

\def\spc#1{\mathcal{#1}}
\def\rank{\mathsf{rank}}

\def\set#1{\mathsf{#1}}

\def\Proof{{\bf Proof.}}

\newtheorem{theo}{{Theorem}}
\newtheorem{lem}{{Lemma}}
\newtheorem{prop}{{Proposition}}

\newtheorem{definition}{{Definition}}

\newtheorem{construction}{{Construction}}

\newcommand{\sket}[1]{|#1\>} 
\newcommand{\ket}[1]{|#1\>}
\newcommand{\kket}[1]{\left|#1\middle\rangle\!\right\rangle} 

\newcommand{\bra}[1]{\<#1|}
\newcommand{\bbra}[1]{\left\langle\!\middle\langle#1\right|} 

\newcommand{\sbraket}[2]{\langle#1|#2\rangle}

\newcommand{\ketbra}[1]{\ket{#1}\bra{#1}}

\newcommand{\mc}{\text{\sf min-cut}}

\newcommand{\qmf}{\text{\sf quantum-max-flow}}
\newcommand{\poly}{\text{\sf poly}}
\newcommand{\Temp}{{\sf{Temp}}}

\newcommand{\Rho}{\mathrm{P}}

\usepackage{tikz}
\usetikzlibrary{calc}

\usepackage{array} 
\usepackage{etoolbox}
\usepackage[ruled,linesnumbered]{algorithm2e}

\usetikzlibrary{decorations.pathreplacing, positioning, shapes.misc}
\tikzset{tensor/.style={rectangle,color=black,draw=black,fill=white,thick,
                    inner sep=1pt,minimum size=5mm}}
\tikzset{parameter/.style={rectangle,color=black,draw=black,fill=black!10,thick,
                    inner sep=1pt,minimum size=5mm}}
\tikzset{virtual/.style={rectangle,inner sep=1pt,minimum size=5mm}}
\tikzset{prepare/.style={rounded rectangle, rounded rectangle east arc=none,color=black,draw=black,fill=white,thick,inner sep=0pt,minimum size=5mm}}
\tikzset{measure/.style={rounded rectangle, rounded rectangle west arc=none,color=black,draw=black,fill=white,thick,inner sep=0pt,minimum size=5mm}}
\newenvironment{mathtikz}{\begin{array}{c}\begin{tikzpicture}}{\end{tikzpicture}\end{array}}

\makeatletter

\pgfkeys{/pgf/.cd,
  rectangle corner radius north west/.initial=10pt,
  rectangle corner radius north east/.initial=10pt,
  rectangle corner radius south west/.initial=10pt,
  rectangle corner radius south east/.initial=10pt
}
\newif\ifpgf@rectanglewrc@donecorner@
\def\pgf@rectanglewithroundedcorners@docorner#1#2#3#4#5{%
  \edef\pgf@marshal{%
    \noexpand\pgfintersectionofpaths
      {%
        \noexpand\pgfpathmoveto{\noexpand\pgfpoint{\the\pgf@xa}{\the\pgf@ya}}%
        \noexpand\pgfpathlineto{\noexpand\pgfpoint{\the\pgf@x}{\the\pgf@y}}%
      }%
      {%
        \noexpand\pgfpathmoveto{\noexpand\pgfpointadd
          {\noexpand\pgfpoint{\the\pgf@xc}{\the\pgf@yc}}%
          {\noexpand\pgfpoint{#1}{#2}}}%
        \noexpand\pgfpatharc{#3}{#4}{#5}%
      }%
    }%
  \pgf@process{\pgf@marshal\pgfpointintersectionsolution{1}}%
  \pgf@process{\pgftransforminvert\pgfpointtransformed{}}%
  \pgf@rectanglewrc@donecorner@true
}
\pgfdeclareshape{rectangle with rounded corners}
{
  \inheritsavedanchors[from=rectangle] 
  \inheritanchor[from=rectangle]{north}
  \inheritanchor[from=rectangle]{north west}
  \inheritanchor[from=rectangle]{north east}
  \inheritanchor[from=rectangle]{center}
  \inheritanchor[from=rectangle]{west}
  \inheritanchor[from=rectangle]{east}
  \inheritanchor[from=rectangle]{mid}
  \inheritanchor[from=rectangle]{mid west}
  \inheritanchor[from=rectangle]{mid east}
  \inheritanchor[from=rectangle]{base}
  \inheritanchor[from=rectangle]{base west}
  \inheritanchor[from=rectangle]{base east}
  \inheritanchor[from=rectangle]{south}
  \inheritanchor[from=rectangle]{south west}
  \inheritanchor[from=rectangle]{south east}

  \savedmacro\cornerradiusnw{%
    \edef\cornerradiusnw{\pgfkeysvalueof{/pgf/rectangle corner radius north west}}%
  }
  \savedmacro\cornerradiusne{%
    \edef\cornerradiusne{\pgfkeysvalueof{/pgf/rectangle corner radius north east}}%
  }
  \savedmacro\cornerradiussw{%
    \edef\cornerradiussw{\pgfkeysvalueof{/pgf/rectangle corner radius south west}}%
  }
  \savedmacro\cornerradiusse{%
    \edef\cornerradiusse{\pgfkeysvalueof{/pgf/rectangle corner radius south east}}%
  }

  \backgroundpath{%
    \northeast\advance\pgf@y-\cornerradiusne\relax
    \pgfpathmoveto{}%
    \pgfpatharc{0}{90}{\cornerradiusne}%
    \northeast\pgf@ya=\pgf@y\southwest\advance\pgf@x\cornerradiusnw\relax\pgf@y=\pgf@ya
    \pgfpathlineto{}%
    \pgfpatharc{90}{180}{\cornerradiusnw}%
    \southwest\advance\pgf@y\cornerradiussw\relax
    \pgfpathlineto{}%
    \pgfpatharc{180}{270}{\cornerradiussw}%
    \northeast\pgf@xa=\pgf@x\advance\pgf@xa-\cornerradiusse\southwest\pgf@x=\pgf@xa
    \pgfpathlineto{}%
    \pgfpatharc{270}{360}{\cornerradiusse}%
    \northeast\advance\pgf@y-\cornerradiusne\relax
    \pgfpathlineto{}%
    \pgfpathclose
  }

  \anchor{before north east}{\northeast\advance\pgf@y-\cornerradiusne}
  \anchor{after north east}{\northeast\advance\pgf@x-\cornerradiusne}
  \anchor{before north west}{\southwest\pgf@xa=\pgf@x\advance\pgf@xa\cornerradiusnw
    \northeast\pgf@x=\pgf@xa}
  \anchor{after north west}{\northeast\pgf@ya=\pgf@y\advance\pgf@ya-\cornerradiusnw
    \southwest\pgf@y=\pgf@ya}
  \anchor{before south west}{\southwest\advance\pgf@y\cornerradiussw}
  \anchor{after south west}{\southwest\advance\pgf@x\cornerradiussw}
  \anchor{before south east}{\northeast\pgf@xa=\pgf@x\advance\pgf@xa-\cornerradiusse
    \southwest\pgf@x=\pgf@xa}
  \anchor{after south east}{\southwest\pgf@ya=\pgf@y\advance\pgf@ya\cornerradiusse
    \northeast\pgf@y=\pgf@ya}

  \anchorborder{%
    \pgf@xb=\pgf@x
    \pgf@yb=\pgf@y%
    \southwest%
    \pgf@xa=\pgf@x
    \pgf@ya=\pgf@y%
    \northeast%
    \advance\pgf@x by-\pgf@xa%
    \advance\pgf@y by-\pgf@ya%
    \pgf@xc=.5\pgf@x
    \pgf@yc=.5\pgf@y%
    \advance\pgf@xa by\pgf@xc
    \advance\pgf@ya by\pgf@yc%
    \edef\pgf@marshal{%
      \noexpand\pgfpointborderrectangle
      {\noexpand\pgfqpoint{\the\pgf@xb}{\the\pgf@yb}}
      {\noexpand\pgfqpoint{\the\pgf@xc}{\the\pgf@yc}}%
    }%
    \pgf@process{\pgf@marshal}%
    \advance\pgf@x by\pgf@xa%
    \advance\pgf@y by\pgf@ya%
    \pgfextract@process\borderpoint{}%
    \pgf@rectanglewrc@donecorner@false
    \southwest\pgf@xc=\pgf@x\pgf@yc=\pgf@y
    \advance\pgf@xc\cornerradiussw\relax\advance\pgf@yc\cornerradiussw\relax
    \borderpoint
    \ifdim\pgf@x<\pgf@xc\relax\ifdim\pgf@y<\pgf@yc\relax
      \pgf@rectanglewithroundedcorners@docorner{-\cornerradiussw}{0pt}{180}{270}{\cornerradiussw}%
    \fi\fi
    \ifpgf@rectanglewrc@donecorner@\else
      \southwest\pgf@yc=\pgf@y\relax\northeast\pgf@xc=\pgf@x\relax
      \advance\pgf@xc-\cornerradiusse\relax\advance\pgf@yc\cornerradiusse\relax
      \borderpoint
      \ifdim\pgf@x>\pgf@xc\relax\ifdim\pgf@y<\pgf@yc\relax
       \pgf@rectanglewithroundedcorners@docorner{0pt}{-\cornerradiusse}{270}{360}{\cornerradiusse}%
      \fi\fi
    \fi
    \ifpgf@rectanglewrc@donecorner@\else
      \northeast\pgf@xc=\pgf@x\relax\pgf@yc=\pgf@y\relax
      \advance\pgf@xc-\cornerradiusne\relax\advance\pgf@yc-\cornerradiusne\relax
      \borderpoint
      \ifdim\pgf@x>\pgf@xc\relax\ifdim\pgf@y>\pgf@yc\relax
       \pgf@rectanglewithroundedcorners@docorner{\cornerradiusne}{0pt}{0}{90}{\cornerradiusne}%
      \fi\fi
    \fi
    \ifpgf@rectanglewrc@donecorner@\else
      \northeast\pgf@yc=\pgf@y\relax\southwest\pgf@xc=\pgf@x\relax
      \advance\pgf@xc\cornerradiusnw\relax\advance\pgf@yc-\cornerradiusnw\relax
      \borderpoint
      \ifdim\pgf@x<\pgf@xc\relax\ifdim\pgf@y>\pgf@yc\relax
       \pgf@rectanglewithroundedcorners@docorner{0pt}{\cornerradiusnw}{90}{180}{\cornerradiusnw}%
      \fi\fi
    \fi
  }
}

\makeatother

\begin{document}
\title{Quantum Compression of Tensor Network States}
\author{Ge Bai}
\email{baige@connect.hku.hk}
\affiliation{Department of Computer Science, The University of Hong Kong, Pokfulam Road, Hong Kong}
\author{Yuxiang Yang}
\email{yangyu@phys.ethz.ch}
\affiliation{Institute for Theoretical Physics, ETH Z\"urich, 8093 Z\"urich, Switzerland}
\author{Giulio Chiribella}
\email{giulio@cs.hku.hk}
\affiliation{Department of Computer Science, The University of Hong Kong, Pokfulam Road, Hong Kong}
\affiliation{Department of Computer Science, University of Oxford, Wolfson Building, Parks Road, Oxford, UK }
\affiliation{HKU Shenzhen Institute of Research and Innovation, Kejizhong 2nd Road, Shenzhen, China}
\affiliation{Perimeter Institute For Theoretical Physics, 31 Caroline Street North,  Waterloo N2L 2Y5, Ontario, Canada.}

\begin{abstract}
We   design  quantum compression algorithms for parametric families of tensor network states.
We first establish an upper bound on the amount of memory needed to store an arbitrary state from a given state family.  The bound is determined by the minimum cut of a suitable flow network,  and is related to the flow of information from the manifold of parameters that specify the states to the physical systems in which the states are embodied.
For  given network topology and given edge dimensions, our upper bound is tight when all edge dimensions are powers of the same integer. When this condition is not met, the bound is  optimal up to a multiplicative factor smaller than $1.585$.
 We then provide a  compression algorithm for general state families,  and show that the algorithm runs in polynomial time for matrix product states.


\end{abstract}
\maketitle

\section{Introduction}

Quantum data compression \cite{schumacher1995quantum,jozsa1994new} 
is one of the pillars of quantum information theory. At the foundational level,  it  establishes the qubit as the basic unit of  quantum information.    At the more practical level, it  provides a blueprint for  the efficient transmission of quantum data in future quantum communication networks, with applications to distributed quantum computing \cite{beals2013efficient}
and quantum cloud computing \cite{barz2012demonstration}.

The ultimate limit for compressing sequences of independently prepared quantum states was initially established in the pure state case \cite{schumacher1995quantum} and later extended to mixed states  \cite{lo1995quantum,horodecki1998limits,barnum2001quantum}.
 Universal compression protocols for the scenario where   the average state of each system is unknown,  except for an upper bound on its  von Neumann entropy,   were provided in Ref. \cite{jozsa1998universal}. In recent years, there has been an interest in developing compression protocols for identically prepared systems \cite{plesch2010efficient,chiribella2015universal,yang2016efficient,yang2016optimal,yang2018compression}.
    Such systems  occur in a wide range of tasks, including quantum tomography \cite{d2001quantum,d2003quantum}, quantum cloning \cite{gisin1997optimal,bruss1998optimal}, estimation \cite{helstrom1969quantum,holevo2011probabilistic}, and quantum machine learning  \cite{lloyd2014quantum}. Compression protocols  for identically prepared systems have  found applications in quantum metrology \cite{yang2018quantum} and inspired new results in  quantum state estimation \cite{yang2019attaining}.  An instance of compression for identically prepared systems was experimentally demonstrated  in Ref. \cite{rozema2014quantum}.

  Most of the existing compression protocols  assume that the input systems are in a product state.  However, many relevant scenarios involve correlated systems, whose state cannot be expressed as a tensor product of  single-system states.
    The ability to store correlated  states into a smaller amount of quantum bits is important for the  simulation of many-body quantum systems on small and medium-size quantum computers. For example, Kraus {\em et al.} showed  that  $\log n$ qubits  are enough to simulate several families  of $n$-qubit many-body states \cite{kraus2011compressed,boyajian2013compressed,boyajian2015compressed}. In particular, the result of Ref. \cite{kraus2011compressed} led to an experimental simulation of a 32-spin Ising chain using only 5 qubits \cite{li2014experimental}.  In addition to quantum simulations, many-body states are relevant to quantum metrology, where they can serve as probes for unknown quantum processes \cite{beau2017nonlinear,czajkowski2019many}. In this context, compression protocols for many-body states could  be useful  to transmit  such  probes from one location to another,  or to store them  in a quantum memory until further processing is required.


In this paper we address  the compression of  tensor network states, a broad class  that includes cluster states \cite{briegel2001persistent,raussendorf2001one}, matrix product states (MPS) \cite{fannes1992finitely,perez2006matrix,verstraete2006matrix}, projected entangled pair states (PEPS) \cite{verstraete2004renormalization,verstraete2006criticality},  tree tensor networks \cite{shi2006classical}, and  multi-scale entanglement renormalization ansatz (MERA) states \cite{vidal2008class}.

First, we provide an efficiently computable upper bound of the number of qubits required to compress unknown states from a given  parametric family of tensor network states. The upper bound can be interpreted as a bottleneck for the information flow from the parameters specifying the states to the physical systems in which the states are embodied. For the family of all tensor network states with given network topology and given edge dimensions, this upper bound is tight whenever  all the edge dimensions are powers of the same integer. In general, the upper bound is tight up to a multiplicative factor of at most $\log 3 \approx 1.585$.

Second, we design  a quantum algorithm that implements the compression protocol, and we show that the algorithm runs in polynomial time for families of MPSs.  For more general state families, we provide sufficient conditions for  the algorithm to run in polynomial time.   Informally, the conditions express the fact that  the linear span of the state family  contains a ``sufficiently dense'', yet polynomial-size set of states that can be  efficiently prepared on a quantum computer.

One of the state families considered in our paper involves translationally invariant MPSs \cite{perez2006matrix}, hereafter abbreviated as TIMPS.
We show that a completely  unknown TIMPS of $n$ identical systems with  given bond dimension can be compressed  without errors into a number of logical qubits growing at most as  $O(\log n)$. 
Our result enables a compressed simulation of various models of many-body quantum states, such as the one-dimensional Ising model \cite{ising1925beitrag} and the AKLT model \cite{affleck2004rigorous}.  The logarithmic scaling of the total memory is optimal, as the set of TIMPSs 
includes the set of all identically prepared states, for which the optimal compression protocol is known to require $\Omega  (\log n)$ memory qubits, both for exact  \cite{yang2016efficient} and  approximate compression protocols \cite{yang2016optimal,yang2018compression}.
The same result holds  for  higher dimensional lattices, and for a broader class of tensor network states  for which the correlation tensors are site-independent: a generic site-independent $n$-particle state with a given bond dimension can be perfectly stored  into $O(\log n)$ logical qubits. We also consider  tensor network states  with the property that all tensors except  those on the boundary are constant.   For every subset of systems in the bulk, we show that the exact  compression protocol  satisfies an  area law: the number of logical qubits used to compress the systems in the chosen subset  is proportional to the size of its boundary.

This article is structured as follows. In Section \ref{sec:notations} we introduce the graphical notations for tensor networks. In Section \ref{sec:main} we state our first result on the memory usage of exact compression of tensor network states and apply it to a case-wise study of tensor network state families in Section \ref{sec:examples}. We extend our results from pure states to marginal and mixed states in Section \ref{sec:marginal}.
Section \ref{sec:mps_scheme} provides a compression protocol for MPSs with variable boundary conditions, which can be realised by logarithmic-depth circuits explicitly constructed from the description of the MPS.
In Section \ref{sec:algorithm} we construct a quantum algorithm realising compression protocols for general efficiently preparable states, and discuss its applicability to tensor network states.
Finally, we conclude with discussions on how our results can provide bounds for coding theory in Section \ref{sec:conclusion}.

\section{Preliminaries}\label{sec:notations}
\subsection{Compression of parametric state families}

Consider a quantum system  $\rm P$
 with Hilbert space $\spc H_{\rm P}$, and denote by $S(\map{H_{\rm P}})$ the set of density operators on  $\map{H_{\rm P}}$.
Let   $\{\rho_x\}_{x\in\set X} \subseteq  S(\map H_{\rm P})$ be a parametric family of quantum states,  labeled by a parameter $x$ in a given manifold $\set X$.  
For example,  $x$ could be a parameter that determines  the  Hamiltonian of the  system, and $\rho_x$ could be the ground state of the Hamiltonian parametrized by $x$.

Given a parametric family $\{\rho_x\}_{x\in\set X}$, the goal of compression  is  to store  the states of the family    into  a quantum memory  $\rm M$,  whose dimension is smaller than the dimension of the initial system $\rm P$. A compression protocol for the states  $\{\rho_x\}_{x\in\set X}$ is specified by an encoding channel     $\map{E}: S(\map{H_{\rm P}}) \to S(\map{H_{\rm M}})$,  and by  a decoding channel $\map{D}: S(\map{H}_{\rm M}) \to S(\map{H}_{\rm P})$,  where       $\spc H_{\rm M}$ denotes the Hilbert space of the quantum memory.   Mathematically,   the channels are described by completely positive  trace-preserving linear maps. Both channels $\map E$ and $\map D$ are required to be  independent of the parameter $x$, meaning that the compression operations must  work ``blindly'', without any knowledge of  which state is being compressed.


In the following we will consider  exact compression protocols, that is, protocols satisfying the condition
\begin{align}\label{exact}(\map{D}\circ\map{E})(\rho_x) = \rho_x \qquad \forall x\in\set X \, .
\end{align}
For pure state families, with $\rho_x  =  |\Psi_x\>\<\Psi_x|$ for all $x\in\set X$,  the simplest     compression protocols are defined by   isometries $V: \spc H_{\rm in}  \to \spc H_{\rm M}$ from the input  subspace  $\spc H_{\rm in}  := \Span  \{  |\Psi_x\> \}_{x\in\set X}  \subseteq \spc H_{\rm P}$ to the memory space $\spc H_{\rm M}$.
An optimal compression protocol is a protocol that uses a memory system whose dimension is exactly $d_{\rm  M}   =  \dim ( \spc{  H_{\rm in}})$. In this case,  the isometry $V: \spc H_{\rm in} \to \spc H_{\rm M}$  is actually  a unitary.

In theory, constructing compression protocols for families of pure states is straightforward: one only needs to determine the input subspace $\spc H_{\rm in}$, and to define an isometry $V$ from $\spc H_{\rm in}$ to a memory space $\spc H_{\rm M}  \simeq \spc H_{\rm in}$.   In practice, the efficiency of this construction is an issue.  When the  input system consists of many  particles, constructing the isometry $V$  may  be  computationally unfeasible, because it requires manipulations of exponentially long vectors.  The situation is different when the subspace  $\spc H_{\rm in}$ exhibits some specific structure that can be used to efficiently identify it and to construct the encoding operations.   An example of this situation is the totally symmetric subspace $\spc H_{\rm sym}   = \Span  \{  |\phi\>^{\otimes n}  ~|~  |\phi\>  \in   \spc H_{\rm P} \}$, for which an efficient compression exists \cite{plesch2010efficient} and is based on  the Schur transform \cite{bacon2006efficient,bacon2007quantum}.
In this paper we will identify other scenarios in which the compression operations can be constructed efficiently, taking advantage of the tensor network structure.




\subsection{Graphical notation for tensors}

Here we introduce the  graphical notation used in the rest of the paper. Our notation coincides, up to minor changes, with other notations  used in the literature on tensor networks \cite{cirac2009renormalization,singh2010tensor,singh2011tensor}.
\medskip

{\em Vectors and matrices.} A vector is represented as a box connected to an open edge. A column vector has an outgoing edge, while a row vector  has an ingoing edge. A matrix is represented as a box with both an ingoing edge and an outgoing one. In the following examples,  $\ket{v}$ is a $d$-dimensional column vector,   
$\bra{v}$ is the adjoint of $\ket{v}$,  $\bra{\overline{v}}$ is the transpose of $\ket{v}$, and $A$ is a $d\times d$ matrix.

\begin{align} \ket{v} = \begin{mathtikz}
\node[tensor] (v) at (0,0) {$v$};
\draw[->, thick] (v) -- (-0.75,0);
\end{mathtikz}
, \quad \bra{\overline{v}} = \begin{mathtikz}
\node[tensor] (v) at (0,0) {$v$};
\draw[<-, thick] (v) -- (0.75,0);
\end{mathtikz}
, \quad A = \begin{mathtikz}
\node[tensor] (A) at (0,0) {$A$};
\draw[->, thick] (A) -- (-0.75,0);
\draw[<-, thick] (A) -- (0.75,0);
\end{mathtikz}
\end{align}

For a matrix,  the place where  an arrow is attached to the  box matters. Here we assume that the left side of $A$ corresponds to its row index, and the right side its column index.   For a vector, the attachment position is unimportant, because the vector has only one index.

{\em Multiplication.} An edge connecting two tensors represents a summation over the corresponding index. With this notation, one can  conveniently represent multiplications between matrices and vectors.
\begin{align}
AB\ket{v} = \sum_{i,j,k} A_{ij}B_{jk}v_k \ket{i} =  \begin{mathtikz}
\node[tensor] (A) at (0,0) {$A$};
\node[tensor] (B) at (1,0) {$B$};
\node[tensor] (v) at (2,0) {$v$};
\draw[->, thick] (v) --  node[above] {$k$} (B);
\draw[->, thick] (B) -- node[above] {$j$} (A);
\draw[->, thick] (A) -- node[above] {$i$} (-0.75,0);
\end{mathtikz}
\end{align}
The outgoing open edge indicates that the result of the multiplication  is a column vector.

{\em Tensor product.} A tensor network with several disconnected components is a tensor product of the components (or an outer product of vectors).
\begin{align}
A\ket{v} \otimes B\ket{v} = \begin{mathtikz}
\node[tensor] (A) at (0,0) {$A$};
\node[tensor] (v) at (1,0) {$v$};
\draw[->, thick] (v) -- (A);
\draw[->, thick] (A) -- (-0.75,0);
\node[tensor] (vd) at (1,-1) {$v$};
\node[tensor] (Ad) at (0,-1) {$B$};
\draw[->, thick] (vd) -- (Ad);
\draw[->, thick] (Ad) -- +(-0.75,0);
\end{mathtikz}, \qquad
A\ket{v} \bra{v} A^\dag = \begin{mathtikz}
\node[tensor] (A) at (0,0) {$A$};
\node[tensor] (v) at (1,0) {$v$};
\draw[->, thick] (v) -- (A);
\draw[->, thick] (A) -- (-0.75,0);
\node[tensor] (vd) at (2,0) {$\overline{v}$};
\node[tensor] (Ad) at (3,0) {$A^\dag$};
\draw[<-, thick] (vd) -- (Ad);
\draw[<-, thick] (Ad) -- (3.75,0);
\end{mathtikz} = \begin{mathtikz}
\node[tensor] (A) at (0,0) {$A$};
\node[tensor] (v) at (1,0) {$v$};
\draw[->, thick] (v) -- (A);
\draw[->, thick] (A) -- (-0.75,0);
\node[tensor] (vd) at (1,-1) {$\overline{v}$};
\node[tensor] (Ad) at (0,-1) {$\overline{A}$};
\draw[<-, thick] (vd) -- (Ad);
\draw[<-, thick] (Ad) -- +(-0.75,0);
\end{mathtikz}
\end{align}

{\em Trace.}  The trace of a matrix is represented by connecting its two indices:
\begin{align} \Tr[A] = \begin{mathtikz}
\node[tensor] (A) at (0,0) {$A$};
\draw[->, thick] (A) -- (-0.75,0) -- (-0.75,0.5) -- (0.75,0.5) -- (0.75,0) -- (A);
\end{mathtikz}
\end{align}
In general, a network with no open edges evaluates to a scalar.

{\em Higher-order tensors.}  Higher order tensors can  describe states and linear operations involving multiple systems. To represent them, one uses boxes with more than two edges.  For example, the following graph represents an order-3 tensor $T$, where $\{\ket{i}\},\{\ket{j}\},\{\ket{k}\}$ are orthonormal bases in their corresponding spaces.
\begin{align}\begin{mathtikz}
    \node[tensor] (A1) at (0,0) {$T$};
    \draw[->, thick] (A1) -- node[right] {$i$} +(0,0.75);
    \draw[->, thick] (A1) -- node[above] {$j$} +(-0.75,0);
    \draw[<-, thick] (A1) -- node[above] {$k$} +(0.75,0);
\end{mathtikz} = \sum_{i,j,k} T_{i,j,k} \ket{i}\ket{j}\bra{k}\end{align}
We label the edges by $i,j$ and $k$ to indicate their correspondence to the first, second and third index of $T$, respectively.  In the following, the indices will be sometimes omitted in the graphical notation.

{\em Reversal of edges.} Multiplication by the unnormalised maximally entangled state $\kket{I}: = \sum_i \ket{i}\ket{i}$  or its adjoint $\bbra{I} := \sum_i \bra{i}\bra{i}$ does not alter the elements of a tensor, but it converts a column index to a row index and {\em vice versa}.
We represent a multiplication with $\kket{I}$ by a reversal of the direction of the arrow involved in the multiplication.
\begin{align}T\kket{I} = \begin{mathtikz}
    \node[tensor] (Phi) at (1,0) {$I$} edge [->, thick] (1.75,0);
    \node[tensor] (A1) at (0,0) {$T$} edge [->, thick] (-0.75,0) edge [->, thick] (0,0.75) edge [<-, thick] (Phi);
\end{mathtikz} = \begin{mathtikz}
\node[tensor] (A1) at (0,0) {$T$};
    \draw[->, thick] (A1) -- node[right] {$i$} +(0,0.75);
    \draw[->, thick] (A1) -- node[above] {$j$} +(-0.75,0);
    \draw[->, thick] (A1) -- node[above] {$k$} +(0.75,0);
\end{mathtikz} = \sum_{i,j,k} T_{i,j,k} \ket{i}\ket{j}\ket{k}\end{align}
We always assume that the Hilbert space of each edge comes with a default basis,    so that for each edge, the maximally entangled state is uniquely defined.

{\em Vectorisation.} If we reverse all ingoing edges of a tensor, we obtain a tensor with only outgoing edges, which is a column vector on the tensor product of the  Hilbert spaces corresponding to all the edges. For example,
\begin{align}
\begin{mathtikz}
    \node[tensor] (B) at (0,0) {$B$} edge[->, thick] (0.75,0) edge[->, thick] (-0.75,0);
\end{mathtikz} =
\begin{mathtikz}
    \node[tensor] (B) at (1,0) {$B$};
    \node[tensor] (Phi) at (2,0) {$I$};
    \draw[->, thick] (Phi) -- (B);
    \draw[->, thick] (B) -- (0.25, 0);
    \draw[->, thick] (Phi) -- +(0.75,0);
\end{mathtikz} = (B\otimes I)\kket{I} = \sum_{i,j} B_{ij}\ket{i}\ket{j}\end{align}

In a tensor network, reversing non-open edges does not affect the values assigned to  the whole network. For example,
\begin{align}
\begin{mathtikz}
    \node[tensor] (A) at (0,0) {$A$};
    \node[tensor] (B) at (1,0) {$B$};
    \node[tensor] (v) at (2,0) {$v$};
    \draw[<-, thick] (v) -- (B);
    \draw[->, thick] (B) -- (A);
    \draw[->, thick] (A) -- (-0.75,0);
\end{mathtikz} =
\begin{mathtikz}
    \node[tensor] (A) at (0,0) {$A$};
    \node[tensor] (B) at (1,0) {$B$};
    \node[tensor] (v) at (2,0) {$v$};
    \draw[->, thick] (v) -- (B);
    \draw[->, thick] (B) -- (A);
    \draw[->, thick] (A) -- (-0.75,0);
\end{mathtikz}
\end{align}
for every $A, B$ and $|v\>$.

\subsection{Tensor networks}
Informally, a tensor network is a set of tensors connected with each others.  In the following we introduce a few formal definitions that will become useful later in the paper.

\begin{definition}
A {\em tensor network template} \cite{cui2016quantum}
 is a triple  ${\sf{Temp}}  = (G,  d, V_{\rm filled} )$, where
\begin{itemize}
\item $G  =  (V,E)$ is an oriented\footnote{We recall that an oriented graph is a directed graph in which no edge is bidirected, namely,   for every two vertices $u$ and $v$, at most one of the ordered  pairs $(u,v)$ and $(v,u)$ is an edge in the graph.  } graph, with  set of vertices $V$ and  set of edges $E\subseteq  V  \times V$
\item   $d:  E \to \N_+$ is a function that associates each edge $e$ with an integer  $d(e)$, called the {\em dimension of the edge} $e$
\item  $V_{\rm filled}  \subset  V$ is a subset of vertices, called the {\em filled vertices}, such that each vertex in $V\setminus V_{\rm filled}$  is adjacent to one and only one vertex, and that vertex is in  $V_{\rm filled}$.
\end{itemize}
\end{definition}

\begin{figure}[H]
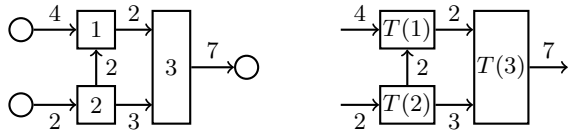

$$
\begin{mathtikz}
    \node[draw=black,thick,circle] (x) at (0,1) {};
    \node[draw=black,thick,circle] (y) at (0,0) {};
    \node[tensor] (A) at (1,1) {1};
    \node[tensor] (B) at (1,0) {2};
    \node[tensor, minimum height=15mm] (C) at (2,0.5) {3};
    \node[draw=black,thick,circle] (out) at (3,0.5) {};
    \node[virtual] (Cup) at ($(C)+(0,0.5)$) {};
    \node[virtual] (Cdown) at  ($(C)-(0,0.5)$) {};
    \draw[->, thick] (x) -- node[above] {4} (A);
    \draw[->, thick] (y) -- node[below] {2} (B);
    \draw[->, thick] (A) -- node[above] {2} (Cup);
    \draw[->, thick] (B) -- node[below] {3} (Cdown);
    \draw[->, thick] (B) -- node[right] {2} (A);
    \draw[->, thick] (C) -- node[above] {7} (out);
\end{mathtikz} \qquad \begin{mathtikz}
    \node (x) at (0,1) {};
    \node (y) at (0,0) {};
    \node[tensor] (A) at (1,1) {$T(1)$};
    \node[tensor] (B) at (1,0) {$T(2)$};
    \node[tensor, minimum height=15mm] (C) at (2.25,0.5) {$T(3)$};
    \node (out) at (3.25,0.5) {};
    \draw[->, thick] (x) -- node[above] {4} (A);
    \draw[->, thick] (y) -- node[below] {2} (B);
    \draw[->, thick] (A) -- node[above] {2} (A-|C.west);
    \draw[->, thick] (B) -- node[below] {3} (B-|C.west);
    \draw[->, thick] (B) -- node[right] {2} (A);
    \draw[->, thick] (C) -- node[above] {7} (out);
\end{mathtikz}$$
    \caption{\label{fig:template}{\em On the left:} A tensor network template $\Temp$. The number on each edge $e\in E$ indicates the dimension $d(e)$ of that edge. Boxes represent vertices in $V_{\rm filled}=\{1,2,3\}$ and circles  represents vertices in $V_{\rm empty}$. {\em On the right:} A tensor network $N=(\Temp,T)$. Each vertex in $V_{\rm filled}$ is assigned a tensor by the mapping $T$. Vertices in $V_{\rm empty}$ are omitted, resulting into open edges. }
\end{figure}

A tensor network is obtained from a tensor network template by filling all vertices in $V_{\rm filled}$ with tensors:
\begin{definition}
A {\em tensor network} is a pair   $N = ({\sf{Temp}},  T)$,  where  ${\sf{Temp}}$ is a tensor network template, and  $T$ is a function mapping filled vertices  $v \in    V_{\rm filled}$ into tensors $T(v)
$, with the order of the tensor $T(v)$ equal to the number of edges incident on $v$.
 Each edge $e$ incident on a filled vertex represents an index of the corresponding tensor, and  the values of the index range from $1$ to $d(e)$.
\end{definition}
Graphically, we will represent a tensor network as a diagram where the filled vertices are represented by boxes,  and the  empty vertices $V_{\rm empty} : =  V\setminus V_{\rm filled}$ are omitted.
  An illustration is shown in Figure \ref{fig:template}.

In the following, we will associate each edge $e\in E$ with a Hilbert space $\spc H_e$ of dimension $d(e)$.  With this notation, the tensor network defines an operator from the total Hilbert space associated to the ingoing edges to the total Hilbert space associated to the outgoing ones:

\begin{definition}\label{def:tnoperator}
Let $E_{\rm in} := \{(u,v)\in E : u\in V_{\rm empty}, v \in V_{\rm filled}\}$ ($E_{\rm out} := \{(u,v)\in E : u \in V_{\rm filled}, v \in V_{\rm empty}\}$) be the set of ingoing (outgoing) edges of a tensor network $N$, and let $\spc H_{\rm in}  :    = \bigotimes_{e\in E_{\rm in}}\spc H_e $ and $\spc H_{\rm out}  :  = \bigotimes_{e\in E_{\rm out}}\spc H_e$ be the corresponding Hilbert spaces.
The {\em tensor network operator} of a tensor network $N$ is a linear operator $N_*:\spc H_{\rm in}\to  \spc H_{\rm out}$ obtained from contracting the tensors $\{ T(v) \}_{v\in  V_{\rm filled}}$ according to the connections specified by the network template.
\end{definition}

For example, the tensor network operator associated to the tensor network in Figure \ref{fig:template} is
\begin{align}
N_*  =  \sum_{i=1}^4\sum_{j=1}^2 \sum_{k=1}^7  \, \left[  \sum_{l=1}^{2} \sum_{m=1}^2 \sum_{n =1}^3   T(1)_{ilm} \,  T(2)_{jln}    \,   T(3)_{mn k} \right]    |k\>  \<i| \otimes \<j| \, .
\end{align}

When a given operator $A$ arises from the contraction of tensors in a tensor network with template $\sf{Temp}$, we say that the operator $A$ is compatible with that template:
\begin{definition}\label{def:tncompatible}
An operator $A:\bigotimes_{e\in E_{\rm in}}\spc H_e\to \bigotimes_{e\in E_{\rm out}}\spc H_e$ is {\em compatible} with the template  ${\sf{Temp}}$  if there exists  a tensor network  $N   = ( {\sf{Temp}},  T )$ such that   $A$ is the tensor network operator associated to $N$, namely $A =  N_*$.
\end{definition}



\subsection{Pure tensor network states}
If a tensor network $N$ has only outgoing edges but no ingoing ones, its operator $N_*$ is  a vector on the tensor product Hilbert space $\spc H_{\rm out}$.  A pure quantum state $|\Psi\>  \in \spc H_{\rm out}$ is a {\em  tensor network state}  if its amplitudes are represented by  a tensor network, namely, $\ket{\Psi} = N_*$ for some tensor network $N$ with no ingoing edges. 

An example of  tensor network states is provided by  the matrix product states (MPSs) \cite{fannes1992finitely,perez2006matrix,verstraete2006matrix}.
MPSs can be used to represent the ground states of one-dimensional chains of particles with local interactions, including the one-dimensional Ising model \cite{ising1925beitrag} and the AKLT model \cite{affleck2004rigorous}. In addition, many MPSs, including the AKLT state, are a resource  for measurement-based quantum computation (MBQC) \cite{gross2007measurement}.

As an  example, consider  the class of MPSs with open boundary conditions \cite{perez2006matrix} and assume for simplicity that every physical system in the MPS has the same dimension.  Any such    MPS  is  specified  by
\begin{enumerate}
    \item $n$ physical systems, each of dimension $d_{\rm p}$
    \item a correlation space $\spc{H}_{{\rm c}}$  of dimension $d_{\rm c}$, called the bond dimension
    \item a set of $d_{\rm c} \times d_{\rm c}$ matrices $\{A^{[k]}_i\}$, labelled by an index $k\in  \{1,\dots, n\}$ and another index $i \in \{1,\dots, d_{\rm p}\}$
    \item two vectors  $\ket{L}$ and $\ket{R}$ in $\spc{H}_{{\rm c}}$, called the  boundary conditions.
\end{enumerate}
Explicitly, the MPS is the $d_{\rm p}^n$-dimensional vector
\begin{align} \label{eq:mps_LRgeneral}
\ket{\Psi_{L,R, A^{[1]}, \dots, A^{[n]}}} = \sum_{i_1,\dots,i_n=1}^{d_{\rm p}} \bra{\overline{L}} A^{[1]}_{i_1}A^{[2]}_{i_2}\dots A^{[n]}_{i_n} \ket{R}  ~  \ket{i_1,i_2,\dots,i_n} \, .
\end{align}
We assume that the norms of the vectors $\ket{L}$ and $\ket{R}$ are chosen in such a way that the overall vector $\ket{\Psi_{L,R, A^{[1]}, \dots, A^{[n]}}}$ is normalised. The MPS (\ref{eq:mps_LRgeneral}) is described by a tensor network of the following form:
\begin{align} \label{eq:mps_LR_graph_0}
\begin{mathtikz}
\node[tensor, minimum width=35mm] (Psi) at (1.5,0) {$\Psi_{L,R, A^{[1]}, \dots, A^{[n]}}$};
\node[virtual] (A1) at (0,0) {};
\node[virtual] (A2) at (1,0) {};
\node (dots) at (2,0.5) {$\cdots$};
\node[virtual] (An) at (3,0) {};
\foreach \i in {1, 2, n}
    \draw[->, thick] (A\i) -- node[right] {$i_{\i}$} +(0,0.75);
\end{mathtikz} =
\begin{mathtikz}
    \node[tensor] (L) at (-1,0) {$L$};
    \node[tensor] (A1) at (0,0) {$A^{[1]}$} edge [<-, thick] (L);
    \node[tensor] (A2) at (1,0) {$A^{[2]}$} edge [->, thick] (A1);
    \node (dots) at (2,0) {$\cdots$} edge [->, thick] (A2);
    \node[tensor] (An) at (3,0) {$A^{[n]}$} edge [->, thick] (dots);
    \node[tensor] (R) at (4,0) {$R$} edge [->, thick] (An);
    \foreach \i in {1, 2, n}
        \draw[->, thick] (A\i) -- node[right] {$i_{\i}$} +(0,0.75);
\end{mathtikz}
\end{align}
where for each $k$, we regard the set $\{A^{[k]}_i\}_{i=1}^{d_{\rm p}}$ as an  order-3 tensor of dimension $d_{\rm p} \times d_{\rm c} \times d_{\rm c}$,  denoted as $A^{[k]}$.
The vertical arrows correspond to the physical systems, while horizontal ones  correspond to  the correlation spaces.

MPSs with periodic boundary conditions \cite{perez2006matrix} are defined by replacing the boundary conditions with the maximally entangled state, as the following:
\begin{align} \label{eq:mps_LR_periodic}
    \ket{\Psi_{A^{[1]}, \dots, A^{[n]}}} \propto \sum_{i_1,\dots,i_n=1}^{d_{\rm p}} \Tr[A^{[1]}_{i_1}A^{[2]}_{i_2}\dots A^{[n]}_{i_n}] \ket{i_1,i_2,\dots,i_n} =  \begin{mathtikz}
    \node[tensor] (A1) at (0,0) {$A^{[1]}$};
    \node[tensor] (A2) at (1,0) {$A^{[2]}$} edge [->, thick] (A1);
    \node (dots) at (2,0) {$\cdots$} edge [->, thick] (A2);
    \node[tensor] (An) at (3,0) {$A^{[n]}$} edge [->, thick] (dots);
    \node[tensor] (Phi) at (1.5,-0.6) {$I$};
    \draw[->, thick] (Phi) -- +(-2,0) |- (A1);
    \draw[->, thick] (Phi) -- +(2,0) |- (An);
    \foreach \i in {1, 2, n}
        \draw[->, thick] (A\i) -- +(0,0.75);
\end{mathtikz}
\end{align}
where $\kket{I} := \sum_{i=1}^{d_{\rm c}} \ket{i}\ket{i}$ represents the unnormalised maximally entangled state.

We will sometimes restrict our attention to {\em site-independent MPSs} (SIMPS) \cite{perez2006matrix}, that is, MPSs where the  matrices $A^{[k]}_i$ are  independent of $k$.
Hence, the set of matrices will be simply denoted  as $\left\{A_i\right\}_{i=1}^{d_{\rm p}}$. In the site-independent case, Equation (\ref{eq:mps_LRgeneral}) becomes
\begin{align} \label{eq:mps_LR}
\ket{\Psi_{L,R,A}} = \sum_{i_1,\dots,i_n=1}^{d_{\rm p}} \bra{\overline{L}} A_{i_1}A_{i_2}\dots A_{i_n}\ket{R} ~ \ket{i_1,i_2,\dots,i_n} \, .
\end{align}
SIMPSs with periodic boundary conditions (\ref{eq:mps_LR_periodic}) are called translationally invariant MPSs \cite{perez2006matrix}, and have the following form.
\begin{align} \label{eq:TIMPS_first}
    \ket{\Psi_{A}} \propto \begin{mathtikz}
    \node[tensor] (A1) at (0,0) {$A$};
    \node[tensor] (A2) at (1,0) {$A$} edge [->, thick] (A1);
    \node (dots) at (2,0) {$\cdots$} edge [->, thick] (A2);
    \node[tensor] (An) at (3,0) {$A$} edge [->, thick] (dots);
    \node[tensor] (Phi) at (1.5,-0.6) {$I$};
    \draw[->, thick] (Phi) -- +(-2,0) |- (A1);
    \draw[->, thick] (Phi) -- +(2,0) |- (An);
    \foreach \i in {1, 2, n}
        \draw[->, thick] (A\i) -- +(0,0.75);
\end{mathtikz}
\end{align}

\section{Memory bound  for the storage of tensor network states}\label{sec:main}

In this section we apply the framework of flow networks to  bound the amount of memory qubits needed to compress a given family of tensor network states. We illustrate this approach for various families of tensor network states, including MPSs and PEPSs.

\subsection{Parametric families of tensor network states}

A parametric family of tensor network states is specified by a tensor network where some of the tensors  depend on the values  of the  parameters.  For example, consider the following  family of SIMPSs with variable boundary conditions
\begin{align}\label{parametrix_LR}
\begin{mathtikz}
\node[tensor, minimum width=35mm] (Psi) at (1.5,0) {$\Psi_{x}$};
\node[virtual] (A1) at (0,0) {};
\node[virtual] (A2) at (1,0) {};
\node (dots) at (2,0.5) {$\cdots$};
\node[virtual] (An) at (3,0) {};
\foreach \i in {1, 2, n}
    \draw[->, thick] (A\i) -- +(0,0.75);
\end{mathtikz} = \begin{mathtikz}
\node[parameter] (L) at (-1,0) {$L_x$};
\node[tensor] (A1) at (0,0) {$A$} edge [->, thick] (L);
\node[tensor] (A2) at (1,0) {$A$} edge [->, thick] (A1);
\node (dots) at (2,0) {$\cdots$} edge [->, thick] (A2);
\node[tensor] (An) at (3,0) {$A$} edge [->, thick] (dots);
\node[parameter] (R) at (4,0) {$R_x$} edge [->, thick] (An);
\foreach \i in {1, 2, n}
   \draw[->, thick] (A\i) -- node[right] {} +(0,0.75);
\end{mathtikz}
\end{align}
for some parameter $ x\in\set X$ and some mappings $L:  x\mapsto \ket{L_x}$  and $R:  x\mapsto  \ket{R_x}$.  Here, the  vectors   $\ket{L_x}$  and $\ket{R_x}$ are  variable, while the tensor $A$ is fixed.

Using vectorisation, the tensor network can be rearranged in such a way that all the variable tensors are column vectors.  The tensor product Hilbert space associated to the edges of the variable tensors will be called the {\em parameter Hilbert space} and will be denoted by $\spc H_{\set X}$.
The tensor product Hilbert space associated to all the physical systems in the network will be called the {\em physical Hilbert space} and will be denoted by $\spc H_{\rm P}$.

With the above notation, every parametric family of tensor network states can be represented as
\begin{align}\label{parametricTN}
\ket{\Psi_x}   =   N_*   \ket{v_x}  \, ,
\end{align}
where  $\ket{v_x}   $ is a vector in  $\spc H_{\set X}$, $N$ is the tensor network consisting of the fixed tensors, and $N_*:  \spc H_{\set X} \to  \spc H_{\rm P}$ is the tensor network operator  associated to $N$.
   An example of this parametrisation is provided in  the following,
\begin{align}\begin{mathtikz}
    \node[tensor] (Psi) {$\Psi_x$};
    \draw[->, thick] (Psi) -- +(0.75,0);
\end{mathtikz} = \begin{mathtikz}
    \node[parameter, minimum height=15mm] (realx) at (0,0.5) {$v_x$};
    \node[virtual] (x) at (0,1) {};
    \node[virtual] (y) at (0,0) {};
    \node[tensor] (A) at (1,1) {$A$};
    \node[tensor] (B) at (1,0) {$B$};
    \node[tensor, minimum height=15mm] (C) at (2,0.5) {$C$};
    \node[virtual] (Cup) at ($(C)+(0,0.5)$) {};
    \node[virtual] (Cdown) at  ($(C)-(0,0.5)$) {};
    \draw[->, thick] (x) -- (A);
    \draw[->, thick] (y) -- (B);
    \draw[->, thick] (A) -- (Cup);
    \draw[->, thick] (B) -- (Cdown);
    \draw[->, thick] (B) -- (A);
    \draw[->, thick] (C) -- +(0.75,0);
    \draw[dashed, thick] ($(Cup)+(0.5,0.5)$) |- node[right] {$N$} ($(B)+(-0.5,-0.5)$) |- ($(Cup)+(0.5,0.5)$);
\end{mathtikz}\end{align}
 where  the operator $N_*$ arises from the tensor network $N$ inside the dashed  frame.

 It is clear from  Equation (\ref{parametricTN})  that  the dimension of the input subspace  $\spc H_{\rm in}   =  \Span \{|\Psi_x\>\}$  is upper bounded by the rank of $N_*$.  Hence, the (logarithm of the) rank of $N_*$   provides an  upper bound on the number of qubits needed for the optimal compression. However, the matrix $N_*$ may generally have an  exponentially large number of columns  and rows, and its rank may not be efficiently computable.  One way to address this problem is to search for efficiently computable upper bounds on the rank of $N_*$, by inspecting the internal  structure of the tensor network  $N$.   In the following subsection we will exploit a connection between tensor networks and flow networks to construct useful bounds on the rank of $N_*$, and therefore, on the number of qubits needed for compression.

\subsection{Flow networks and memory bounds}
We now provide a construction that associates tensor network templates with flow networks, and provide a memory bound valid for all families of tensor network states with a given template.

A flow network \cite{cormen2009introduction} $\widetilde N=  (G,c,s,t)$ consists of  a directed graph $G=(V,E)$, with  set of vertices $V$ and  set of edges $E$, a function $c:  E \to  \R_{0+} :  =  \{ x\in\R ~|~ x\ge0\}$, associating each  edge $e\in E$ with a non-negative number $c(e)$, called its capacity, and two distinguished vertices, $s$ and $t$,  called the source  and the sink, respectively.

A flow network can be intuitively understood as a pipe system with edges being pipes and vertices being junctions. Fluid enters in the pipe system from  the source $s$ and exits at the sink $t$. At any time, the flux in each pipe is no more than the capacity of the pipe, while at each junction, the total amount of fluid is conserved, meaning that  the total flux going into the junction equals to the total flux going out. This idea is captured by the mathematical notion of flow. A  {\em flow}  $f:  E \to  \R_{0+}$ in a flow network  is an assignment of non-negative numbers  to the edges of the network, subject to the conditions
\begin{enumerate}
\item for every edge $e$, the flow is upper bounded by the capacity, namely $f(e)\le c(e)$
\item for every vertex $v$ other than the source and the sink, the total flow entering in the vertex $v$ is equal to the total flow exiting from it, namely $\sum_{u\in  V ,  \,  (u,v)  \in  E}  f(u,v)  =  \sum_{u\in  V , \,   (v,u)    \in  E}  f(v,u)$.
\end{enumerate}
In the pipe system analogy, $f(e)$ represents the amount of fluid that is flowing through the pipe $e$.

The {\em value of the flow} $f$, denoted by $f_s$, is the total flow exiting from the source, namely
\begin{align}
f_s  :=\sum_{v  \in  V ,  \,    (s,v)  \in  E}  \,  f(s,v)  \, .
\end{align}
The maximum of $f_s$ over all possible flows is called the {\em max-flow} of the network, and is denoted as
\begin{align}
{\text{\sf max-flow}}  (\widetilde N)  : = \max_f  \, f_s \, .
\end{align}

Intuitively, $f_s$ represents the maximum amount of fluid that can enter into a pipe system. This amount can be upper bounded by considering that all the fluid entering from the source has  to exit  from the sink, and, in order to do so, it has to pass through the pipes between the source and the sink.  A  {\em cut}   of the flow network is a cross-section of the pipe system that separates the source and the sink. Formally, a cut is a partition of the vertices into two disjoint subsets $C_s\subset V$ and $C_t\subset V$, with $s\in C_s$ and $t\in C_t$. We write the cut as $C= (C_s,C_t)$.  An edge $(u,v)$ is called a {\em cut edge} if $u $ belongs to $C_s$ and $v$ belongs to $C_t$.

Since all the fluid entering from the source has to pass through the cut $C$, the total capacity of the pipes associated to cut edges poses an upper bound to the flux.
Explicitly, the {\em capacity of the cut} $(C_s,C_t)$, denoted as $c(C_s,C_t)$ is the sum of the capacities of the cut edges, namely
\begin{align}
c(C_s,C_t)   :=  \sum_{u\in C_s, \,  v\in C_t , \,  (u,v) \in E}  \,  c(u,v) \, ,
\end{align}
and one has the upper bound
\begin{align}
{\text{\sf max-flow}}  (\widetilde N)  \le c(C_s,C_t)
\end{align}
for every possible cut $(C_s,C_t)$.  The best bound is obtained by choosing the cut with minimum capacity.    The minimum of the capacity $c(C_s,C_t)$ over all possible cuts $(C_s,C_t)$ is called the {\em min-cut}, and is denoted by
\begin{align}
\text{\sf min-cut}  (  \widetilde N)   :=  \min_{(C_s,C_t)}  \,  c(C_s,C_t) \, .
\end{align}

The max-flow min-cut theorem states that $\text{\sf max-flow}(\widetilde N)= \text{\sf min-cut}(\widetilde N)$ \cite{cormen2009introduction}. Intuitively, this shows that the maximum flux a pipe system can carry from $s$ to $t$ is {\em exactly} equal to the capacity of the  minimal cross-section of the pipes.

\begin{figure}[H]
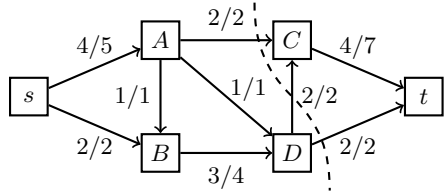

$$\begin{mathtikz}
    \node[tensor] (B) at (0,-0.75) {$B$};
    \node[tensor] (A) at (0,0.75) {$A$};
    \node[tensor] (C) at (1.75,0.75) {$C$};
    \node[tensor] (D) at (1.75,-0.75) {$D$};
    \node[tensor] (s) at (-1.75,0) {$s$};
    \node[tensor] (t) at (3.5,0) {$t$};
    \draw[->, thick] (s) -- node[above] {$4/5$} (A);
    \draw[->, thick] (s) -- node[below, yshift=-1] {$2/2$} (B);
    \draw[->, thick] (A) -- node[left] {$1/1$} (B);
    \draw[->, thick] (A) -- node[above] {$2/2$} (C);
    \draw[->, thick] (B) -- node[below, yshift=-1] {$3/4$} (D);
    \draw[->, thick] (A) -- node[right, xshift=-2, yshift=3] {$1/1$} (D);
    \draw[<-, thick] (C) -- node[right] {$2/2$} (D);
    \draw[->, thick] (C) -- node[above] {$4/7$} (t);
    \draw[->, thick] (D) -- node[below, yshift=-1] {$2/2$} (t);
    \draw [thick, dashed] (1.25,1.25) to[out=-90,in=135] (1.75,0) to[out=-45,in=90] (2.25,-1.25);
\end{mathtikz}$$
\caption{\label{fig:flow} A flow network.  The numbers on each edge indicate the flow $f(e)$ and the capacity $c(e)$ of the edge, in the form $f(e)/c(e)$. The dashed line indicates a cut $(C_s,C_t)$, with vertices to its left belonging to $C_s$, vertices to its right belonging to $C_t$.
}
\end{figure}

An example of a flow network is shown in Figure \ref{fig:flow}.

\begin{figure}[H]
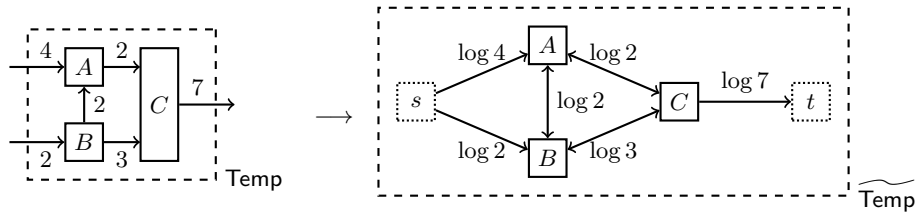

$$
\begin{array}{ccc}
 \begin{mathtikz}
    \node[virtual] (x) at (-0.25,1) {};
    \node[virtual] (y) at (-0.25,0) {};
    \node[tensor] (A) at (1,1) {$A$};
    \node[tensor] (B) at (1,0) {$B$};
    \node[tensor, minimum height=15mm] (C) at (2,0.5) {$C$};
    \node[virtual] (Cup) at ($(C)+(0,0.5)$) {};
    \node[virtual] (Cdown) at  ($(C)-(0,0.5)$) {};
    \draw[->, thick] (x) -- +(0.5,0) -- node[above] {4} (A);
    \draw[->, thick] (y) -- +(0.5,0) -- node[below] {2} (B);
    \draw[->, thick] (A) -- node[above] {2} (Cup);
    \draw[->, thick] (B) -- node[below] {3} (Cdown);
    \draw[->, thick] (B) -- node[right] {2} (A);
    \draw[->, thick] (C) -- node[above] {7} +(0.75,0) -- +(1,0);
    \draw[dashed, thick] ($(Cup)+(0.75,0.5)$) |- node[right] {${\sf{Temp}}$} ($(B)+(-0.75,-0.5)$) |- ($(Cup)+(0.75,0.5)$);
\end{mathtikz} & \longrightarrow &
\begin{mathtikz}
    \node[tensor] (W2) at (0,-0.75) {$B$};
    \node[tensor] (W1) at (0,0.75) {$A$};
    \node[tensor] (W3) at (1.75,0) {$C$};
    \node[tensor, densely dotted] (s) at (-1.75,0) {$s$};
    \node[tensor, densely dotted] (t) at (3.5,0) {$t$};
    \draw[->, thick] (s) -- node[above] {$\log 4$} (W1);
    \draw[->, thick] (s) -- node[below, yshift=-1] {$\log 2$} (W2);
    \draw[<->, thick] (W1) -- node[right] {$\log 2$} (W2);
    \draw[<->, thick] (W2) -- node[below, yshift=-1] {$\log 3$} (W3);
    \draw[<->, thick] (W1) -- node[above] {$\log 2$} (W3);
    \draw[->, thick] (W3) -- node[above] {$\log 7$} (t);
    \draw[dashed, thick] ($(t)+(0.5,1.25)$) |- node[right] {$\widetilde {\sf{Temp}}$} ($(s)+(-0.5,-1.25)$) |- ($(t)+(0.5,1.25)$);
\end{mathtikz}
\end{array}
$$
\caption{\label{fig:flownetwork} Conversion of a  tensor network template into a flow network. {\em On the left:}
the tensor network template ${\sf{Temp}}$. The number on each edge indicates $d(e)$, the dimension of the Hilbert space assigned to each edge. {\em On the right:} the flow network $\widetilde{{\sf{Temp}}}$. The number on each edge indicates its capacity $c(e)$.}
\end{figure}

The analogy of the pipe system will be useful to understand the intuitive content of our results, where we use flow networks to model the ``flow of information'', rather than the flow of a material fluid.  We imagine information flowing into the ingoing edges and out of the outgoing edges of the tensor network, and each edge has a capacity equal to the maximum amount of information it can carry, which is $\log d(e)$ qubits for an edge $e$ with dimension $d(e)$.
Given a tensor network template, we relate it to a flow network that can be constructed in the following way:

\begin{construction}\label{cons:t2f}
Let ${\sf{Temp}}  = (G,  d,  V_{\rm filled})$  be a tensor network template,  and let $V_{\rm empty}  =  V\setminus V_{\rm filled}$ the set of empty vertices.
The flow network associated to the template ${\sf{Temp}}$, denoted by $\widetilde{\sf{Temp}} =  (\widetilde G,  c, s,t)$, is constructed through  the following prescriptions:
\begin{enumerate}
  \item Add the vertices $s$ and $t$ to $V$.
     \item Replace each ingoing  edge  $(u,v)$ (with $u\in  V_{\rm empty}$ and $v  \in  V_{\rm filled}$) with an edge $(s, v)$. Define the capacity of the edge $(s,v)$ as $c (s,v)  :=  \log d(  u,v)$.
    \item Replace each outgoing edge $(v,u)$ (with $v  \in  V_{\rm filled}$ and    $u\in  V_{\rm empty}$) with an edge $(v,t)$.  Define the capacity of the edge $(v,t)$ as $c (v,t)  :=  \log d(  v,u)$.
    \item Remove all the vertices in $V_{\rm empty}$.
\item For every internal edge $(v,w)$ (with both $v$ and $w$ in $V_{\rm filled}$)  include also $(w,v)$ in the set of edges.  Define the capacities $c(v,w)  =  c(w,v)   =  \log d(v,w)$.
 \end{enumerate}
\end{construction}
An illustration of the above procedure is provided in Figure \ref{fig:flownetwork}.

\begin{figure}[H]
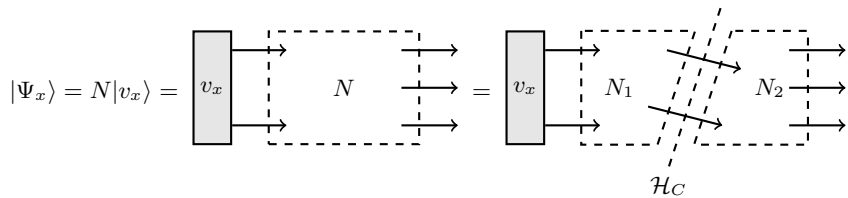

$$
\ket{\Psi_x} = N\ket{v_x} = \begin{mathtikz}
    \draw[use as bounding box, white] (-0.25,-1) rectangle (3.25,2);
    \node[parameter, minimum height=15mm] (x) at (0,0.5) {$v_x$};
    \node[tensor, dashed, minimum height=15mm, minimum width=20mm] (N) at (1.75,0.5) {$N$};
    \draw[->, thick] ($(x.east)+(0,0.5)$) -- ($(N.west)+(0.25,0.5)$);
    \draw[->, thick] ($(x.east)+(0,-0.5)$) -- ($(N.west)+(0.25,-0.5)$);
    \draw[->, thick] ($(N.east)+(-0.25,0.5)$) -- +(0.75,0);
    \draw[->, thick] ($(N.east)+(-0.25,-0.5)$) -- +(0.75,0);
    \draw[->, thick] ($(N.east)+(-0.25,0)$) -- +(0.75,0);
\end{mathtikz} = \begin{mathtikz}
    \draw[use as bounding box, white] (-0.25,-1) rectangle (4.25,2);
    \node[parameter, minimum height=15mm] (x) at (0,0.5) {$v_x$};
    \node[virtual, minimum height=15mm, minimum width=10mm] (S) at (1.25,0.5) {$N_1$};
    \draw[dashed, thick] (S.north west) -- (S.south west) -- (S.south east) -- ($(S.north east)+(0.5,0)$) -- (S.north west);
    \node[virtual, minimum height=15mm, minimum width=10mm] (T) at (3.25,0.5) {$N_2$};
    \draw[dashed, thick] (T.north west) -- ($(T.south west)-(0.5,0)$) -- (T.south east) -- (T.north east) -- (T.north west);
    \draw[->, thick] ($(x.east)+(0,0.5)$) -- ($(S.west)+(0.25,0.5)$);
    \draw[->, thick] ($(x.east)+(0,-0.5)$) -- ($(S.west)+(0.25,-0.5)$);
    \draw[->, thick] ($(T.east)+(-0.25,0.5)$) -- +(0.75,0);
    \draw[->, thick] ($(T.east)+(-0.25,-0.5)$) -- +(0.75,0);
    \draw[->, thick] ($(T.east)+(-0.25,0)$) -- +(0.75,0);
    \draw[->, thick] ($(S.east)+(0.125,0.5)$) -- ($(T.west)+(0.125,0.25)$);
    \draw[->, thick] ($(S.east)+(-0.125,-0.25)$) -- ($(T.west)+(-0.125,-0.5)$);
    \coordinate (start) at ($(S.north east)!0.75!(T.north west)$);
    \coordinate (end) at ($(S.south east)!0.25!(T.south west)$);
    \draw[dashed, thick] ($(start)!-0.2!(end)$) -- ($(start)!1.2!(end)$) node[below] (label) {$\map{H}_C$};
\end{mathtikz}
$$
\caption{\label{fig:cut}\label{fig:N1N2}The cut $C$ divides the network $N$ into two subnetworks $N_1$ and $N_2$. The combined Hilbert space of all cut edges is $\map{H}_C$.
}
\end{figure}

Consider the tensor network $N  =   ({\sf{Temp}}, T)$ associated to the fixed tensors in the given state family  $\{|\Psi_x\>\}$, and let  $\widetilde {{\sf{Temp}}}$ be the flow associated to the template    ${\sf{Temp}}$ via  Construction \ref{cons:t2f}.
 Every cut in the flow network $\widetilde {{\sf{Temp}}}$ defines a factorisation of the operator $N_*$ as $N_*  =    N_{2*}   N_{1*}$, where  $N_{1*}$ and $N_{2*}$ are the operators of the two subnetworks $N_1$ and $N_2$ on the two sides of the cut, as illustrated in
  Figure \ref{fig:N1N2}.
 Then, one has the bound
 \begin{align}
\rank (N_*)  \le \min  \{ \rank (N_{1*}), \rank  (N_{2*})\} \le d_C \, ,
 \end{align}
 where $d_C$ is the dimension of  the Hilbert space  $\spc H_{C}$ associated to the edges in the cut.  Hence, $\lceil \log(d_{\rm C} ) \rceil$ qubits are sufficient to compress the state family.   Recalling that the logarithm of $d_C$ is the capacity of the cut, we obtain the following:

\begin{theo} \label{thm:tncompression}
For every cut $C=(C_s,C_t)$ of $\widetilde {{\sf{Temp}}}$, 
the state family $\{\ket{\Psi_x}\}_{x \in \set{X}}$ can be compressed without errors into 
$\lceil c(C) \rceil$ memory qubits.  In particular, the state family can be compressed into
\begin{align}
Q_{\rm cut}  = \lceil  \text{\sf  min-cut}   (\widetilde {{\sf{Temp}}})\rceil
\end{align}
memory qubits, where $  \text{\sf  min-cut}   (\widetilde {{\sf{Temp}}})  =  \min_C  \log d_{\rm C}$ is the minimum cut of the flow network $\widetilde {{\sf{Temp}}}$.
\end{theo}

Intuitively, Theorem \ref{thm:tncompression}  tells us that, since the maximum amount of information that can flow in the network is upper bounded by the capacity of the min-cut, the amount of memory  needed to store this information is also upper bounded by the capacity of the min-cut.
 The point of Theorem \ref{thm:tncompression} is  that, while  the  calculation of $\rank (N_*)$ may not be computationally feasible, the minimum cut can be found efficiently using known algorithms such as the {Relabel-To-Front algorithm}  \cite{cormen2009introduction}, which runs  in $O(|V|^3)$ time.

In Section \ref{sec:examples}, we will provide explicit examples of minimum cuts for some relevant  families of tensor network states.
 Before that, we will discuss the optimality of $Q_{\rm cut}$ as an upper bound on the number of memory qubits needed for compression.

\subsection{Optimality for fixed tensor network templates}


The amount of memory used by the best compression protocol is intuitively related to the flow of quantum information from the parameters specifying the quantum state to the physical systems in which the states are embodied.
To make this intuition precise, suppose that we want to compress a known, but otherwise generic  family of tensor network states with network template ${\sf{Temp}}$, that is, a family of the form $\{  N_*  |v_x\>  \}_{x\in\set X}$, where $N_*$ is a  tensor network operator compatible with the template ${\sf{Temp}}$, and $\{  |v_x\>\}$ is a generic  set of (suitably normalised) vectors in the parameter space $\spc H_{\set X}$.   In the worst case over $\{ |v_x\>\}$ and $N_*$, it is easy to see that the minimum number of memory qubits  necessary for compression is   $\lceil   \log  \rank (N_*)   \rceil$.   Indeed, the vectors $\{  |v_x\>\}$ could form a spanning set for the parameter space   $\spc H_{\set X}$, so that the dimension of the input space $\spc H_{\rm in}  =  \Span  \{  N_*  |v_x\>\}_{x\in\set X}$ is exactly equal to the rank of $N_*$. It is then immediate to conclude that every exact compression protocol will require at least $\lceil   \log  \rank (N_*)   \rceil$ memory qubits.    Taking   the worst case over $N_*$,  we obtain the following
\begin{prop}\label{prop:maxflow}
The minimum  number of memory qubits  required for the exact compression of a generic state family of tensor network states with template  ${\sf{Temp}}$ is
\begin{align}\label{Qmin}
Q_{\min}   = \left \lceil {\text{\sf quantum-max-flow}}  (\widetilde {{\sf{Temp}}}) \right\rceil \,,
\end{align}
where ${\text{\sf quantum-max-flow}}  (\widetilde {{\sf{Temp}}}) $ is the {\em quantum max-flow} \cite{cui2016quantum}, defined as
\begin{align}
{\text{\sf quantum-max-flow}}  (\widetilde {{\sf{Temp}}})   :  =  \max_{N_*}  \,    \log  \rank (N_*) \,,
\end{align}
 the maximum being  over all tensor network operators $N_*$  compatible with the template ${\sf{Temp}}$.\footnote{Note that the   quantum max-flow adopted here is the logarithm of the quantum max-flow  defined  in Ref. \cite{cui2016quantum}.}
 \end{prop}

 Now, an important question is whether the compression protocols of  Theorem \ref{thm:tncompression}
can reach the  minimum number of qubits (\ref{Qmin}),   in the worst case over all state families compatible with a given network template. In other words, the question is  whether the equality  $\lceil  {\text{\sf quantum-max-flow}}   (\widetilde {{\sf{Temp}}})  \rceil   =   \lceil  \text{\sf  min-cut}   (\widetilde {{\sf{Temp}}})   \rceil$ holds.   Such equality would follow from a quantum version of the  max-flow min-cut theorem \cite{ford2015flows}, which would state  the equality  $ {\text{\sf quantum-max-flow}}   (\widetilde {{\sf{Temp}}})   =   \text{\sf  min-cut}   (\widetilde {{\sf{Temp}}}) $.  Remarkably, Ref. \cite{cui2016quantum} shows that such quantum version does not always hold, and in general the quantum max-flow is only a lower bound on the min-cut
\begin{align}
{\text{\sf quantum-max-flow}}   (\widetilde {{\sf{Temp}}})   \le   \text{\sf  min-cut}   (\widetilde {{\sf{Temp}}}) \, .
\end{align}
Nevertheless, the  equality holds in the case where all dimensions are powers of the same integer \cite{cui2016quantum}.  In this case, the validity of the quantum max-flow-min-max theorem implies the following   optimality property:

\begin{prop}\label{thm:optimality}
Let ${{\sf{Temp}}} $  be a  network template with $d(e)   =  b^{n(e)}$ for some fixed integer $b$ and for some integer-valued function $n:   E \to  \N $.    Then,  $Q_{\rm cut}= Q_{\rm min}$, meaning that the number of qubits used in Theorem \ref{thm:tncompression} is minimum in the worst case over all state families with the given network template.
\end{prop}

Proposition \ref{thm:optimality} guarantees that, under the assumption that each dimension $d(e)$ is an integer power of $b$, the amount of qubits used in  Theorem \ref{thm:tncompression} 
 is optimal for the least compressible family of tensor network states compatible with the given template. In the general case, we show that the amount of qubit used by the  compression protocol  of Theorem \ref{thm:tncompression} is at most $\log 3$ times the minimum number needed for compression. This result, provided in the following Proposition,  is based on a general relation between  the min-cut and the quantum-max-flow  of a generic tensor network:
 \begin{prop}\label{thm:optimality2}
 For every network template ${{\sf{Temp}}}  =  (G  ,  d,  V_{\rm filled})$, one has the bound
 \begin{align}
  {\text {\sf min-cut}}  (\widetilde {\sf Temp})   \le (\log3)~   \qmf(\widetilde {{\sf{Temp}}}) \, .
 \end{align}
 As a consequence, one has the bound
 \begin{align}
 Q_{\rm cut}  \le  ( \log3) ~  Q_{\rm min}  + 1  \, ,
 \end{align}
 which implies that, asymptotically, the number of qubits used in  the  compression protocol  of Theorem \ref{thm:tncompression}  is at most $\log 3$ times the minimum number $Q_{\rm min}$.
\end{prop}
The proof is provided in Appendix \ref{app:optimality2}.  In conclusion,  the number of qubits used in Theorem \ref{thm:tncompression}  is either minimum (if all dimensions are power of the same integer), or within a factor $\log 3 \approx 1.585$ of the minimum (if some dimensions are not power of the same integer).
Equipped with this result, in the next section we will analyse the number of qubits needed to compress various families of tensor network states.

\section{Exponentially compressible families of tensor network states}\label{sec:examples}
In this section we apply the  memory bound of Theorem \ref{thm:tncompression}  to various families of tensor network states. In most of these examples, the amount of memory qubits required to store the states is exponentially smaller than the original number of physical particles in which the states are embodied.

\subsection{MPSs with variable boundary conditions}\label{ss:mps}\


Consider the following family of MPSs with variable boundary conditions:
\begin{align} \label{eq:mps_LR_graph_2}
\ket{\Psi_{L,R}} = \sum_{i_1,\dots,i_n=1}^{d_{\rm p}} \bra{\overline{L}} A^{[1]}_{i_1}A^{[2]}_{i_2}\dots A^{[n]}_{i_n}\ket{R}   ~  \ket{i_1,i_2,\dots,i_n} =
\begin{mathtikz}
    \node[parameter] (L) at (-1,0) {$L$};
    \node[tensor] (A1) at (0,0) {$A^{[1]}$} edge [<-, thick] (L) edge [->, thick] (0,0.75);
    \node[tensor] (A2) at (1,0) {$A^{[2]}$} edge [->, thick] (A1) edge [->, thick] (1,0.75);
    \node (dots) at (2,0) {$\cdots$} edge [->, thick] (A2);
    \node[tensor] (An) at (3,0) {$A^{[n]}$} edge [->, thick] (dots) edge [->, thick] (3,0.75);
    \node[parameter] (R) at (4,0) {$R$} edge [->, thick] (An);
    \draw[dashed, thick] (3.5,0.5) -| (-0.5, -0.5) -| (3.5,0.5);
\end{mathtikz}
\end{align}
Here,  the tensors $A^{[1]},A^{[2]},\dots, A^{[n]}$ are fixed and known, and the free parameters are  the components of the vectors $\ket{L}$ and $\ket{R}$.  We consider the case where $d_{\rm c}= O(\poly(n))$, namely the the bond dimension $d_{\rm c}$ grows at most polynomially with $n$, which is true for MPS approximations of ground states one-dimensional gapped Hamiltonians \cite{hastings2007area,arad2013area,huang2014area}. In cases of the one-dimensional Ising model \cite{ising1925beitrag} and the AKLT model \cite{affleck2004rigorous}, $d_{\rm c}$ is even constant. In  Equation  (\ref{eq:mps_LR_graph_2}), we regard the tensors surrounded by the dashed line as the tensor network $N$, and  we write $\ket{\Psi_{L,R}} = N_*(\ket{L}\otimes\ket{R})$.

\begin{figure}[H]
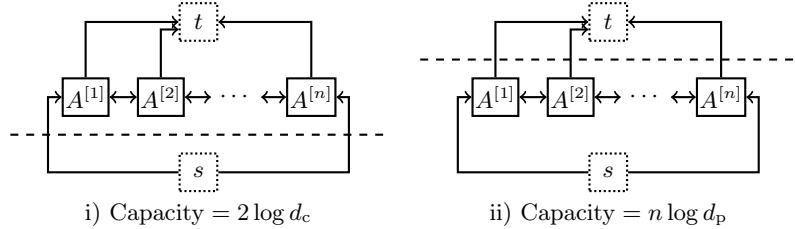

$$
\begin{array}{cc}
    \begin{mathtikz}
        \node[tensor, densely dotted] (s) at (1.5,-1) {$s$};
        \node[tensor, densely dotted] (t) at (1.5,1) {$t$};
        \node[tensor] (A1) at (0,0) {$A^{[1]}$};
        \draw[<-, thick] (A1) -- (-0.5,0) |- (s);
        \draw[->, thick] (A1) |- (t);
        \node[tensor] (A2) at (1,0) {$A^{[2]}$} edge [<->, thick] (A1);
        \draw[->, thick] (A2) |- ($(t.west)+(0,-0.1)$);
        \node (dots) at (2,0) {$\cdots$} edge [<->, thick] (A2);
        \node[tensor] (An) at (3,0) {$A^{[n]}$} edge [<->, thick] (dots);
        \draw[->, thick] (An) |- (t);
        \draw[<-, thick] (An) -- (3.5,0) |- (s);
        \draw[dashed, thick] (-1, -0.5) -- (4,-0.5);
    \end{mathtikz} & \begin{mathtikz}
        \node[tensor, densely dotted] (s) at (1.5,-1) {$s$};
        \node[tensor, densely dotted] (t) at (1.5,1) {$t$};
        \node[tensor] (A1) at (0,0) {$A^{[1]}$};
        \draw[<-, thick] (A1) -- (-0.5,0) |- (s);
        \draw[->, thick] (A1) |- (t);
        \node[tensor] (A2) at (1,0) {$A^{[2]}$} edge [<->, thick] (A1);
        \draw[->, thick] (A2) |- ($(t.west)+(0,-0.1)$);
        \node (dots) at (2,0) {$\cdots$} edge [<->, thick] (A2);
        \node[tensor] (An) at (3,0) {$A^{[n]}$} edge [<->, thick] (dots);
        \draw[->, thick] (An) |- (t);
        \draw[<-, thick] (An) -- (3.5,0) |- (s);
        \draw[dashed, thick] (-1, 0.5) -- (4,0.5);
    \end{mathtikz} \\
    \text{i) Capacity}=2\log d_{\rm c} & \text{ii) Capacity}=n\log d_{\rm p}
\end{array}$$
\caption{\label{fig:mps_LR_graph_cut} Flow network associated to the states (\ref{eq:mps_LR_graph_2}), and the two candidate minimum cuts.  Except the edges connected to the source $s$ or the sink $t$, all edges are made bidirectional. The dashed lines indicate the  minimum cuts. When $n$ is sufficiently large, as $d_{\rm c}$ is assumed to grow polynomially with $n$, $2\log d_{\rm c} \leq n\log d_{\rm p}$, and the one on the left is the minimum cut.}

\end{figure}

In order to apply Theorem \ref{thm:tncompression}, the first step is to convert $N$ to a flow network, as illustrated in  Figure \ref{fig:mps_LR_graph_cut}.  Then,  Theorem \ref{thm:tncompression}  guarantees that the states $\{\ket{\Psi_{L,R}}\}$ can be compressed into a number of qubits equal to the capacity of the minimum cut.
For the flow network in  Figure \ref{fig:mps_LR_graph_cut}, there are two candidates for the minimum cut $(C_s,C_t)$:
\begin{itemize}
\item[i)] $C_s$ contains only $s$, while $C_t$ contains all the other vertices.  This cut has capacity $2\log d_{\rm c}$.
\item[ii)] $C_t$ contains only $t$, while $C_s$ contains all other vertices. This cut has capacity $n\log d_{\rm p}$.
\end{itemize}
For all other cuts,  the cut edges necessarily contain at least two edges of capacity $\log d_{\rm c}$ and one edge of capacity $\log d_{\rm p}$, leading to a capacity larger than that of i).
For sufficiently large $n$, as we have assumed $d_{\rm c}$ is no larger than a polynomial of $n$, we have $2\log d_{\rm c} \leq n\log d_{\rm p}$, and i) is the minimum cut.
Therefore the states $\{\ket{\Psi_{L,R}}\}$ can be compressed into a number of $\lceil 2\log d_{\rm c}\rceil $ qubits, which is $O(\log n)$ assuming $d_{\rm c}= O(\poly(n))$.


Note that although  we considered  MPSs with open boundary conditions, our memory bound also applies   to  other cases.    For example, it applies to MPS with periodic conditions (\ref{eq:mps_LR_periodic}).
More generally, the bound holds for any  set of states $\{\ket{\Psi_B}\}$ of the form $\ket{\Psi_{B}} = N_*\ket{B}$, where $\ket{B} \in \map{H}_{{\rm c}}\otimes \map{H}_{{\rm c}}$ is a generic vector on the joint Hilbert space of the boundary conditions.  

In Section  \ref{sec:mps_scheme} we will provide an explicit compression protocol that achieves the memory bound  $\lceil 2\log d_{\rm c}\rceil $ and can be implemented efficiently on a quantum computer.


\subsection{
Site-independent MPS}\label{ss:SIMPS}

Another important family of MPSs is the family of site-independent MPSs \cite{perez2006matrix}.  Suppose that  we know nothing about an  MPS except that it is site-independent (\ref{eq:mps_LR}), has a constant bond dimension  $d_{\rm c}$, and has a constant physical dimension $d_{\rm p}$. A constant bond dimension can be observed in some systems where the interactions are local, such as the one-dimensional Ising model \cite{ising1925beitrag} and the AKLT model \cite{affleck2004rigorous}.  A generic state of this form can be expressed as
\begin{align} \label{eq:mps_LR_fixed}
    \ket{\Psi_{A,L,R}} = \sum_{i_1,\dots,i_n=1}^{d_{\rm p}} \bra{\overline{L}} A_{i_1}A_{i_2}\dots A_{i_n}\ket{R} ~ \ket{i_1,i_2,\dots,i_n} =  \begin{mathtikz}
    \node[parameter] (L) at (-1,0) {$L$};
    \node[parameter] (A1) at (0,0) {$A$} edge [<-, thick] (L) edge [->, thick] (0,0.75);
    \node[parameter] (A2) at (1,0) {$A$} edge [->, thick] (A1) edge [->, thick] (1,0.75);
    \node (dots) at (2,0) {$\cdots$} edge [->, thick] (A2);
    \node[parameter] (An) at (3,0) {$A$} edge [->, thick] (dots) edge [->, thick] (3,0.75);
    \node[parameter] (R) at (4,0) {$R$} edge [->, thick] (An);
\end{mathtikz}
\end{align}

Note that the entries of the tensor $A$ are  free parameters, like the entries of  the vectors $\ket{L}$ and $\ket{R}$.
We now provide an alternative tensor network representation of the state $|\Psi_{A,L,R}\>$.   To this purpose, we can convert the tensor $A$ into a vector  of dimension $d_{\rm c}^2d_{\rm p}$,  denoted by $\ket{A} \in \spc{H}_{d_{\rm c}^2d_{\rm p}}$. The vectorisation
is implemented by inserting  copies of the unnormalised maximally entangled state $\kket{I} := \sum_{i=1}^{d_{\rm c}} \ket{i}\ket{i}$, as in the following picture:
\begin{align} \label{eq:mpsall_compression1}
    \ket{\Psi_{A,L,R}} = \begin{mathtikz}
    \node[parameter] (L) at (-1,0) {$L$};
    \node[parameter] (A1) at (0,0) {$A$} edge [->, thick] (0,1.5);
    \node[parameter] (A2) at (1,0) {$A$} edge [->, thick] (1,1.5);
    \node[virtual] (dots) at (2,0) {$\cdots$};
    \node[parameter] (An) at (3,0) {$A$} edge [->, thick] (3,1.5);
    \node[parameter] (R) at (4,0) {$R$};
    \foreach \i in {1,2,3,4,5} \node[tensor] (P\i) at (-1.5+\i,0.75) {$I$};
    \draw[->, thick] (L) -| ($(P1.south)+(-0.125,0)$);
    \draw[->, thick] (A1) -| ($(P1.south)+(0.125,0)$);
    \draw[->, thick] (A1) -| ($(P2.south)+(-0.125,0)$);
    \draw[->, thick] (A2) -| ($(P2.south)+(0.125,0)$);
    \draw[->, thick] (A2) -| ($(P3.south)+(-0.125,0)$);
    \draw[->, thick] (dots) -| ($(P3.south)+(0.125,0)$);
    \draw[->, thick] (dots) -| ($(P4.south)+(-0.125,0)$);
    \draw[->, thick] (An) -| ($(P4.south)+(0.125,0)$);
    \draw[->, thick] (An) -| ($(P5.south)+(-0.125,0)$);
    \draw[->, thick] (R) -| ($(P5.south)+(0.125,0)$);
\end{mathtikz} = \bbra{I}^{\otimes (n+1)} ( \ket{L} \otimes \ket{A}^{\otimes n} \otimes \ket{R})
\end{align}

Now, note that the vector $\ket{A}^{\otimes n}$ belongs to  the symmetric subspace of $\spc{H}_{d_{\rm c}^2d_{\rm p}}^{\otimes n}$, which has dimension ${n+d_{\rm c}^2d_{\rm p}-1 \choose d_{\rm c}^2d_{\rm p}-1}$. Let $\map{S}$ be a  Hilbert space of dimension  ${n+d_{\rm c}^2d_{\rm p}-1 \choose d_{\rm c}^2d_{\rm p}-1}$. Then,  there exists an isometry $V_{\rm S}$ such that, for any $\ket{A}$, there exists a vector $\ket{S_A} \in \map{S}$ satisfying $\ket{A}^{\otimes n} = V_{\rm S}\ket{S_A}$. Using this fact, we can replace $\ket{A}^{\otimes n}$ by $V_{\rm S}\ket{S_A}$, thus obtaining a new tensor network representation with smaller minimum cut:
\begin{align} \label{eq:mpsall_compression2}
    \ket{\Psi_{A,L,R}} = \begin{mathtikz}
    \node[parameter] (S) at (-1.25,-0.75) {$S_A$};
    \node[parameter] (L) at (-1.25,0) {$L$};
    \node[virtual] (A1) at (0,-0.75) {} edge [->, thick] (0,1.5);
    \node[virtual] (A2) at (1,-0.75) {} edge [->, thick] (1,1.5);
    \node[virtual] (dots) at (2,0) {$\cdots$};
    \node[virtual] (An) at (3,-0.75) {} edge [->, thick] (3,1.5);
    \node[parameter] (R) at (4.25,0) {$R$};
    \node[tensor, minimum width = 40mm] (V) at (1.5,-0.75) {$V_{\rm S}$};
    \foreach \i in {1,2,3,4,5} \node[tensor] (P\i) at (-1.5+\i,0.75) {$I$};
    \draw[->, thick] (L) -| ($(P1.south)+(-0.125,0)$);
    \draw[<-, thick] ($(P1.south)+(0.125,0)$) -- +(0,-1);
    \draw[<-, thick] ($(P2.south)+(-0.125,0)$) -- +(0,-1);
    \draw[<-, thick] ($(P2.south)+(0.125,0)$) -- +(0,-1);
    \draw[<-, thick] ($(P3.south)+(-0.125,0)$) -- +(0,-1);
    \draw[<-, thick]  ($(P3.south)+(0.125,0)$) -- +(0,-1);
    \draw[<-, thick]  ($(P4.south)+(-0.125,0)$) -- +(0,-1);
    \draw[<-, thick] ($(P4.south)+(0.125,0)$) -- +(0,-1);
    \draw[<-, thick] ($(P5.south)+(-0.125,0)$) -- +(0,-1);
    \draw[->, thick] (R) -| ($(P5.south)+(0.125,0)$);
    \draw[->, thick] (S) -- (V);
    \draw[dashed, thick] ($(L)+(-0.5,0.375)$) -| ($(L)+(0.375,0.25)$) -- ($(S)+(0.375,-0.5)$) -| ($(R)+(-0.375,0.25)$) |- ($(R)+(0.5,0.375)$);
\end{mathtikz} = \bbra{I}^{\otimes (n+1)} (\ket{L} \otimes V_{\rm S}\ket{S_A} \otimes \ket{R})
\end{align}

For sufficiently large $n$, the minimum cut  is illustrated by  the dashed line in Equation (\ref{eq:mpsall_compression2}). The cut edges are the outgoing edges of $L$, $R$ and $S_A$,  and their  combined dimension is $\dim\spc{H}_C = d_{\rm c}^2{n+d_{\rm c}^2d_{\rm p}-1 \choose d_{\rm c}^2d_{\rm p}-1}$. Hence, Theorem  \ref{thm:tncompression} implies that the total number of qubits to encode the states $\{\ket{\Psi_{A,L,R}}\}$ is  $\left\lceil \log \dim\spc{H}_C \right\rceil \leq \left\lceil (d_{\rm c}^2d_{\rm p}-1) \log (n+d_{\rm c}^2d_{\rm p}-1) + 2\log d_{\rm c} \right\rceil$.



\subsection{Memory bounds for other families of tensor network states}\label{ss:summary}

In  Appendix \ref{ss:families} we derive memory bounds for several families of tensor network states, including PEPS  (either with variable boundary conditions or site-independent) and MPS/PEPS  generated by an unknown unitary gate acting identically on  the physical particles, a scenario that is relevant to the use of MPS/PEPS in quantum metrology. The results  of Appendix \ref{ss:families} are summarised in Table \ref{tab:main}.

\preto\tabular{\setcounter{magicrownumbers}{0}}
\newcounter{magicrownumbers}
\newcommand\rownumber{\refstepcounter{magicrownumbers}\arabic{magicrownumbers}}
\renewcommand{\arraystretch}{1.5}
\begin{table}[H]
    \caption{\label{tab:main}Memory bounds for tensor network state families.}
    \begin{tabular}{|c|>{\centering}m{6cm}|c|c|c|}
        \hline
        Case & State & Expression & Parameters& Memory (qubits, rounding up)\\
        \hline
        \rownumber\label{row:MPS_LR} & MPSs with variable boundary conditions & $\ket{\Psi_{L,R}}$ (\ref{eq:mps_LR_graph_2}) & $L,R$ & $2\log d_{\rm c}$\\
        \hline
        \rownumber\label{row:MPS} & Site-independent MPSs  & $\ket{\Psi_{A,L,R}}$ (\ref{eq:mps_LR_fixed}) & $A,L,R$ & $(d_{\rm c}^2d_{\rm p}-1)\log (n+d_{\rm c}^2d_{\rm p}-1) + 2\log d_{\rm c}$\\
        \hline
        \rownumber\label{row:PEPS_B} & $n\times m$ PEPSs  with variable boundary condition & $\sket{\Psi_{B}^{\rm (PEPS)}}$ (Fig. \ref{fig:peps2})& $B$ & $(2n+2m)\log d_{\rm c}$\\
        \hline
        \rownumber\label{row:PEPS} & Site-independent  $n\times m$ PEPSs & $\sket{\Psi_{A,B}^{\rm (PEPS)}}$ (Fig. \ref{fig:peps}) & $A,B$ & \begin{minipage}{6cm}\centering\vspace{0.2em}$(d_{\rm c}^4d_{\rm p}-1)\log (nm+d_{\rm c}^4d_{\rm p}-1)$\\$ + (2n+2m)\log d_{\rm c}$\vspace{0.2em}\end{minipage}\\
        \hline
        \rownumber\label{row:Ug} & Fixed $n$-system state under $U_g^{\otimes n}$ &$U_g^{\otimes n} \ket{\Psi_0}$ & $g$ & \begin{minipage}{6cm}\centering\vspace{0.1em}$\frac{d_{\rm p}^2+d_{\rm p}-2}{2}\log(n+d_{\rm p}-1)$\vspace{0.1em}\end{minipage}\\
        \hline
        \rownumber\label{row:UgTNS} & Tensor network state under $U_g^{\otimes n}$ & $U_g^{\otimes n} N_* \ket{v_x}$ & $g,x$ & \begin{minipage}{6cm}\vspace{0.2em}\begin{minipage}{6cm}\centering$\frac{d_{\rm p}^2+d_{\rm p}-2}{2}\log(n+d_{\rm p}-1) + \mc(\widetilde{{\sf{Temp}}})$\end{minipage}\\ where $\widetilde{\sf{Temp}}$ is the flow network associated with $N$ \vspace{0.2em}\end{minipage}\\
        \hline
        \rownumber\label{row:UgMPS} & MPSs with variable boundary conditions under $U_g^{\otimes n}$ & $U_g^{\otimes n} \ket{\Psi_{L,R}}$ & $L,R,g$ & $\frac{d_{\rm p}^2+d_{\rm p}-2}{2}\log(n+d_{\rm p}-1) + 2\log d_{\rm c}$\\
        \hline
        \rownumber\label{row:UgPEPS} & $n\times m$ PEPSs with variable boundary condition under $U_g^{\otimes nm}$ & $U_g^{\otimes nm}\sket{\Psi_{B}^{\rm (PEPS)}}$ & $B,g$ & $\frac{d_{\rm p}^2+d_{\rm p}-2}{2}\log(nm+d_{\rm p}-1) + (2n+2m)\log d_{\rm c}$\\
        \hline
    \end{tabular}
\end{table}


Cases \ref{row:MPS_LR} and \ref{row:PEPS_B} deal with the compression of multipartite states where the tensors responsible for the  correlations  between particles are known, while the boundary condition is unknown. The scaling of the memory size manifests an area law: the number of qubits needed to encode the state is proportional to the size of the boundary.  The area law can be immediately read out from the graphical representation of the states, as Theorem \ref{thm:tncompression} states that  the memory size equals to the minimum cut between the system and the variable terms, and in this case the variable terms are only on the boundary.

Cases \ref{row:MPS} and \ref{row:PEPS} consider a site-independent multipartite system with fixed bond dimension. A logarithmic scaling can be observed: the memory size is $O(\log n)$ for a system of $n$ particles. The same scaling is also observed in the compression of identical uncorrelated systems \cite{yang2018compression}.
Cases \ref{row:Ug}, \ref{row:UgTNS}, \ref{row:UgMPS} and \ref{row:UgPEPS} exemplify a tensor network state under an unknown global transformation.
The total memory usage equals to a fixed term $\frac{d_{\rm p}^2+d_{\rm p}-2}{2}\log(n+d_{\rm p}-1)$ plus the memory for the tensor network state. The fixed term can be interpreted as the amount of information contained in the unknown transformation $U_g^{\otimes n}$.


\section{Local compression of bipartite  states}\label{sec:marginal}


In this section we  extend Theorem \ref{thm:tncompression} to the scenario where some of the physical systems are inaccessible, and the task is to compress the accessible part, while maintaining the correlations in the overall system.

Consider a family of pure states $\{\ket{\Psi_x}\} \subset \spc H_{\rm P} \otimes \map{H}_{\rm E}$ of the composite system ${\rm P}\otimes {\rm E}$, consisting of a physical system ${\rm P}$ (with Hilbert space $\spc H_{\rm P}$) and of its environment ${\rm E}$ (with Hilbert space $\map{H}_{\rm E}$).
Here we are interested in compression protocols where the encoding and decoding operations act only on system ${\rm P}$, but still allow one to recover the joint state $\ket{\Psi_x}$. Our goal is to find channels $\map{E}:S(\spc H_{\rm P}) \to S(\spc{H}_{\rm M})$ and $\map{D}:S(\spc{H}_{\rm M}) \to S(\spc H_{\rm P})$ that satisfy
\begin{align}\label{eq:marginal_goal}
    (\map{D}\circ\map{E} \otimes \map{I}_{\rm E})(\ketbra{\Psi_x}) = \ketbra{\Psi_x},~\forall x \in \set{X},
\end{align}
where $\map{I}_{\rm E}$ is the identity map on system ${\rm E}$. An optimal pair of channels $(\map E,\map D)$ is a pair that  minimises  the memory size, namely the dimension of $\spc{H}_{\rm M}$.

We call the above task  {\em local compression} of the states  $\{\ket{\Psi_x}\} \subset \spc H_{\rm P} \otimes \map{H}_{\rm E}$.  Operationally, local compression is important in the situation where Alice and Bob share a state of  the composite system ${\rm P}\otimes {\rm E}$, of which Alice holds part $\rm P$, while Bob holds part $\rm E$. In this scenario, it is interesting to ask how Alice can store  her part of her system in  a quantum memory, while ensuring that the correlations with Bob's system are preserved.


{

A local  compression protocol can be constructed from a partial isometry $V: \spc H_{\rm P} \to \spc{H}_{\rm M}$ that satisfies the following:
\begin{align}\label{eq:marginal_goal2}
    (V^\dag V \otimes I_{\rm E})\ket{\Psi_x} = \ket{\Psi_x},~\forall x \in \set{X} ,
\end{align}
where $I_{\rm E}$ is the identity operator on system $\rm E$. 
}

\begin{figure}[H]
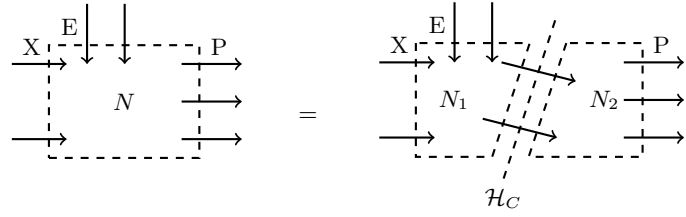

    $$\begin{array}{ccc}
        \begin{mathtikz}
            \node[virtual, minimum height=15mm] (x) at (0,0.5) {};
            \node[tensor, dashed, minimum height=15mm, minimum width=20mm] (N) at (1.75,0.5) {$N$};
            \draw[->, thick] ($(x.east)+(0,0.5)$) -- node[above,xshift=-1mm] {$\rm X$} ($(N.west)+(0.25,0.5)$);
            \draw[->, thick] ($(x.east)+(0,-0.5)$) -- ($(N.west)+(0.25,-0.5)$);
            \draw[->, thick] ($(N.east)+(-0.25,0.5)$) -- node[above,xshift=1mm] {${\rm P}$} +(0.8,0);
            \draw[->, thick] ($(N.east)+(-0.25,-0.5)$) -- +(0.8,0);
            \draw[->, thick] ($(N.east)+(-0.25,0)$) -- +(0.8,0);
            \node[virtual, minimum height=15mm, minimum width=10mm] (S) at (1.25,0.5) {};
            \draw[<-, thick] ($(S.east)+(0,0.5)$) -- +(0,0.8);
            \draw[<-, thick] ($(S.east)+(-0.5,0.5)$) -- node[left,yshift=1mm] {${\rm E}$} +(0,0.8);
            \node[virtual] (placeholder) at (3.5,-0.75) {};
        \end{mathtikz} & = & \begin{mathtikz}
            \node[virtual, minimum height=15mm] (x) at (0,0.5) {};
            \node[virtual, minimum height=15mm, minimum width=10mm] (S) at (1.25,0.5) {$N_1$};
            \draw[dashed, thick] (S.north west) -- (S.south west) -- (S.south east) -- ($(S.north east)+(0.5,0)$) -- (S.north west);
            \node[virtual, minimum height=15mm, minimum width=10mm] (T) at (3.25,0.5) {$N_2$};
            \draw[dashed, thick] (T.north west) -- ($(T.south west)-(0.5,0)$) -- (T.south east) -- (T.north east) -- (T.north west);
            \draw[->, thick] ($(x.east)+(0,0.5)$) --  node[above,xshift=-1mm] {$\rm X$} ($(S.west)+(0.25,0.5)$);
            \draw[->, thick] ($(x.east)+(0,-0.5)$) -- ($(S.west)+(0.25,-0.5)$);
            \draw[->, thick] ($(T.east)+(-0.25,0.5)$) -- node[above,xshift=1mm] {${\rm P}$} +(0.8,0);
            \draw[->, thick] ($(T.east)+(-0.25,-0.5)$) -- +(0.8,0);
            \draw[->, thick] ($(T.east)+(-0.25,0)$) -- +(0.8,0);
            \draw[->, thick] ($(S.east)+(0.125,0.5)$) -- ($(T.west)+(0.125,0.25)$);
            \draw[->, thick] ($(S.east)+(-0.125,-0.25)$) -- ($(T.west)+(-0.125,-0.5)$);
            \draw[<-, thick] ($(S.east)+(0,0.5)$) -- +(0,0.8);
            \draw[<-, thick] ($(S.east)+(-0.5,0.5)$) -- node[left,yshift=1mm] {${\rm E}$} +(0,0.8);
            \coordinate (start) at ($(S.north east)!0.75!(T.north west)$);
            \coordinate (end) at ($(S.south east)!0.25!(T.south west)$);
            \draw[dashed, thick] ($(start)!-0.2!(end)$) -- ($(start)!1.2!(end)$) node[below] (label) {$\map{H}_C$};
        \end{mathtikz}
    \end{array}
    $$
\caption{\label{fig:cor:marginal} 
Example of cut between the physical systems and the parameter space/environment.  The capacity of the cut limits  the information flow from both the parameter space and the environment to the physical systems.}
\end{figure}

In order to  generalize  Theorem \ref{thm:tncompression} to the scenario of local compression, we need to cope with the presence of the inaccessible environment  ${\rm E}$. The key idea is to regard the environment not as an output of the tensor network, but  as another source of information, in addition to the parameter Hilbert space in which  the parameter $x$ is encoded.   Mathematically, this change of perspective corresponds to a reversal of the edges associated to the environment, which become inputs, instead of outputs.   After the edges have been reversed, we apply Theorem \ref{thm:tncompression}, and search for the minimum cut that separates physical systems from the  parameter space  and  from the environment, as shown in Figure \ref{fig:cor:marginal}.

This approach leads to  an upper bound on the number of memory qubits needed for local compression,  provided in  the following proposition:

\begin{prop}\label{thm:marginal}
Let $\{\ket{\Psi_x}\}$ be a parametric family  of  pure states of the form $\ket{\Psi_x} = N_*\ket{v_x}$, where    $N_*: \spc{H}_{\set X} \to \map{H}_{\rm P} \otimes \map{H}_{\rm E}$ is a fixed tensor network operator and  $\ket{v_x} \in \spc{H}_{\set X}$ is some  (not necessarily normalised) vector, parameterised by parameter $x$. Let $N=  ({{\sf{Temp}}},  T)$ be the tensor network associated to the operator $N_*$, each of whose outgoing edges corresponds to either a physical system ({\em i.e.} a subsystem of ${\rm P}$) or a part of the environment ({\em i.e.} a subsystem of ${\rm E})$.
Let ${\sf{Temp}}'$ be the tensor network template obtained from reversing all open edges of ${\sf{Temp}}$ that are associated to the environment.
Let $\widetilde {{\sf{Temp}}'}$ be the flow network corresponding to the template ${\sf{Temp}}'$ via Construction \ref{cons:t2f}.
For every cut $C=(C_s,C_t)$ of $\widetilde {{\sf{Temp}}'}$, the marginal state on system ${\rm P}$ of the state family $\{\ket{\Psi_x}\}_{x \in \set{X}}$ can be compressed without errors into $\lceil c(C) \rceil$ qubits.
\end{prop}
The proof is provided in Appendix \ref{app:marginal}.

As an example, consider the scenario   where Alice holds the leftmost $n$ systems of an MPS and Bob holds the rest $n'$ systems, and Alice's task is to store her systems in a quantum memory, while preserving the correlations with Bob's systems.
\begin{align} \label{eq:mps_AB}
    \ket{\Psi_{L,R}} =  \begin{mathtikz}
    \node[parameter] (L) at (-1,0) {$L$};
    \node[tensor] (A1) at (0,0) {$A^{[1]}$} edge [<-, thick] (L) edge [->, thick] (0,0.75);
    \node (dots) at (1,0) {$\cdots$} edge [->, thick] (A1);
    \node[tensor] (A2) at (2,0) {$A^{[n]}$} edge [->, thick] (dots) edge [->, thick] (2,0.75);
    \node[tensor] (A3) at (3.25,0) {$A^{[n+1]}$} edge [->, thick] (A2) edge [->, thick] (3.25,0.75);
    \node (dots2) at (4.5,0) {$\cdots$} edge [->, thick] (A3);
    \node[tensor] (A4) at (5.75,0) {$A^{[n+n']}$} edge [->, thick] (dots2) edge [->, thick] (5.75,0.75);
    \node[parameter] (R) at (7,0) {$R$} edge [->, thick] (A4);
    \draw[thick, densely dotted] (2.5,-0.5) -- +(0,2);
    \node[virtual] (Alice) at (1,1.5) {Alice};
    \node[virtual] (Bob) at (4,1.5) {Bob};
    \coordinate (br) at (0,-0.375);
    \draw [decorate,decoration={brace,amplitude=10pt,mirror},xshift=0pt,yshift=-3pt]
        (A1.west |- br) -- (A2.east |- br) node [black,midway,yshift=-18pt]
        {$n$};
    \draw [decorate,decoration={brace,amplitude=10pt,mirror},xshift=0pt,yshift=-3pt]
        (A3.west |- br) -- (A4.east |- br) node [black,midway,yshift=-18pt]
        {$n'$};
\end{mathtikz}
\end{align}

We assume Alice does not know the boundary conditions $\ket{L}$ and $\ket{R}$, and therefore  her goal is to find channels $\map{E}: S(\spc H_{\rm P}) \to S(\spc{H}_{\rm M})$ and $\map{D}: S(\spc{H}_{\rm M}) \to S(\spc H_{\rm P})$ that satisfy
\begin{align}\label{eq:marginal_goal3}
    (\map{D}\circ\map{E} \otimes \map{I}_{\rm B})(\ketbra{\Psi_{L,R}}) = \ketbra{\Psi_{L,R}},~\forall \ket{L}, \ket{R}
\end{align}
for some Hilbert space $\spc{H}_{\rm M}$ whose dimension should be minimised.

\begin{figure}[H]
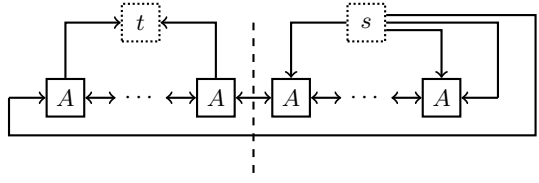

$$\begin{mathtikz}
    \node[tensor, densely dotted] (s) at (4,1) {$s$};
    \node[tensor, densely dotted] (t) at (1,1) {$t$};
    \coordinate (L) at (-0.75,0);
    \node[tensor] (A1) at (0,0) {$A$} edge [<-, thick] (L);
    \node (dots) at (1,0) {$\cdots$} edge [<->, thick] (A1);
    \node[tensor] (A2) at (2,0) {$A$} edge [<->, thick] (dots);
    \node[tensor] (A3) at (3,0) {$A$} edge [<->, thick] (A2);
    \node (dots2) at (4,0) {$\cdots$} edge [<->, thick] (A3);
    \node[tensor] (A4) at (5,0) {$A$} edge [<->, thick] (dots2);
    \coordinate (R) at (5.75,0) {} edge [->, thick] (A4);
    \draw[thick, dashed] (2.5,-1) -- +(0,2);
    \draw[->, thick] (A1) |- (t);
    \draw[->, thick] (A2) |- (t);
    \draw[<-, thick] (A3) |- (s);
    \draw[<-, thick] (A4) |- ($(s.east)+(0,-0.1)$);
    \draw[thick] (R) |- (s);
    \draw[thick] (L) |- ($(R)+(0.5,-0.5)$) |- ($(s.east)+(0,0.1)$);
\end{mathtikz}$$
\caption{\label{fig:ABMPS}Flow network for the compression of $\ket{\Psi_{L,R}}$. Except the edges connected to $s$ or $t$, all edges are made bidirectional. The dashed line indicates the cut.}
\end{figure}

Using Proposition \ref{thm:marginal}, we convert the tensor network in Equation (\ref{eq:mps_AB}) to into the flow network in Figure \ref{fig:ABMPS}, with the cut indicated by the dashed line. There are two cut edges, each has dimension $d_{\rm c}$, and therefore Alice can still compress her state into $\lceil\log\dim\map{H}_C\rceil = \lceil 2\log d_{\rm c}\rceil$ qubits, as in the case of compression of an MPS with variable boundary conditions (Section \ref{ss:mps}). The compression protocol will be presented explicitly in Section \ref{sec:mps_scheme}.

Proposition \ref{thm:marginal} automatically provide upper bounds on the amount of memory needed to compress mixed tensor network states.  Any such state  $\rho_x \in S(\spc H_{\rm P})$ can be regarded as the marginal of a pure tensor network state involving an environment, namely  $\Tr_{\rm E}[\ketbra{\Psi_x}] = \rho_x$ for some pure state $\ket{\Psi_x} \in \spc H_{\rm P} \otimes \map{H}_{\rm E}$.  Clearly, a local compression protocol for the purifications $\{|\Psi_x\>\}$ is also a compression protocol for the mixed states $\{\rho_x\}$, as one can see by taking the partial trace over the environment on both sides of the local compression condition  (\ref{eq:marginal_goal}).

\section{Compression protocol for matrix product states with variable boundary conditions} \label{sec:mps_scheme}


Here we construct an explicit  compression protocol for the family of all MPSs with variable boundary conditions. The idea is to perform a local compression on each pair of adjacent physical systems locally, and to iterate the protocol until we cannot reach a smaller memory size.
For simplicity of presentation, we assume the number of physical systems $n$ is a power of 2. We consider two adjacent physical systems, the $i$-th and the $(i+1)$-th, and regard the others as the environment.

\begin{figure}[H]
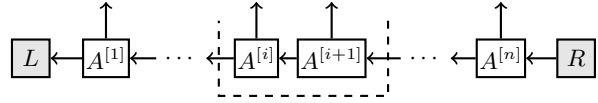

$$
    \begin{mathtikz}
    \node[parameter] (L) at (-1,0) {$L$};
    \node[tensor] (A1) at (0,0) {$A^{[1]}$} edge [->, thick] (L) edge [->, thick] (0,0.75);
    \node (dots) at (1,0) {$\cdots$} edge [->, thick] (A1);
    \node[tensor] (A2) at (2,0) {$A^{[i]}$} edge [->, thick] (dots) edge [->, thick] (2,0.75);
    \node[tensor] (A3) at (3,0) {$A^{[i+1]}$} edge [->, thick] (A2) edge [->, thick] (3,0.75);
    \node (dots2) at (4.25,0) {$\cdots$} edge [->, thick] (A3);
    \node[tensor] (A4) at (5.25,0) {$A^{[n]}$} edge [->, thick] (dots2) edge [->, thick] (5.25,0.75);
    \node[parameter] (R) at (6.25,0) {$R$} edge [->, thick] (A4);
    \draw[thick, dashed] (3.75,0.5) |- (1.5,-0.5) -- (1.5,0.5);
\end{mathtikz}
$$
\caption{\label{fig:cutAA}We regard the $i$-th and the $(i+1)$-th systems as the physical systems to which the compression protocol will be applied. The other systems are regarded as the environment.}
\end{figure}

By Proposition \ref{thm:marginal}, there exists a partial   isometry $V_{i,i+1}$ that faithfully encodes the $i$-th and the $(i+1)$-th physical systems into a single system of dimension $d_{\rm c}^2$.  Note that, {\em per se}, $V_{i,i+1}$ may not be a useful compression operation, because the dimension $d_{\rm c}^2$ may be larger than the dimension  $d_{\rm p}^2$ of the two input systems. Nevertheless, we now show that a concatenation of partial isometries like $V_{i,i+1}$ can squeeze the initial state into the minimum number of qubits, equal to  $\lceil 2 \log d_{\rm c}\rceil $.

Explicitly, the partial   isometry $V_{i,i+1}$ satisfies the local compression condition (\ref{eq:marginal_goal2}), which  reads
\begin{align}
(V_{i,i+1}^\dag V_{i,i+1} \otimes I_{1,\dots,i-1,i+2,\dots,n})\ket{\Psi_{L,R}} = \ket{\Psi_{L,R}} \, ,
\end{align}
where $I_{1,\dots,i-1,i+2,\dots,n}:=\bigotimes_{k=1}^{i-1}I_k \otimes\bigotimes_{k=i+2}^n I_k$ is the identity operator on all systems except the $i$-th and the $(i+1)$-th.   Applying the partial isometries $V_{1,2}$,  $V_{3,4}$, $...$, $V_{n-1,n}$ in parallel,  we obtain the relation
\begin{align} (V_{1,2}^\dag V_{1,2} \otimes V_{3,4}^\dag V_{3,4} \otimes \dots \otimes V_{n-1,n}^\dag V_{n-1,n})\ket{\Psi_{L,R}} = \ket{\Psi_{L,R}}  \qquad \forall |L\> , \, \forall |R\> \,.
\end{align}
This condition means that the product isometry  $V^{(1)}  :  = V_{1,2} \otimes V_{3,4} \otimes \dots \otimes V_{n-1,n}$ defines an exact  compression protocol that stores  $n$ systems (each of dimension $d_{\rm p}$) into $n/2$ systems (each of dimension $d_{\rm c}^2$).

The construction can be iterated, because the output of the isometry $V^{(1)}$  is itself an MPS. This can be verified by
 defining the tensors
\begin{align}
\begin{mathtikz}
    \node[tensor] (A1) at (0,0) {$A^{[i,i+1]}$};
    \draw[->, thick] (A1) -- +(0,0.75);
    \draw[->, thick] (A1) -- +(-1,0);
    \draw[<-, thick] (A1) -- +(1,0);
\end{mathtikz} := \begin{mathtikz}
    \node[tensor] (A0) at (0,0) {$A^{[i]}$};
    \node[tensor] (A1) at (1,0) {$A^{[i+1]}$};
    \node[tensor, minimum width=15mm] (V) at (0.5,1) {$V_{i,i+1}$};
    \draw[thick,->] (A1) -- (A0);
    \draw[thick,->] (A0) -- (A0 |- V.south);
    \draw[thick,->] (A1) -- (A1 |- V.south);
    \draw[thick,->] (A0) -- +(-0.75,0);
    \draw[thick,<-] (A1) -- +(0.75,0);
    \draw[thick,->] (V) -- +(0,0.75);
\end{mathtikz}\end{align}
so that the output state $ \ket{\Psi_{L,R}^{(2)}} := V^{(1)}\ket{\Psi_{L,R}}$ can be expressed in the MPS form
\begin{align}\ket{\Psi_{L,R}^{(2)}} = \begin{mathtikz}
    \node[parameter] (L) at (-1.25,0) {$L$};
    \node[tensor] (A1) at (0,0) {$A^{[1,2]}$} edge [->, thick] (L);
    \node[tensor] (A2) at (1.25,0) {$A^{[3,4]}$} edge [->, thick] (A1);
    \node (dots) at (2.5,0) {$\cdots$} edge [->, thick] (A2);
    \node[tensor] (An) at (3.75,0) {$A^{[n-1,n]}$} edge [->, thick] (dots);
    \node[parameter] (R) at (5,0) {$R$} edge [->, thick] (An);
    \foreach \i in {1,2,n} \draw[->, thick] (A\i) -- +(0,0.75);
\end{mathtikz}\end{align}
Crucially, the bond dimension is still $d_{\rm c}$.

Now,  we can again apply Proposition \ref{thm:marginal} to each adjacent pair of $d_{\rm c}^{2}$-dimensional systems, and compress them into a single $d_{\rm c}^2$-dimensional system, using  partial isometries $V_{1,4},V_{5,8},\dots,V_{n-3,n}$.
Also in this case,  Proposition \ref{thm:marginal} guarantees that the $(n/2)$-particle   $\ket{\Psi_{L,R}^{(2)}}$ is encoded faithfully into the $(n/4)$-particle  state $|\Psi_{L,R}^{(4)} \>  :  =  V^{(2)}  \ket{\Psi_{L,R}^{(2)}}$, $V^{(2)}:  = V_{1,4}\otimes V_{5,8}\otimes \cdots \otimes V_{n-3,n}$.

\begin{figure}[H]
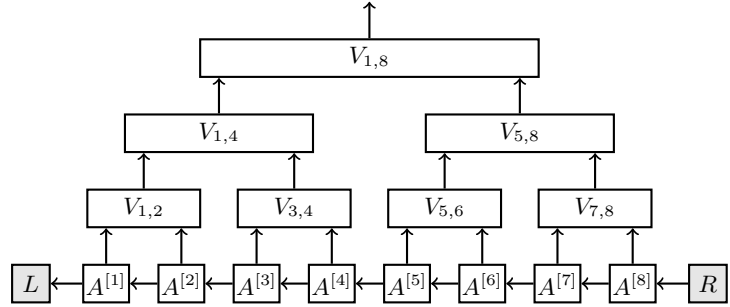

$$\begin{mathtikz}
    \node[parameter] (A0) at (0,0) {$L$};
    \foreach \i in {1,2,...,8}
        \node[tensor] (A\i) at (\i,0) {$A^{[\i]}$};
    \node[parameter] (A9) at (9,0) {$R$};
    \def\lasti{0};
    \foreach \i[remember=\i as \lasti] in {1,2,...,9}
        \draw[->,thick] (A\i) -- (A\lasti);
    \foreach \i[evaluate=\i as \ip using \i+1, evaluate=\i as \pos using \i+0.5] in {1,3,...,8}
        \node[tensor, minimum width=15mm] (A\i{}L1) at (\pos,1) {$V_{\i,\pgfmathprintnumber{\ip}}$};
    \foreach \i[evaluate=\i as \ip using \i+3, evaluate=\i as \pos using \i+1.5] in {1,5,...,8}
        \node[tensor, minimum width=25mm] (A\i{}L2) at (\pos,2) {$V_{\i,\pgfmathprintnumber{\ip}}$};
    \foreach \i[evaluate=\i as \ip using \i+7, evaluate=\i as \pos using \i+3.5] in {1}
        \node[tensor, minimum width=45mm] (A\i{}L3) at (\pos,3) {$V_{\i,\pgfmathprintnumber{\ip}}$};
    \draw[->,thick] (A1{}L3) -- +(0,0.75);
    \foreach \i in {1,2,...,8}
        \draw[->,thick] (A\i) -- (A\i|-A1{}L1.south);
    \foreach \i in {1,3,...,7}
        \draw[->,thick] (A\i{}L1) -- (A\i{}L1|-A1{}L2.south);
    \foreach \i in {1,5,...,7}
        \draw[->,thick] (A\i{}L2) -- (A\i{}L2|-A1{}L3.south);
\end{mathtikz}$$
\caption{\label{fig:MPS8}Tensor network for $\ket{\Psi_{L,R}^{(8)}}$, which is also the circuit structure to perform the encoding operation. Each partial isometry $V_{i,j}$ acts on a space with dimension no more than $\max\{d_{\rm c}^4,d_{\rm p}^2\}$ and outputs a system with dimension $d_{\rm c}^2$.}
\end{figure}

Iterating this pairwise encoding  for a total of $\log n$ times, we can faithfully compress the input state into $\lceil  2 \log d_{\rm c} \rceil$ qubits.   An illustration of the compression protocol for $n=8$ is provided in Figure \ref{fig:MPS8}.

The encoding can be realised by a quantum circuit of depth $O(\log n)$, implementing partial isometries $V_{i,j}$ shown above. 
Since each partial isometry has size no larger than $d_{\rm c}^2 \times \max\{d_{\rm c}^4,d_{\rm p}^2\}$ and the circuit uses $n-1$ such partial isometries in total, the overall complexity of the encoding operations is $O(\poly(n))$ (assuming $d_{\rm c}$ is polynomial in $n$), meaning that this construction is efficient in the number of physical systems.      The same argument applies to the decoding circuit, which can be obtained from the encoding circuit by reversing each gate.

Note that the above technique also applies to the local compression of  MPSs, corresponding to the scenario where only a subset of the physical system is accessible.

\section{Compression algorithm for pure states in low-dimensional subspaces}\label{sec:algorithm}

\subsection{The algorithm}

Here we outline a general quantum algorithm for compressing  families of pure states  lying in a low-dimensional subspace  of a high-dimensional quantum system.  The idea of the algorithm is  to train a quantum machine to perform the desired compression operations, by showing to the machine how such operations should act on a fiducial set of input states.

\begin{figure}[h]
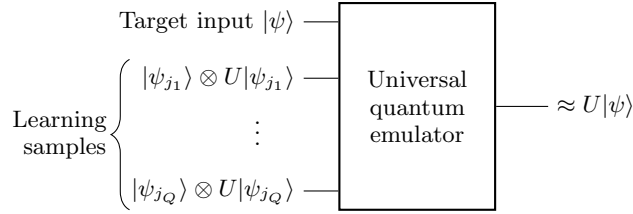

$$\begin{mathtikz}
    \node[text width=30mm,align=right] (I1) at (-1,0) {Target input $\sket{\psi}$};
    \node[text width=30mm,align=right] (I2) at (-1,-0.75) {\hfill$\ket{\psi_{j_1}}  \otimes U\ket{\psi_{j_1}}$};
    \node[align=right] (I3) at (0,-1.375) {$\vdots$};
    \node[text width=30mm,align=right] (I4) at (-1,-2.25) {\hfill$\ket{\psi_{j_Q}} \otimes   U\ket{\psi_{j_Q}}$};
    \node[tensor, minimum height=27.5mm] (E) at (2.125,-1.125) {\begin{minipage}{2cm}\centering Universal quantum emulator\end{minipage}};
    \foreach \i in {1,2,4} \draw (I\i) -- (I\i-|E.west);
    \node[align=left] (O) at (4.5,-1.125) {$\approx U\ket{\psi}$};
    \draw (E.east|-O) -- (O);
    \draw [decorate,decoration={brace,amplitude=7pt,mirror},xshift=0pt,yshift=0pt]
    (-1.7,-0.5) -- (-1.7,-2.5) node [black,midway,xshift=-30pt,text width=15mm,align=right]
    {Learning samples};
\end{mathtikz}$$
\caption{\label{fig:emulator} {\bf Universal quantum emulator.}  The machine learns how to approximately perform an unknown unitary gate  $U$  on a generic  input state $|\psi\>$ by consuming $Q$ pairs of input-output states.
 }
\end{figure}

 The algorithm is based on the {\em universal quantum emulator} of Marvian and Lloyd \cite{marvian2016universal}, a quantum circuit that  ``learns'' how to implement a completely unknown unitary gate $U$ from a  set of examples, as illustrated in Figure  \ref{fig:emulator}.    To implement the gate $U$ on a state $|\psi\>$,  the emulator  consumes  $Q$ pairs of input-output states, of the form  $   ( |\psi_{j_k}\>,    U |\psi_{j_k}\>  ) $ with  $k  \in  \{1,\dots, Q\}$. Each input state $|\psi_{j_k}\>$ is taken from a  set of $m$  possible inputs $\{ |\psi_j\> \}_{j=1}^m$, with $m\le Q$.

For large $Q$,  the output of the emulator converges to the desired output state $U|\psi\>$ provided that
 \begin{enumerate}
 \item  the input state $|\psi\>$ belongs to the subspace    $\spc K  :=\Span  \{ |\psi_j\> \}_{j=1}^m$, and
 \item  the quantum channel  $\map R  :  S  (\spc K)  \to S  (\spc K) $ defined by
 \begin{align}\label{mapR}
 \map R  (\rho)   :  =    \frac 1 m  \,       \sum_{j=1}^m  \big(   I_{\spc K}  -  2 |\psi_j\>\<\psi_j|\big)  \,  \rho \,  \big(   I_{\spc K}  - 2 |\psi_j\>\<\psi_j|\big)  \, ,  \qquad \forall \rho  \in  S (\spc K)
 \end{align}
where $I_{\spc K}$ is the identity on $\spc K$, is {\em mixing} \cite{burgarth2007generalized},  meaning that every input state $\rho$ converges to a fixed state  $\rho_0$ (in this particular case $\rho_0= I/d$) under a large number of repeated applications of the channel; in formula,   $\lim_{k\to \infty}  \map R^k  (\rho)  = \rho_0$.
\end{enumerate}
The second condition is equivalent to the statement that the channel $\map R$ has one and
 only one eigenvalue  on  the unit circle   \cite{burgarth2013ergodic}.  Since all the eigenvalues of a quantum channel are inside the unit circle, this implies that the eigenvalue $\lambda$ with the second largest modulus satisfies the condition $|\lambda  |<1$.  Equivalently, this means that   the spectral gap  $\gamma_{\map R}  :  =  1-  |\lambda|$ is non-zero.

The number of input-output pairs needed to  approximate  the gate $U$ depends on the error tolerance    $\varepsilon$, on the dimension of the subspace spanned by the input states $\{  |\psi_j\> \}$, denoted by $r$,  and on the spectral gap $\gamma_{\map R}$. Specifically, Marvian and Lloyd  \cite{marvian2016universal} show that $Q$ grows as
\begin{align}\label{Q}
    Q = O\Big(r^2\varepsilon^{-1}\gamma_{\map R}^{-2}\log^2(r\varepsilon^{-1})\Big) \, .
\end{align}


To emulate the encoding, which is an isometry instead of a unitary, we choose a unitary $U$  acting on the composite system ${\rm P} \otimes {\rm M}$,  and satisfying the condition
\begin{align}\label{U}
U \, (  |\Psi_x\>\otimes |W_0\> )  =   |\Psi_0\> \otimes V|\Psi_x\>  \, ,
\end{align}
where $|\Psi_0\>$  ($|W_0\>$) is a  fixed state of the physical (memory) system, and $V:  \spc H_{\rm in} \to  \spc H_{\rm M}$ is an isometry that encodes the input states into the memory system.     To train the emulator, we use input (output) states of the form
\begin{align}\ |\psi_j\>   =   |\Phi_j\> \otimes |W_0\> \, ,  \qquad   \Big(  U|\psi_j\>   =  |\Psi_0\>  \otimes V|\Phi_j\>  \Big) \, ,
\end{align}
where the states $\{ |\Phi_j\>\}_{j=1}^m$, hereafter called the {\em fiducial states},  span   the input subspace $\spc H_{\rm in}  := \Span \{|\Psi_x\>\}_{x\in\set X}$.
  In general, the states $\{ |\Phi_j\>\}_{j=1}^m$  may or may not be a subset of the state family we are trying to compress.

The isometry $V$ is constructed from the Gram matrix $G_{jk}   =   \<\Phi_j| \Phi_k\>$ via  the following procedure:
\begin{enumerate}
\item Compute the rank of $G$, denoted by  $r$, and set $\spc H_{\rm M}   = \C^r$. The calculation of the rank can be done, {\em e.g.} by diagonalising $G$ and counting the non-zero eigenvalues.  Note that $r $ is equal to the dimension of $\Span  \{|\Phi_j\>\}_{j=1}^m$, which, by construction, is equal to the dimension of the input subspace $\spc H_{\rm in}$.
\item Construct an $r  \times  m$ matrix $W$ such that $W^\dag W  =  G$.  This can be done by diagonalising  $G$ as $G=X\Lambda X^\dag$, setting $W=\sqrt{\Lambda}X^\dag$, and removing the zero rows from $W$.
\item For $j \in  \{ 1, \dots,  m\}$, define $ V  |\Phi_j\>   :  =   W  |j\>$ .
\end{enumerate}
The above definition is well-posed and uniquely determines the linear operator $V$ within the subspace spanned by the fiducial states $\{ |\Phi_j\>\}_{j=1}^m$.  Note that  $V$ is an isometry: indeed, for every vector $|\Psi\>  = \sum_{j}   \,  c_j  |\Phi_j\>  \in  \spc H_{\rm in}$, one has
\begin{align}
\nonumber \left \|   V |\Psi\>  \right\|^2  &  = \sum_{j,k}  \,  \overline c_{j}  c_k  \,  \<\Phi_j |  V^\dag V |   \Phi_k\>   \\
\nonumber & = \sum_{j,k} \,  \overline c_{j}  c_k  \,  \< j|  W^\dag W   |k  \>  \\
\nonumber & = \sum_{j,k} \,  \overline c_{j}  c_k  \,  \< j|  G   |k  \>  \\
\nonumber & = \sum_{j,k} \,  \overline c_{j}  c_k  \,  \< \Phi_j| \Phi_k  \>  \\
&  =  \| |  \Psi\>\|^2  \, .
\end{align}

To train the emulator, we will use input-output pairs of the form $(|\Phi_j\>\otimes |W_0\>  ,  |\Psi_0\>\otimes |W_j\>)$, with
$|W_j\> :  =  W|j\> $.  Now, recall that the number of input-output pairs needed by the emulator depends on the spectral gap of the channel $\map R$ in Equation (\ref{mapR}), with states   $|\psi_j\> =  |\Psi_j\>\otimes |W_0\>$.   Since  the states $  |\Psi_j\>\otimes |W_0\> $ are unitarily equivalent to the states $ |\Psi_0\>  \otimes |W_j\>$,
the spectral gap of the channel $\map R  :  S (\spc H_{\rm P}) \to S (\spc H_{\rm P})$ is equal to the spectral gap of the channel $\map R': S (\spc H_{\rm M}) \to S (\spc H_{\rm M})$  defined by
\begin{align}\label{mapRprime}
\map R'  (\rho)   :  =    \frac 1 m  \,       \sum_{j=1}^m  \big(   I_{r}  -  2|W_j\>\<W_j|\big)  \,  \rho \,  \big(   I_{r}  -  2|W_j\>\<W_j|\big)  \, ,  \qquad \forall \rho  \in  S (\C^r)   \, ,
\end{align}
where $I_r$ is the identity operator on $\C^r$.  This observation is important because channel $\map R'$ acts only on the memory space, and therefore its spectral gap  involves the diagonalisation of a low-dimensional  matrix.   With the knowledge of the spectral gap, we can keep under control the error in the emulator protocol, and determine how many input-output pairs are needed to attain the desired level of accuracy in the implementation of the gate $U$.

All together, the algorithm can be summarised as follows:

\begin{algorithm}[H]
    \Indentp{0.5em}
    \SetInd{0.5em}{1em}
    \SetKwInput{Preprocessing}{Preprocessing}
    \SetKwInput{Input}{Input}
    \SetKwInput{Output}{Output}
    \SetKwInput{Emulation}{Emulation}
    \SetKwFor{Whenever}{whenever}{do}{end}
    \Input{Quantum state to be compressed $\ket{\Psi} \in \spc{H}_{\rm P}$ and classical description of fiducial set $\{  |\Phi_j\>\}_{j=1}^m$ }
    \Output{Approximation of the compressed state $V|\Psi\> \in \spc{H}_{\rm M}$}
      \BlankLine
    \Preprocessing{}
Compute the Gram matrix  $G_{jk} = \sbraket{\Phi_{j}}{\Phi_{k}}$\;\label{step:gram} 
Compute the rank $r = \rank(G)$ and set $\spc H_{\rm M}  =  \C^r$\; \label{step:rank}
  Find an $r \times m$ matrix $W$ satisfying $W^\dag W = G$\; \label{step:W}
  Compute the spectral gap of the channel $\map R$ in Equation (\ref{mapR}).  \label{step:gap}

    \BlankLine
  \Emulation{}
    Run the universal quantum emulator with input state $\ket{\Psi}\otimes\ket{W_0}$\;
    \ForEach{emulator's request for the $j$-th input-output pair}{
        Prepare $\ket{\Phi_{j}}$\;\label{step:preparepsi}
        Prepare $\ket{W_j}  :=  W|j\>$\;\label{step:preparew}
        Prepare $\ket{\Psi_0}$ and $\ket{W_0}$\;
        Input the pair $(\ket{\Phi_{j}}\otimes\ket{W_0}, \ket{\Psi_0}\otimes\ket{W_j})$ into the emulator\;\label{step:pair}
    }
    Discard the first system of  the emulator's output.
    \caption{Encoding operation for state family $\{\ket{\Psi_x}\}$}\label{alg:encoding}
\end{algorithm}

The above algorithm implements  an approximation of the encoding channel  $\map E  (\rho)   =  \Tr_{\rm P}  [  U (\rho \otimes |W_0\>\<W_0|)  U^\dag]$ to any desired  accuracy.    The same construction applies to the decoding channel  $\map D (\rho)   =  \Tr_{\rm M}  [  U^\dag (|\Psi_0\>\<\Psi_0|   \otimes  \rho )  U]$,  by simply exchanging the role of the input and output of the quantum emulator. The algorithm reaches the optimal memory size for exact compression, because the memory space has dimension $r$, which is exactly equal to the dimension of the input subspace $\spc{H}_{\rm in}$.

\subsection{Running time}

Here we analyse the running time of the general quantum  compression algorithm, providing sufficient conditions for its efficient implementation.   We will  measure the size of the input physical system $\rm P$ in terms of the number of logical qubits needed to represented it, namely $  n  :  =    \lceil  \log \dim \spc{H}_{\rm P}  \rceil$.

The running time of Step \ref{step:gram} (calculation of the Gram matrix) depends on the structure of the fiducial states. The calculation of the matrix element $G_{jk} =  \<\Phi_j|\Phi_k\>$   can be implemented efficiently  for various families of tensor network states, such as MPSs \cite{cozzini2007quantum,verstraete2008matrix,cirac2009renormalization,eisert2013entanglement} and MERAs \cite{vidal2008class}.  The   number of matrix elements  is $O(m^2)$, where $m$ is the size of the fiducial set.   Hence, the efficient implementation of Step \ref{step:gram} requires  $m$ to  be at most polynomial in $n$.  In the following, we will always assume  $m=  O(\poly(n))$.   Of course, this implies that the subspace containing the input states has polynomial  dimension $d_{\rm in}  =  O(\poly(n))$, namely it is exponentially smaller than the total Hilbert space  $\spc H_{\rm P}$.


Under the assumption  $m=  O(\poly(n))$, Steps \ref{step:rank} and \ref{step:W} (calculation of the rank and construction of the matrix $W$) can be implemented in polynomial time by diagonalising the Gram matrix $G$, {\em e.g.} with  the QR algorithm \cite{demmel1997applied}.  Note that one has $r   =  d_{\rm in}   = O( \poly (n))$, meaning that  the memory system has polynomial dimension.

Step \ref{step:gap}, the calculation of the spectral gap, can be implemented in polynomial time by diagonalising the $r^2\times r^2$ matrix describing the channel $\map R'$.

The emulation part has running  time $T_{\rm tot}   =   Q \, T_{\rm prep}+   T_{\rm emulator}$, where $Q$ is  the number of input-output pairs used by the emulator,  $T_{\rm prep}$ is  the time complexity of preparing each input-output pair,  and $T_{\rm emulator}$ is the running time of the emulator.

The running time  of the emulator  can be bounded as $T_{\rm emulator} = O(n \, Q\log Q  )$ \cite{marvian2016universal}. The complexity of preparing the input-output pair  $(|\Phi_j\>\otimes |W_0\>  ,  |\Psi_0\>\otimes |W_j\>)$ is essentially the complexity of preparing the fiducial state $|\Phi_j\>$.  Indeed, $|\Psi_0\>$ can be chosen to be any efficiently preparable state, {\em i.e.} any state preparable in $O(\poly (n))$ time. The states $|W_j\>$ and $|W_0\>$ are efficiently preparable by construction,  because they are vectors with  a polynomial number of efficiently computable entries. Hence, the preparation time $T_{\rm prep}$  is polynomial if and only if each fiducial state  $|\Phi_j\>$ can be prepared in polynomial time.
This condition is satisfied whenever the fiducial states $\{  |\Phi_j\>\}_{j=1}^m$ are MPSs \cite{cirac2009renormalization,eisert2013entanglement} or MERAs \cite{vidal2008class}.

By Equation (\ref{Q}), under the assumption $m=  O(\poly (n))$, the    number of input-output examples required by the emulator is polynomial in $n$ and $\varepsilon^{-1}$ if 
the inverse spectral gap $\gamma_{\map R}^{-1}$ is at most polynomial in $n$.

 In summary, the compression operations can be implemented in polynomial time if 
 \begin{enumerate}
 \item \label{cond:fid} the number of fiducial states is at most polynomial, $m=  O(\poly (n)))$,
 \item \label{cond:overlap} the overlap of any two fiducial states can be computed in polynomial time,
 \item \label{cond:prepare} each fiducial state can be prepared in polynomial time, and
 \item \label{cond:gap} the inverse spectral gap is at most polynomial, $\gamma_{\map R}^{-1}  =  O(\poly(n))$.
 \end{enumerate}

 Conditions \ref{cond:fid} and \ref{cond:overlap} are relatively straightforward for tensor network states. As we have seen in Section \ref{sec:examples}, many families of tensor network states are contained in subspaces of dimension $O(\poly(n))$, making it easy to satisfy Condition \ref{cond:fid}.
 In addition, the overlap between two tensor network states can be efficiently computed in many physically relevant cases ({\em e.g.}  MPSs and MERAs).  In those cases, if the input subspace $\spc{H}_{\rm in}$ has polynomial dimension (as it must be in order to satisfy Condition \ref{cond:fid}),  then every state in $\spc{H}_{\rm in}$ is a linear combination of polynomial number of tensor network states, and the  overlap between any two states in $\spc{H}_{\rm in}$ can be computed in polynomial time.

Condition \ref{cond:prepare} is satisfied when the fiducial states are efficiently preparable tensor network states, such as MPSs or MERAs, which can be prepared through  sequences of isometries \cite{cirac2009renormalization,eisert2013entanglement,vidal2008class}.  However, it is not automatically  satisfied when the fiducial states are generic vectors in $\spc H_{\rm in}$.   The problem is that, in general,  a linear combination of efficiently preparable states may not be  an efficiently preparable state.
For MPSs, however, this condition is satisfied:

\begin{lem}\label{lem:MPSspan}
Let $\{ |\Sigma_k\>\}_{k=1}^t$ be a polynomial-size set of MPSs that span the input space $\spc H_{\rm in}$, and let  $\{c_k\}_{k=1}^t$ be a set of coefficients such that the linear combination  $|\Psi\>  =  \sum_k c_k\,  |\Sigma_k\>$ is a unit vector.  If the initial states have bond dimension $d_{\rm c}$, then the state $|\Psi\>$  is an  MPS with bond dimension $t\,  d_{\rm c}$ and can be prepared in polynomial time.
\end{lem}
The proof is provided in Appendix \ref{app:basis}.
For more general families of tensor network states, other than MPSs, a sufficient condition for the efficient preparability of the fiducial states will be given in the next section.

Finally,   Condition \ref{cond:gap} can be satisfied by a suitable choice of fiducial states, as we show in the following.

 In general, Condition \ref{cond:gap} is satisfied  by  choosing the fiducial set to be ``sufficiently dense'' in the input subspace.  An example of such choice is provided in the following.
   Let $\{|1\>,\dots, | r\>\}$ be a fixed basis for the input subspace $\spc H_{\rm in}$.    For every $l \in  \{ 1,\dots,  r\} $, we define
\begin{align}\label{vectors}
|\Psi_{l, x, \pm} \> :  =  \frac {  |l\>  \pm  |l\oplus 1\>}{\sqrt 2} \, , \quad |\Psi_{l,y,\pm}\>:  =   \frac {  |l\>  \pm i |l\oplus 1\>}{\sqrt 2} , \quad |\Psi_{l,z,+}\>:  =
|l\> ,  \quad |\Psi_{l,z,-}\>  :  = |l\oplus 1\> \, ,\end{align}
where $\oplus$ denotes addition modulo $r$, and we adopt the convention $|0\>:  = |r\>$.
\begin{lem}\label{prop:gap}
The spectral gap of the  channel $\map{R}$ in Equation (\ref{mapR}) associated to the states $\{  |\Psi_{l, \alpha , s} \>\}_{  l\in  \{1,\dots,  r\},  \alpha  \in  \{x,y,z\},  s\in  \{ +, -\}}$  in Equation (\ref{vectors})   is $\gamma_{\map R} = 8 [\sin   (\pi/r) ]^2  / (3 r) $.
\end{lem}

The proof is provided in Appendix \ref{app:gap}.    Lemma \ref{prop:gap} guarantees that the fiducial set $\{  |\Psi_{l, \alpha , s} \>\}_{  l\in  \{1,\dots,  r\},  \alpha  \in  \{x,y,z\},  s\in  \{ +, -\}}$ gives rise to a channel with inverse spectral gap growing at most as $O(r^3)$,    where $r$ is the dimension of the input subspace. Since the input subspace is assumed to be of polynomial dimension (Condition \ref{cond:fid}), this result guarantees that the inverse spectral gap is at most polynomial.

Lemmas \ref{lem:MPSspan} and \ref{prop:gap} imply that every family of MPSs that can be compressed into a logarithmic number of  qubits can be compressed in polynomial time  on a quantum computer.
\begin{theo}\label{theo:MPS}
Let   $\{  |\Psi_x\>\}_{x\in  \set X}$ be a parametric family of  $n$-particle MPSs with fixed  bond dimension $d_{\rm c}$.   If the input subspace  $\spc H_{\rm in}  =  \Span\{  |\Psi_x\>\}_{x\in  \set X}$ has polynomial dimension $r  =  \poly  (n)$, then the states $\{  |\Psi_x\>\}_{x\in  \set X}$ can be compressed into $\lceil  \log r \rceil$ qubits with error $\varepsilon$ in  polynomial time  $\poly (n,  \varepsilon^{-1})$.
\end{theo}

\Proof   ~Let $\{  |\Sigma_k\>\}_{k=1}^t$ be a subset of the states $\{  |\Psi_x\>\}_{x\in  \set X}$, with the properties that  {\em (i)} $\{  |\Sigma_k\>\}_{k=1}^t$ spans the input subspace, and {\em (ii)} the number of states $t$ is polynomial in $n$.  Such a set exists because, by hypothesis, $\spc H_{\rm in}$ has polynomial dimension. Then, let $\{  |l\>  \}_{l=1}^r$ be the orthonormal basis of $\spc H_{\rm in}$ obtained by applying the Gram-Schmidt procedure to the set $\{  |\Sigma_k\>\}_{k=1}^t$.  Explicitly,  $|1\> : = |\Sigma_1\>$,  $  |2\>    :  =  (  |\Sigma_2\>    -   \<  \Sigma_1|\Sigma_2\> \, |\Sigma_1\>  )/\sqrt{  1-  |\<\Sigma_1|\Sigma_2\>|^2}\, , \dots,  |t\>   :  =  (|\Sigma_t\>   -  \sum_{k<t}  \,  \<  k  |\Sigma_t\> \,  |k\>)/ \sqrt{1- \sum_{k<t}   |\<k|  \Sigma_t\>|^2}$.  By construction, each vector $|l\>$ is a linear combination of MPSs, and the expansion coefficients can be computed from  the scalar products $\< \Sigma_k|\Sigma_l\>$.  Since the states  $\{  |\Sigma_k\>\}_{k=1}^t$  are MPSs, the scalar products can be computed efficiently, and Lemma \ref{lem:MPSspan} implies that the linear combinations $\{  |l\>\}_{l=1}^r$ can be prepared in polynomial time.     From the basis $\{  |l\>\}_{l=1}^r$, one can then construct the fiducial states defined in Equation (\ref{vectors}).  Since the fiducial states are linear combinations of at most 2 basis vectors, they can all be prepared in polynomial time (again, due to Lemma \ref{lem:MPSspan}).  Moreover, Lemma  \ref{prop:gap} guarantees that the  channel   $\map{R}$ associated to the states (\ref{vectors})  has inverse spectral gap of polynomial size. Hence, all Conditions
 \ref{cond:fid},  \ref{cond:overlap}, \ref{cond:prepare}, and \ref{cond:gap} are satisfied, implying that the compression algorithm \ref{alg:encoding} runs in polynomial time on the states $\{  |\Psi_x\>\}_{x\in  \set X}$. \qed

\medskip

Theorem \ref{theo:MPS} guarantees that most relevant families of MPSs can be compressed efficiently on a quantum computer.
For other state families,  a sufficient condition for compressibility in polynomial time is given by the following proposition:

 \begin{prop}\label{prop:efficientlyprep}
  Let   $\{  |\Psi_x\>\}_{x\in  \set X}$ be a parametric family of $n$-particle tensor network states with a given network template.
 If the input subspace $\spc H_{\rm in} = \Span \{\ket{\Psi_x}\}_{x\in \set X}$ contains a spanning set  $\{ |\Sigma_k\>  \}_{k=1}^t$ with the following properties
 \begin{enumerate}[label=(\roman*)]
\item \label{cond:spol}  the number of states $t$ is at most polynomial in $n$,
\item \label{cond:sprep} each state $|\Sigma_k\>$ is efficiently preparable by a coherent process $\ket{k} \mapsto |k\> \otimes   |\Sigma_k\>$,
\item \label{cond:sgram} the  Gram matrix   $S_{kl}  := \<\Sigma_k|\Sigma_l\>$  is efficiently computable,
\item  \label{cond:seig}  the minimum non-zero eigenvalue of the  Gram matrix   $S$ is at least inverse-polynomial in $n$,
\end{enumerate}
 then the states   $\{\ket{\Psi_x}\}_{x\in \set X}$  can be compressed  into
$\lceil \log \dim \spc{H}_{\rm in} \rceil$ qubits with error $\varepsilon$, using a quantum algorithm that runs in  polynomial time $O(\poly(n,\varepsilon^{-1}))$.
\end{prop}

 The proof is provided   in Appendix \ref{app:efficientlyprep}. Note that Proposition \ref{prop:efficientlyprep} is not specific to tensor network states, and applies broadly to every parametric family of states confined in a low-dimensional subspace of the total Hilbert space.

\section{Conclusions}\label{sec:conclusion}

We designed  compression protocols for parametric families of tensor network states, in which some of the tensors depend on the parameters, while some others are constant.   Physically, the variable tensors can be  associated to systems that carry  unknown parameters, or to  inaccessible degrees of freedom of the environment.   Given a tensor network with constant and variable tensors, one can construct a flow network, where the variable tensors are connected to the source, and the physical systems are connected to the sink.  In such a network, every cut identifies an exact deterministic compression protocol that compresses every state in the parametric family into a quantum memory of dimension equal to the size of the cut. In addition to quantifying the amount of memory needed to store tensor network states, we provided a general  quantum compression algorithm, and we identified sufficient conditions for the algorithm to run in polynomial time, showing that they are satisfied by all families of MPSs.

{
Our results can be applied to site-independent tensor network states of  $n$ quantum systems, showing that every such state can be compressed  without error into a memory of $O(\log n)$ qubits. This scaling is optimal, because the set of site-independent MPSs contains as a subset the set of all identically-prepared states, which is known to require $O(\log n)$ qubits, both for exact  \cite{yang2016efficient} and  approximate compression protocols \cite{yang2016optimal,yang2018compression}.
The  optimal prefactor in the logaritmic scaling of the memory for general site-independent tensor network states remains to be determined.

 Our results can also  be used to provide upper bounds on the  amount of information one  can encode into multipartite system using tensor network codes, such as the toric code \cite{kitaev1997quantum,kitaev2003fault,bravyi1998quantum} and holographic codes \cite{pastawski2015holographic,latorre2015holographic}. For example, our method shows that a toric code with circumference $L$  and with variable boundary conditions can be faithfully compressed into $L$ qubits (a toric code is a PEPS with $d_{\rm c}=2$ \cite{verstraete2006criticality}). As a consequence, we can deduce that the number of qubits one can encode with the toric code is at most $L$. This result is consistent with the construction by  Bravyi and Kitaev \cite{bravyi1998quantum}, which shows that one  can encode up to  $L/2-1$  qubits.  The discrepancy between this value and our bound is mostly due to that in Ref. \cite{bravyi1998quantum}, the boundary condition of a toric code is not arbitrary, while we consider  arbitrary boundary conditions. More generally, we showed that  tensor network states with variable boundary and constant interior satisfy an area law,  according to which the number of qubits needed to compress these  states is proportional to the size of the boundary.  }

In this work we mainly focused on  exact compression.   Since noise and imperfections are unavoidable in every realistic  implementation, an important avenue of future research is to extend our results to approximate compression protocols. 
Tolerating a small compression error could offer great savings in terms of the amount of memory needed to store families of tensor network states.
In the case of uncorrelated systems, it was observed  that tolerating any non-zero error decreases the memory size discontinuously \cite{yang2016efficient,yang2016optimal}.
  Extending the study of this phenomenon to correlated systems is an interesting open question for future research.


\medskip

	{\bf Acknowledgements.}  	   This work is supported by the  National Natural Science Foundation of China through grant 11675136, the Hong Kong Research Grant Council through Grant No. 17300317 and 17300918,   the HKU Seed Funding for Basic Research, the Foundational Questions Institute through grant FQXi-RFP3-1325,  the John Templeton Foundation through grant  60609, Quantum Causal Structures,  the Croucher Foundation, the Swiss National Science Foundation via the National Center for Competence in Research ``QSIT" as well as project No.\ 200020\_165843, and the ETH Pauli Center for Theoretical Studies.
    This publication was made possible through the support of a grant from the John Templeton Foundation. The opinions expressed in this publication are those of the
authors and do not necessarily reflect the views of the John Templeton Foundation.
	
\bibliography{mps}

\appendix

\section{Proof of Proposition \ref{thm:optimality2}} \label{app:optimality2}

For a generic tensor network template ${{\sf{Temp}}}  =  (G  ,  d,  V_{\rm filled})$,   define the new template  ${{\sf{Temp}}}_2= (G, d_2,V_{\rm filled})$ by setting $d_2 (e):  =  2^{  \lfloor \log d(e)  \rfloor}$, and let $C^*=(C^*_s,C^*_t)$ be  a minimum cut of the flow network $\widetilde {{\sf{Temp}}}_2$.

Then, we have the following chain of inequalities:
\begin{align}
   {\text {\sf min-cut}}  (\widetilde {\sf Temp})      &  \le    c   (C^*)  \\
\nonumber &    =        \sum_{u \in C^*_s, v\in C^*_t}   \log d(u,v)    \\
\nonumber & \le     \left(   \max_e \frac{ \log d (u,v)}{\lfloor \log d(u,v)  \rfloor}  \right)   \,        \sum_{u \in C^*_s, v\in C^*_t}   \lfloor \log d(u,v)  \rfloor   \\
\nonumber &\le  \left( \max_{n  \in \N}  \frac   {\log (n)}{\lfloor \log(n) \rfloor} \right)
  \,  \sum_{u \in C^*_s, v\in C^*_t} \lfloor \log d(u,v)  \rfloor \\
  \nonumber &  =   \log 3  \cdot   {\text{\sf min-cut}}  (\widetilde {{\sf{Temp}}}_2) \\
  \nonumber &=   \log 3  \cdot   \qmf(\widetilde {{\sf{Temp}}}_2) \\
  \nonumber &\le  \log 3  \cdot   \qmf(\widetilde {{\sf{Temp}}})  \, ,
 \end{align}
the  last inequality following  from the fact that the set of  tensor network operators with  template $\widetilde {{\sf{Temp}}}_2$ is included in the set of tensor network operators  with template  $\widetilde {{\sf{Temp}}}$. This proves the upper bound
  $   {\text {\sf min-cut}}  (\widetilde {\sf Temp})   \le  \log 3  \cdot   \qmf(\widetilde {{\sf{Temp}}}) $. \qed

\section{Memory bounds for various tensor network families}
\label{ss:families}
\subsection{PEPS with variable boundary conditions}\label{ss:PEPS}
\begin{figure}[H]
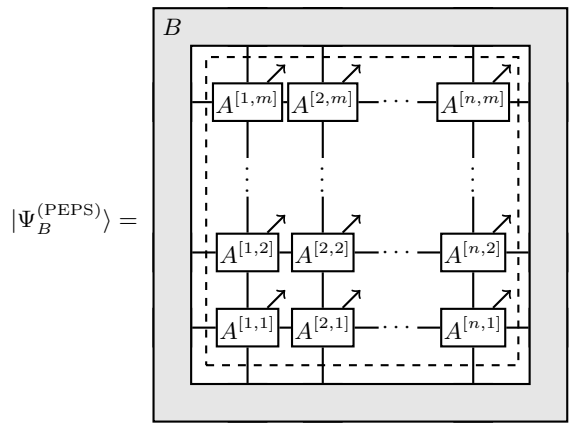

$$\sket{\Psi_{B}^{\rm (PEPS)}} = \begin{mathtikz}
    \def\ny{{"1","2","","n"}}
    \def\nx{{"1","2","","m"}}
    \node[virtual] (A22) at (2,2) {};
    \foreach \i in {0,1,3}
    {
        \foreach \j in {0,1,3}
        {
            \node[tensor] (A\i\j) at (\i,\j) {$A^{[\pgfmathparse{\ny[\i]}\pgfmathresult,\pgfmathparse{\nx[\j]}\pgfmathresult]}$};
            \draw[->, thick] (A\i\j) -- +(0.5,0.5);
        }
    }
    \foreach \i in {-1,0,1,3,4}
    {
        \node[virtual, inner sep=0pt] (A\i2) at (\i,2) {\rotatebox{90}{$\cdots$}};
        \node[virtual, inner sep=0pt] (A2\i) at (2,\i) {$\cdots$};
    }
    \foreach \i in {0,1,3}
    {
        \node[parameter] (B\i) at (\i,-1) {};
        \node[parameter] (T\i) at (\i,4) {};
        \draw[thick] (B\i) -- (A\i0) -- (A\i1) -- (A\i2) -- (A\i3) -- (T\i);
        \node[parameter] (L\i) at (-1,\i) {};
        \node[parameter] (R\i) at (4,\i) {};
        \draw[thick] (L\i) -- (A0\i) -- (A1\i) -- (A2\i) -- (A3\i) -- (R\i);
    }
    \draw[dashed, thick] (-0.55,-0.5) -| (3.6, 3.6) -| (-0.55, -0.5);
    \draw[parameter, even odd rule]
            (-1.25,-1.25) rectangle (4.25,4.25) (-0.75,-0.75) rectangle (3.75,3.75);
    \node[virtual] (B) at (-1,4) {$B$};
\end{mathtikz}$$
\caption{\label{fig:peps2} PEPS on a square lattice. The basic module $A^{[i,j]}$ is an order-5 tensor, while the tensor $B$ is a variable describing the boundary condition. The directions of non-open edges are omitted. The dashed line indicates the cut, and the edges crossing it are the cut edges.}
\end{figure}

PEPSs are a  higher dimensional analog of MPSs \cite{verstraete2004renormalization,verstraete2006criticality}. A PEPS is defined by a lattice of tensors, where each tensor has edges connected to its neighbors.
Consider a set of 2-dimensional  PEPSs in which all the tensors are fixed except those on the boundary, as in Figure \ref{fig:peps2}.
Here, each tensor $A^{[i,j]}$  in the figure is a fixed  order-5 tensor, and the shaded loop is a tensor $B$ describing a variable boundary condition.

As in the MPS case, we regard the tensors $A^{[i,j]}$ on the lattice as the tensor network $N$, and by properly choosing the edge directions, $N$ defines a linear operator from the systems on the boundary to the physical systems. We call this linear operator $N_*$, and write the PEPS as $\sket{\Psi_B^{\rm (PEPS)}} = N_*\ket{B}$, where $\ket{B}$ is a vectorised version of the tensor $B$ describing the boundary condition.   Then,  we convert $N$ to a flow network and look for its minimum cut.  Assuming that the bond dimension  $d_{\rm c}$ is a constant, while   the lattice has size $n\times m$ for large $n$ and $m$, {the optimal cut consists of the source (which replaces the tensor $B$) on one side, and of the sink and the tensors $A^{[i,j]}$ on the other side}, as shown in Figure \ref{fig:peps2}.  The cut edges contain $2n+2m$ number of $d_{\rm c}$-dimensional systems, with combined dimension $d_{\rm c}^{2n+2m}$.  Using Theorem \ref{thm:tncompression}, we conclude that the states $\{\sket{\Psi_B^{\rm (PEPS)}}\}$ can be compressed into $\lceil (2n+2m)\log d_{\rm c}\rceil$ qubits.

This result is consistent  with the area law for PEPSs, which indicates that the amount of information contained in a two-dimensional region  is upper bounded by a term proportional to its perimeter, in this case $2n+2m$. More generally, this  result is an instance of the bulk-boundary correspondence in Ref. \cite{cirac2011entanglement}, which shows that the bulk (namely physical systems) and boundary of a PEPS are related by an isometry.  Our result can be seen as a special case of the ``holographic compression'' of Ref. \cite{wilming2018single}, which states that a state with area law can be approximately compressed into a memory proportional to the boundary size. In the special case of PEPSs, our construction shows that the compression is exact.

\subsection{Site-independent PEPSs}\label{ss:TIPEPS}
\begin{figure}[H]
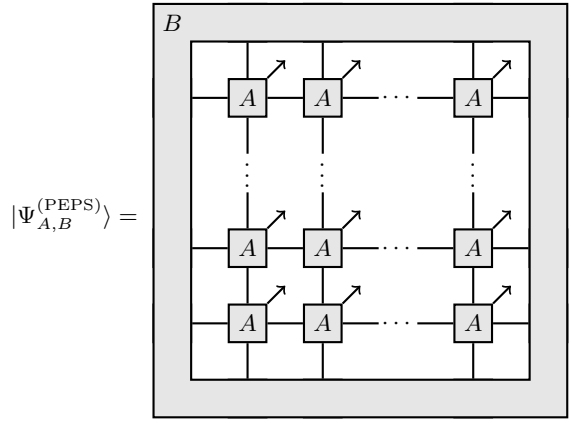

$$\sket{\Psi_{A,B}^{\rm (PEPS)}} = \begin{mathtikz}
    \node[virtual] (A22) at (2,2) {};
    \foreach \i in {0,1,3}
    {
        \foreach \j in {0,1,3}
        {
            \node[parameter] (A\i\j) at (\i,\j) {$A$};
            \draw[->, thick] (A\i\j) -- +(0.5,0.5);
        }
    }
    \foreach \i in {-1,0,1,3,4}
    {
        \node[virtual, inner sep=0pt] (A\i2) at (\i,2) {\rotatebox{90}{$\cdots$}};
        \node[virtual, inner sep=0pt] (A2\i) at (2,\i) {$\cdots$};
    }
    \foreach \i in {0,1,3}
    {
        \node[parameter] (B\i) at (\i,-1) {};
        \node[parameter] (T\i) at (\i,4) {};
        \draw[thick] (B\i) -- (A\i0) -- (A\i1) -- (A\i2) -- (A\i3) -- (T\i);
        \node[parameter] (L\i) at (-1,\i) {};
        \node[parameter] (R\i) at (4,\i) {};
        \draw[thick] (L\i) -- (A0\i) -- (A1\i) -- (A2\i) -- (A3\i) -- (R\i);
    }
    \draw[parameter, even odd rule]
            (-1.25,-1.25) rectangle (4.25,4.25) (-0.75,-0.75) rectangle (3.75,3.75);
    \node[virtual] (B) at (-1,4) {$B$};
\end{mathtikz}$$
\caption{\label{fig:peps} Site-independent PEPS on a square lattice. The basic module $A$ is an order-5 tensor, while the tensor $B$ describes the boundary condition. The directions of non-open edges are omitted.}
\end{figure}

The method in Section \ref{ss:SIMPS} for MPSs can be generalised to  arbitrary systems that are ``finitely correlated'',  in the sense that they have  a finite bond dimension.  
For example, Figure \ref{fig:peps} shows a site-independent PEPS defined on a square lattice.   Each $A$ in the figure is an order-5 tensor, and the   loop  is a tensor $B$ describing the boundary condition.

Now, suppose that the state  $|\Psi_{A,B}^{\rm (PEPS)}\>$ is a generic site-independent PEPS, defined  on a $n\times m$ square lattice,   with fixed bond dimension $d_{\rm c}$ and physical dimension $d_{\rm p}$.   By  vectorisation, the tensor $A$ is transformed into a vector $\ket{A}$  in a vector space of dimension $d_{\rm c}^4d_{\rm p}$. The dimension of the symmetric subspace of $nm$ copies of $d_{\rm c}^4d_{\rm p}$-dimensional vectors is ${nm+d_{\rm c}^4d_{\rm p}-1 \choose d_{\rm c}^4d_{\rm p}-1}$.
With the same argument as in the SIMPS case, $\ket{A}^{\otimes nm} = V_{\rm S}\ket{S_A}$, where $V_{\rm S}$ is an isometry, and $\ket{S_A}$ is a vector in space $\spc S$ with dimension ${nm+d_{\rm c}^4d_{\rm p}-1 \choose d_{\rm c}^4d_{\rm p}-1}$. We obtain the following tensor network representation for $|\Psi_{A,B}^{\rm (PEPS)}\>$:

\begin{align} \label{eq:pepsall_compression2}
    \ket{\Psi_{A,B}^{\rm (PEPS)}} = \begin{mathtikz}
    \node[parameter] (S) at (-1.5,-0.75) {$S_A$};
    \node[parameter, minimum width = 15mm] (B) at (-2,0) {$B$};
    \node[virtual] (A2) at (4,-0.75) {} edge [->, thick] (4,1.5);
    \node[virtual] (An) at (6,-0.75) {} edge [->, thick] (6,1.5);
    \node[tensor, minimum width = 70mm] (V) at (2.75,-0.75) {$V_{\rm S}$};
    \node[tensor, minimum width = 35mm] (P1) at (-1,1.25) {$I^{\otimes (2n+2m)}$};
    \node[tensor] (P2) at (2.25,1.25) {$I^{\otimes (2nm-n-m)}$};
    \draw[->, thick] ($(B.north)+(-0.5,0)$) -| ($(P1.south)+(-1.5,0)$);
    \draw[->, thick] ($(B.north)+(0.5,0)$) -| ($(P1.south)+(-0.5,0)$);
    \draw[<-, thick] ($(P1.south)+(0.5,0)$) -- +(0,-1.5);
    \draw[<-, thick] ($(P1.south)+(1.5,0)$) -- +(0,-1.5);
    \node[virtual] (dotsP11) at ($(P1.south)+(-1,-0.25)$) {$\cdots$};
    \node[virtual] (dotsP12) at ($(P1.south)+(1,-0.75)$) {$\cdots$};
    \draw[<-, thick] ($(P2.south)+(-0.75,0)$) -- +(0,-1.5);
    \node[virtual] (dotsP2) at ($(P2.south)+(0,-0.75)$) {$\cdots$};
    \draw[<-, thick] ($(P2.south)+(0.75,0)$) -- +(0,-1.5);
    \node[virtual] (dots) at (5,0.25) {$\cdots$};
    \draw[->, thick] (S) -- (V);
    \draw[dashed, thick] ($(B)+(-1.25,0.375)$) -| ($(S)+(0.5,-0.5)$);
\end{mathtikz} = \bbra{I}^{\otimes (2nm+n+m)} (\ket{B} \otimes V_{\rm S}\ket{S_A}) \,,
\end{align}
where we have rearranged the outgoing edges of $V_{\rm S}$ such that the tensor labeled $I^{\otimes (2n+2m)}$ denotes the connections between $V_{\rm S}$ and the boundary condition $B$, and $I^{\otimes (2nm-n-m)}$ corresponds to the connections between neighboring copies of $A$. The open edges correspond to the $nm$ number of physical systems.

$\ket{S_A}$ has dimension ${nm+d_{\rm c}^4d_{\rm p}-1 \choose d_{\rm c}^4d_{\rm p}-1} \leq (nm+d_{\rm c}^4 d_{\rm p}-1)^{d_{\rm c}^4 d_{\rm p}-1}$, and $\ket{B}$ has dimension $d_{\rm c}^{2n+2m}$. Applying   Theorem \ref{thm:tncompression} to  the cut  illustrated with the dashed line in Equation (\ref{eq:pepsall_compression2}), we obtain that the states $\sket{\Psi_{A,B}^{\rm (PEPS)}}$ can be compressed into  into $\left\lceil (d_{\rm c}^4 d_{\rm p}-1)\log (nm+d_{\rm c}^4 d_{\rm p}-1)+(2n+2m)\log d_{\rm c} \right\rceil$ qubits.

The same argument can be applied to a lattice    of $n$  site-independent correlated systems, each of which  has physical dimension $d_{\rm p}$ and interacts with $k$ neighbours.    In this case, a generic state  on the lattice can be compressed into
$\left\lceil d_{\rm c}^kd_{\rm p} \log(n+d_{\rm c}^k d_{\rm p}-1) + b\log d_{\rm c} \right\rceil$ qubits, where $b$ is the boundary size, namely the number of correlation systems across the boundary ($b=0$ for closed lattices, like {\em e.g.} the torus).

\subsection{
Multipartite states generated from a fixed state with the action of identical local unitary transformations}

In this section and the next, we study the compression of state families generated from the action of identical local unitary transformations, namely a state in the form $U_g^{\otimes n}\ket{\Psi_x}$. We first consider the case where the initial state is fixed and known, and generalise it in the next section to the case where the initial state is a tensor network state family.
Consider a parametric family of states of the form
\begin{align}
\ket{\Psi_g}  =  U_g^{\otimes n} \ket{\Psi_0} \, ,
\end{align}
where  $\ket{\Psi_0} \in \map{H}^{\otimes n}$ is a fixed pure state  on $n$ identical  systems, $g$ is an element of a group $G$, and   $U_g$ is a unitary operator belonging to a unitary representation of the group $G$.
  For example, the above states could describe the ground states of a system of $n$ spins immersed in a uniform magnetic field of known intensity and unknown direction. All these states can be obtained from a fixed state (say, corresponding to a magnetic field oriented in the $z$ direction) by rotating the direction of each spin by the same amount.  A compression protocol would give a way to store the state of the spins in a quantum memory without knowing the direction.

\begin{figure}[H]
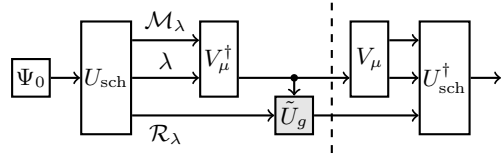

$$\begin{mathtikz}
    \node[tensor] (phi0) at (-0.5,0) {$\Psi_0$};
    \node[tensor, minimum height = 15mm] (sch1) at (0.5,0) {$U_{\rm sch}$};
    \draw[->, thick] (phi0) -- (sch1);
    \node[tensor, minimum height = 10mm] (mu) at (2,0.25) {$V_\mu^\dag$};
    \node[parameter] (U) at (3,-0.5) {$\tilde{U}_g$};
    \draw[->, thick] (sch1.east) -- node[auto] {$\lambda$} ($(mu.west)-(0,0.25)$);
    \draw[->, thick] ($(sch1.east)+(0,0.5)$) -- node[auto] {$\map{M}_\lambda$} ($(mu.west)+(0,0.25)$);
    \draw[->, thick] ($(sch1.east)-(0,0.5)$) -- node[auto,swap] {$\map{R}_\lambda$} (1.75,-0.5) -- ($(U.west)$);
    \node[tensor, minimum height = 10mm] (mu2) at (4,0.25) {$V_\mu$};
    \node[tensor, minimum height = 15mm] (sch2) at (5,0) {$U_{\rm sch}^\dag$};
    \draw[->, thick] ($(mu.east)-(0,0.25)$) -- node (lambdamid) {} ($(mu2.west)-(0,0.25)$);
    \fill (lambdamid.center) circle [radius=0.05];
    \draw[->, thick] (lambdamid.center) -- (U);
    \draw[->, thick] ($(mu2.east)+(0,0.25)$) -- ($(sch2.west)+(0,0.5)$);
    \draw[->, thick] ($(mu2.east)-(0,0.25)$) -- ($(sch2.west)+(0,0)$);
    \draw[->, thick] ($(U.east)$) -- ($(sch2.west)-(0,0.5)$);
    \draw[->, thick] (sch2) -- +(0.75,0);
    \draw[dashed, thick] (3.5,-1) -- (3.5,1);
\end{mathtikz}$$
\caption{\label{fig:sw_network1} Tensor network for $\{U_g^{\otimes n}\ket{\Psi_0}\}$. The dash line indicates the minimum cut.}
\end{figure}

To better characterise the structure of the transform $U_g^{\otimes n}$, we use the Schur-Weyl duality \cite{fulton2013representation}. The Schur-Weyl duality decomposes the Hilbert space $\map{H}^{\otimes n}$ into the following form:
\begin{align} \label{eq:swduality}
    \map{H}^{\otimes n} \simeq \bigoplus_{\lambda \in \map{Y}_{n,d}} (\map{R}_\lambda \otimes \map{M}_\lambda) \,,
\end{align}
where $\map{Y}_{n,d}$ is the set of Young diagrams with $n$ boxes and at most $d$ rows, and $\map{R}_\lambda$ and $\map{M}_\lambda$ are certain subspaces indexed by $\lambda$. We denote the unitary transformation from the original $n$-tensor space to the decomposition as $U_{\rm sch}: \map{H}^{\otimes n} \to \bigoplus_{\lambda \in \map{Y}_{n,d}} (\map{R}_\lambda \otimes \map{M}_\lambda)$, which is known as the Schur transform. One property of the decomposition (\ref{eq:swduality}) is that, $U_g^{\otimes n}$ acts trivially on each subspace $\map{M}_\lambda$. Therefore we can decompose $U_g^{\otimes n}$ with respect to this decomposition as
\begin{align} \label{eq:sw_ug}
    U_{\rm sch}U_g^{\otimes n}U_{\rm sch}^\dag = \sum_{\lambda \in \map{Y}_{n,d}} \ketbra{\lambda} \otimes U_{g,\lambda} \otimes I_{\map{M}_\lambda} \,,
\end{align}
where $\{\ket{\lambda}\}_{\lambda \in \map{Y}_{n,d}}$ is an orthonormal basis that indexes the direct sum, $U_{g,\lambda}$ is a unitary on $\map{R}_\lambda$, and $I_{\map{M}_\lambda}$ is the identity on $\map{M}_\lambda$. To match the decomposition of $U_g^{\otimes n}$, we also decompose $\ket{\Psi_0}$ as
\begin{align} \label{eq:sw_phi}
    U_{\rm sch}\ket{\Psi_0} = \sum_{\lambda \in \map{Y}_{n,d}} \xi_\lambda \ket{\lambda} \otimes \ket{r_\lambda} \otimes \ket{\mu_\lambda} \,,
\end{align}
where $\ket{r_\lambda}$ and $\ket{\mu_\lambda}$ are states in $\map{R}_\lambda$ and $\map{M}_\lambda$, respectively, and $\sum_\lambda |\xi_\lambda|^2 = 1$. Multiplying Equation (\ref{eq:sw_ug}) with Equation (\ref{eq:sw_phi}), we have
\begin{align}
    U_{\rm sch}U_g^{\otimes n} \ket{\Psi_0} = \sum_{\lambda \in \map{Y}_{n,d}} \xi_\lambda \ket{\lambda} \otimes U_{g,\lambda}\ket{r_\lambda} \otimes \ket{\mu_\lambda} \,.
\end{align}
Note that $\ket{\Psi_0}$ is a known fixed state, and therefore $\ket{\mu_\lambda}$ is known and fixed. We can then construct an isometry $V_\mu: \Span\{\ket\lambda\} \to \bigoplus_\lambda \map{M}_\lambda$ that encodes the states $\{\ket{\mu_\lambda}\}_{\lambda \in \map{Y}_{n,d}}$ as
\begin{align} \label{eq:Vmu}
V_\mu\ket{\lambda} := \ket{\lambda}\otimes\ket{\mu_\lambda}, ~ \forall \lambda \in \map{Y}_{n,d} \,.
\end{align}
Defining $\ket{\psi_g} := \sum_{\lambda \in \map{Y}_{n,d}} \xi_\lambda \ket{\lambda} \otimes U_{g,\lambda}\ket{r_\lambda} \in \bigoplus_\lambda \map{R}_\lambda$, we have $U_{\rm sch}^\dag V_\mu\ket{\psi_g} = U_g^{\otimes n} \ket{\Psi_0}$.
We  draw the tensor network generating the states $\{U_g^{\otimes n}\ket{\Psi_0}\}$ in Figure \ref{fig:sw_network1}. 
In Figure \ref{fig:sw_network1}, $\tilde{U}_g := \sum_\lambda \ket{\lambda} \bra{\lambda} \otimes U_{g,\lambda}$ is a compressed version of $U_g^{\otimes n}$. The T-shape intersection is a  copying operation on  the index system $\ket{\lambda}$, defined as
\begin{align}
\begin{mathtikz}
        \fill (0,0) circle [radius=0.05];
        \draw[->, thick] (-0.5,0) -- (0.5,0);
        \draw[->, thick] (0,0) -- (0,-0.5);
    \end{mathtikz} := \sum_{\lambda \in \map{Y}_{n,d}} \ket{\lambda}\ket{\lambda}\bra{\lambda}
\end{align}
The region to the left of the dashed line reads $\tilde{U}_g \left(\sum_\lambda \ket{\lambda}\ketbra{\lambda}\right) V_{\mu}^\dag U_{\rm sch} \ket{\Psi_0}$, which equals to $\ket{\psi_g}$. The partial isometry $V_\mu^\dag$ serves as a ``coherent erasure'' of the system $\map{M}_\lambda$. We first perform a coherent erasure on the multiplicity system $\map{M}_\lambda$ using $V_\mu^\dag$ and reprepare the system later using $V_\mu$. This effectively reduces the minimum cut (dashed line) of the tensor network. The cut edges constitutes the Hilbert space $\bigoplus_{\lambda \in \map{Y}_{n,d}} \map{R}_\lambda$.

Note that we are slightly  abusing the notations for tensor network, as  the two cut edges constitute a space that is not the tensor product space of each edge. The dimension of the edge labeled by $\map{R}_\lambda$ depends on $\lambda$. To avoid ambiguity, we have to restrict the upper edge to take values only in the basis $\{\ket\lambda\}$.

We vectorise the variable tensor $\tilde{U}_g$ as $\ket{\tilde{U}_g}$, and regard all other tensors as the tensor network $N$. Then we have $\ket{\Psi_g} = N_* \ket{\tilde{U}_g}$. Choosing the cut as the dashed line in Figure \ref{fig:sw_network1}, according to Theorem \ref{thm:tncompression}, the states $\{U_g^{\otimes n}\ket{\Psi_0}\}$ can be compressed into a memory of dimension equal to the dimension of $\bigoplus_{\lambda \in \map{Y}_{n,d}} \map{R}_\lambda$, which is

\begin{align}
    \mathbf{dim} \bigoplus_{\lambda \in \map{Y}_{n,d}} \map{R}_\lambda = \sum_{\lambda \in \map{Y}_{n,d}} \mathbf{dim} \map{R}_\lambda \leq (n+d-1)^{(d^2+d-2)/2}
\end{align}
namely a memory of no more than $\left\lceil\frac{d^2+d-2}{2}\log(n+d-1)\right\rceil$ qubits. The last inequality comes from 
Lemma 3 in Ref. \cite{yang2016efficient} shown below, with $r=d$.

\begin{lem} \label{lem:dlambda}
 The total dimension of all the representation spaces corresponding to Young diagrams with no more than
$r$ rows is upper bounded as
\begin{align}
    \sum_{\lambda \in \map{Y}_{n,r}} \mathbf{dim} \map{R}_\lambda \leq (n+d-1)^{(2dr-r^2+r-2)/2}
\end{align}
\end{lem}

\subsection{
Parametric tensor network state family under identical local unitary transformations}


In the previous section, we have discussed about the compression of states obtained from a fixed multipartite state under unknown local unitary transformations.
Here we consider the generalisation in which the unknown  transformations are applied to a parametric family of tensor network states.
  This generalisation could be used to treat the case of $n$    of interacting spins with unknown couplings immersed in a uniform magnetic field of unknown direction.

Explicitly, we consider tensor network states of the form
\begin{align}
\ket{\Psi_{x,g} }   =  U_g^{\otimes n}    N_*  \ket{v_x}   \, ,
\end{align}
 where $\ket{v_x}  \in \spc{H}_{\set X} $  is a  vector  in a suitable parameter   space $\spc{H}_{\set X}$, and $U_g$ is an unknown unitary transformation, representing the action of a group element $g\in  G$ on each physical system.

To use Theorem \ref{thm:tncompression}, our goal is to construct a tensor network that generates the family $\{U_g^{\otimes n}   N_* \ket{v_x}  \}_{g\in G,x\in\set{X}}$, with the property that the corresponding flow network  has small minimum cut.  We do the construction in two steps: we first consider a smaller state family and construct its corresponding tensor network, and
then extend the network so that it generates our target state family $\{U_g^{\otimes n} N_* \ket{v_x}  \}_{g\in G, \, x\in\set{X}}$.

Let $m = \dim\spc{H}_{\set X}$. Choose $m$ values of the parameters $x_1,\dots,x_m$ such that $\{\ket{v_{x_1}},\dots,\ket{v_{x_m}}\}$ is a basis of $\spc{H}_{\set X}$. The smaller family we consider is $\{U_g^{\otimes n} \ket{\Psi_{x_i}}\}_{g\in G, i=1,\dots,m}$, with $\ket{\Psi_{x_i}}  :  =  N_* \ket{v_{x_i}}$.  This family is an extension of the family in the previous section, where instead of fixing the initial state, the initial state is chosen from $m$ alternatives.


For any $x_i$, using the Schur transform, we can decompose $\ket{\Psi_{x_i}}$ as
\begin{align}
    U_{\rm sch}\ket{\Psi_{x_i}} = \sum_{\lambda\in\map{Y}_{n,d_{\rm p}}} \xi_\lambda^{(x_i)} \ket{\lambda} \otimes \ket{r_\lambda^{(x_i)}} \otimes \ket{\mu_\lambda^{(x_i)}}
\end{align}

If $\ket{\mu_\lambda^{(x_i)}}$ is known, as in the previous section, we can construct an isometry $V_\mu$ (\ref{eq:Vmu}). However, in this case $\ket{\mu_\lambda^{(x_i)}}$ is unknown and depends on the value of $i$. Thanks to the fact that $i$ takes a finite number of values (1 to $m$), we can construct one isometry for every value of $i$, in other words, an isometry controlled by $i$. As a result, we redefine $V_\mu$ as
\begin{align}
    V_\mu := \sum_{i=1}^m \left(\sum_{\lambda\in\map{Y}_{n,d_{\rm p}}} \ketbra{\lambda} \otimes \ket{\mu_\lambda^{(x_i)}}\right) \bra{i}
\end{align}
where $\{\ket{i}\}$ is a basis of an $m$-dimensional control system. And $V_\mu^\dag$ is defined as
\begin{align}
    V_\mu^\dag := \sum_{i=1}^m \left(\sum_{\lambda\in\map{Y}_{n,d_{\rm p}}} \ketbra{\lambda} \otimes \bra{\mu_\lambda^{(x_i)}}\right) \bra{i}
\end{align}
where the control system is not transposed.

We then draw the tensor network that generates  $\{U_g^{\otimes n} \ket{\Psi_{x_i}}\}_{g\in G, i=1,\dots,m}$, which is similar to Figure \ref{fig:sw_network1} with additional control systems for $V_\mu$ and $V_\mu^\dag$.
\begin{figure}[H]
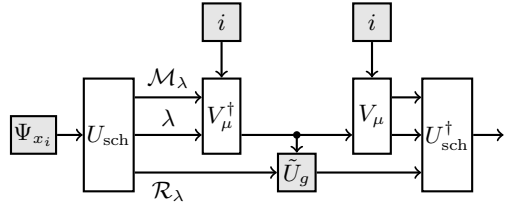

    $$\begin{mathtikz}
        \node[parameter] (phi0) at (-0.5,0) {$\Psi_{x_i}$};
        \node[tensor, minimum height = 15mm] (sch1) at (0.5,0) {$U_{\rm sch}$};
        \draw[->, thick] (phi0) -- (sch1);
        \node[parameter] (x1) at (2,1.5) {$i$};
        \node[tensor, minimum height = 10mm] (mu) at (2,0.25) {$V_\mu^\dag$} edge[<-, thick] (x1);
        \node[parameter] (U) at (3,-0.5) {$\tilde{U}_g$};
        \node[parameter] (x2) at (4,1.5) {$i$};
        \node[tensor, minimum height = 10mm] (mu2) at (4,0.25) {$V_\mu$} edge[<-, thick] (x2);
        \node[tensor, minimum height = 15mm] (sch2) at (5,0) {$U_{\rm sch}^\dag$};
        \draw[->, thick] (sch1.east) -- node[auto] {$\lambda$} ($(mu.west)-(0,0.25)$);
        \draw[->, thick] ($(sch1.east)+(0,0.5)$) -- node[auto] {$\map{M}_\lambda$} ($(mu.west)+(0,0.25)$);
        \draw[->, thick] ($(sch1.east)-(0,0.5)$) -- node[auto,swap] {$\map{R}_\lambda$} (1.75,-0.5) -- ($(U.west)$);
        \draw[->, thick] ($(mu.east)-(0,0.25)$) -- node (lambdamid) {} ($(mu2.west)-(0,0.25)$);
        \fill (lambdamid.center) circle [radius=0.05];
        \draw[->, thick] (lambdamid.center) -- (U);
        \draw[->, thick] ($(mu2.east)+(0,0.25)$) -- ($(sch2.west)+(0,0.5)$);
        \draw[->, thick] ($(mu2.east)-(0,0.25)$) -- ($(sch2.west)+(0,0)$);
        \draw[->, thick] ($(U.east)$) -- ($(sch2.west)-(0,0.5)$);
        \draw[->, thick] (sch2) -- +(0.75,0);
    \end{mathtikz}$$
    \caption{\label{fig:sw_network2} Tensor network for $\{U_g^{\otimes n} \ket{\Psi_{x_i}}\}_{g\in G, i=1,\dots,m}$. $\Psi_{x_i}$ and copies of $i$ are unknown as well as $\tilde{U}_g$.}
\end{figure}

For any fixed $i$, this network reduces to Figure \ref{fig:sw_network1}, and generates the state $U_g^{\otimes n}\ket{\Psi_{x_i}}$. 
{Therefore any state in the family $\{U_g^{\otimes n} \ket{\Psi_{x_i}}\}_{g\in G, i=1,\dots,m}$ can be generated by this network. We regard $\ket{\Psi_{x_i}}$, two copies of $\ket{i}$ and $\tilde{U}_g$ as the parameters, and write $U_g^{\otimes n} \ket{\Psi_{x_i}} = M_* \ket{\Psi_{x_i}}\ket{i}\ket{i}\ket{\tilde{U}_g}$, where $M_*$ is the linear operator represented by the constant tensors in network, and $\ket{\tilde{U}_g}$ is the vectorised version of $\tilde{U}_g$.}

\begin{figure}[H]
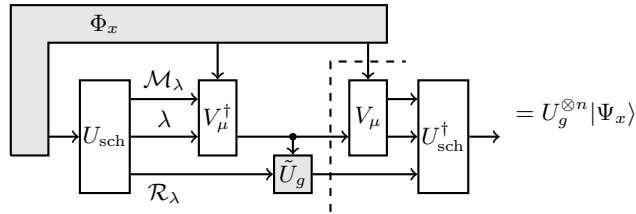

    $$\begin{mathtikz}
        \node[virtual] (phi0) at (-0.5,0) {};
        \node[tensor, minimum height = 15mm] (sch1) at (0.5,0) {$U_{\rm sch}$};
        \draw[->, thick] (phi0) -- (sch1);
        \node[virtual] (x1) at (2,1.5) {};
        \node[tensor, minimum height = 10mm] (mu) at (2,0.25) {$V_\mu^\dag$} edge[<-, thick] (x1);
        \node[parameter] (U) at (3,-0.5) {$\tilde{U}_g$};
        \node[virtual] (x2) at (4,1.5) {};
        \draw[parameter] ($(phi0)-(0.25,0.25)$) |- ($(x2)+(0.25,0.25)$) -- ($(x2)+(0.25,-0.25)$) -| ($(phi0)+(0.25,-0.25)$) -- ($(phi0)-(0.25,0.25)$);
        \node[virtual] (ent) at (0.5,1.5) {$\Phi_x$};
        \node[tensor, minimum height = 10mm] (mu2) at (4,0.25) {$V_\mu$} edge[<-, thick] (x2);
        \node[tensor, minimum height = 15mm] (sch2) at (5,0) {$U_{\rm sch}^\dag$};
        \draw[->, thick] (sch1.east) -- node[auto] {$\lambda$} ($(mu.west)-(0,0.25)$);
        \draw[->, thick] ($(sch1.east)+(0,0.5)$) -- node[auto] {$\map{M}_\lambda$} ($(mu.west)+(0,0.25)$);
        \draw[->, thick] ($(sch1.east)-(0,0.5)$) -- node[auto,swap] {$\map{R}_\lambda$} (1.75,-0.5) -- ($(U.west)$);
        \draw[->, thick] ($(mu.east)-(0,0.25)$) -- node (lambdamid) {} ($(mu2.west)-(0,0.25)$);
        \fill (lambdamid.center) circle [radius=0.05];
        \draw[->, thick] (lambdamid.center) -- (U);
        \draw[->, thick] ($(mu2.east)+(0,0.25)$) -- ($(sch2.west)+(0,0.5)$);
        \draw[->, thick] ($(mu2.east)-(0,0.25)$) -- ($(sch2.west)+(0,0)$);
        \draw[->, thick] ($(U.east)$) -- ($(sch2.west)-(0,0.5)$);
        \draw[->, thick] (sch2) -- +(0.75,0);
        \draw[dashed, thick] (3.5,-1) |- (4.5,1);
    \end{mathtikz} = U_g^{\otimes n}\ket{\Psi_{x}}$$
    \caption{\label{fig:sw_network3} Tensor network for $\{U_g^{\otimes n} \ket{\Psi_x}\}_{g\in G,x\in\set{X}}$. Now the vector $\ket{\Phi_x}$ serves as a parameter, and its three outgoing edges corresponds to the three systems in Equation (\ref{eq:Psiiii}). The dashed line indicates the minimum cut.}
\end{figure}

Now we consider the original state family $\{U_g^{\otimes n} \ket{\Psi_x}\}_{g\in G,x\in\set{X}}$. Take any state $U_g^{\otimes n} \ket{\Psi_x} = U_g^{\otimes n} N_* \ket{v_x}$ from the family, we can decompose $\ket{v_x}$ in the basis $\{\ket{v_{x_i}}\}$ as $\ket{v_x} = \sum_i \alpha_i \ket{v_{x_i}}$. Then the state $U_g^{\otimes n} \ket{\Psi_x}$ can be written as a superposition of states in the smaller family:
\begin{align}
    U_g^{\otimes n} \ket{\Psi_x} = \sum_i \alpha_i U_g^{\otimes n} \ket{\Psi_{x_i}}
\end{align}
This indicates that we can generate $U_g^{\otimes n} \ket{\Psi_x}$ via the linear operator $M_*$ with a superposition of the parameters. Defining
\begin{align}\label{eq:Psiiii}
    \ket{\Phi_x} := \sum_i \alpha_i \ket{\Psi_{x_i}}\ket{i}\ket{i}
\end{align}
we have
\begin{align}
    U_g^{\otimes n} \ket{\Psi_x} = M_* (\ket{\Phi_x} \otimes \ket{\tilde{U}_g})
\end{align}
which shows that the family $\{U_g^{\otimes n} \ket{\Psi_x}\}_{g\in G,x\in\set{X}}$ can be generated by the following tensor network in Figure \ref{fig:sw_network3}.

The lower two cut edges constitutes the space $\bigoplus_{\lambda \in \map{Y}_{n,d}} \map{R}_\lambda$, which has dimension no more than $(n+d-1)^{(d^2+d-2)/2}$ (Lemma \ref{lem:dlambda}). The uppermost cut edge corresponds to the control system $\Span\{\ket{i}\}_{i=1}^m$ with dimension equal to $\dim\spc{H}_{\set X}$, and the combined dimension of all cut edges is $(n+d-1)^{(d^2+d-2)/2} \dim \spc{H}_{\set X}$. Using Theorem \ref{thm:tncompression}, we then obtain the memory size for compression the states $\{U_g^{\otimes n} \ket{\Psi_x}\}_{g\in G,x\in\set{X}}$, as stated in the following proposition.


\begin{prop} \label{prop:UnTN}
Consider a family of tensor network states $\{\ket{\Psi_x}\}_{x\in\set{X}} \subset \spc{H}_{d}^{\otimes n}$ with parameter space $\spc{H}_{\set X}$ (\ref{parametricTN}). The state family generated by applying an unknown unitary transformation on all physical systems simultaneously, namely $\{U_g^{\otimes n} \ket{\Psi_x}\}_{g\in G,x\in\set{X}}$, can be compressed without error into a memory of no more than $\left\lceil\frac{d^2+d-2}{2}\log(n+d-1) + \log \dim \spc{H}_{\set X}\right\rceil$ qubits.
\end{prop}

Note that to minimise the memory usage in Proposition \ref{prop:UnTN}, one may choose an alternative parametrisation of the original family: $\ket{\Psi_x} = N'_* \ket{v'_x}$, where $\ket{v'_x}$ resides in a space $\spc{H}'_{\set X}$ that is smaller than $\spc{H}_{\set X}$. Specifically, the minimum cut $C$ divides $N$ into two subnetworks $N_1$ and $N_2$, so that $N_*$ is a concatenation of the corresponding linear operators, namely $N_*=N_{2*}N_{1*}$, as shown in Figure \ref{fig:N1N2}.
Then by defining $\ket{v'_x}=N_{1*}\ket{v_x}$, we have $\ket{\Psi_x} = N_{2*}\ket{v'_x}$, and the new parameter space is $\spc{H}_C$, the combined Hilbert space of all cut edges. The dimension of $\spc{H}_C$ can be smaller than the original parameter space $\spc{H}_{\set X}$. With the new parametrisation, when we apply Proposition \ref{prop:UnTN} for the states $\{\ket{\Psi_x}\}_{g\in G,x\in\set{X}}$, we use a memory of $\left\lceil\frac{d^2+d-2}{2}\log(n+d-1) + c(C)\right\rceil$ qubits, where $c(C)=\log \dim \spc{H}_{C}$ is the capacity of the minimum cut.

We now consider MPSs and PEPSs. In Section \ref{ss:mps} and Appendix \ref{ss:PEPS}, we showed that an MPS or PEPS with unknown boundary condition can be written as $\ket{\Psi_B} = N_*\ket{B}$, where $\ket{B}\in\spc{H}_{\set X}$ is a vector describing the boundary condition, and $N_*$ is a linear operator. $\spc{H}_{\set X}$ has dimension $d_{\rm c}^2$ for MPSs and $d_{\rm c}^{2n+2m}$ for PEPSs. We can then directly apply Proposition \ref{prop:UnTN} to the states $\{U_g^{\otimes n}\ket{\Psi_B}\}$. An MPS with variable boundary conditions under unknown transformation $U_g^{\otimes n}$ can be compressed into a memory of
$\left\lceil\frac{d_{\rm p}^2+d_{\rm p}-2}{2}\log(n+d_{\rm p}-1) + 2\log d_{\rm c}\right\rceil$ qubits, while a PEPS on a square lattice with variable boundary condition under unknown transformation $U_g^{\otimes nm}$ can be compressed into a memory of
$\left\lceil\frac{d_{\rm p}^2+d_{\rm p}-2}{2}\log(nm+d_{\rm p}-1) + (2n+2m)\log d_{\rm c}\right\rceil$ qubits.


\section{Proof of Proposition \ref{thm:marginal}} \label{app:marginal}

As illustrated in Figure \ref{fig:cor:marginal}, let $N_1$ and $N_2$ be the subnetworks of $N$ induced by the cut $C$. 
Then $N_1$ defines a linear operator $N_1: \spc{H}_{\set X} \to \map{H}_{\rm E} \otimes \map{H}_C$, $N_2$ defines a linear operator $N_2: \map{H}_C \to \spc H_{\rm P}$, and we have $N=(N_2\otimes I_{\rm E})N_1$, where $I_{\rm E}$ is the identity operator on $\map{H}_{\rm E}$.

\begin{figure}[H]
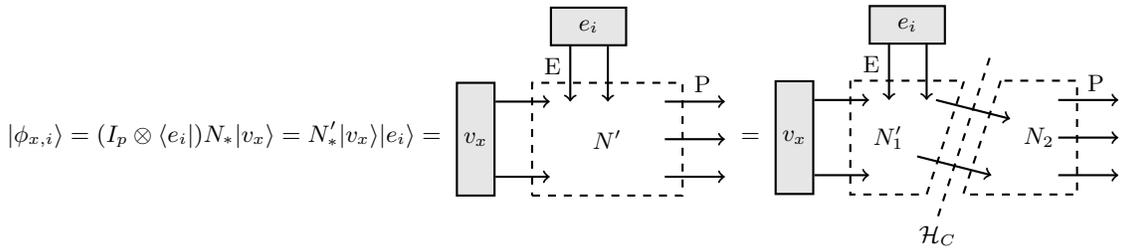

    $$
    \ket{\phi_{x,i}} = (I_p\otimes \bra{e_i})N_*\ket{v_x} = N'_*\ket{v_x}\ket{e_i} = \begin{mathtikz}
        \node[parameter, minimum height=15mm] (x) at (0,0.5) {$v_x$};
        \node[tensor, dashed, minimum height=15mm, minimum width=20mm] (N) at (1.75,0.5) {$N'$};
        \draw[->, thick] ($(x.east)+(0,0.5)$) -- ($(N.west)+(0.25,0.5)$);
        \draw[->, thick] ($(x.east)+(0,-0.5)$) -- ($(N.west)+(0.25,-0.5)$);
        \draw[->, thick] ($(N.east)+(-0.25,0.5)$) -- node[above,xshift=1mm] {$\rm P$} +(0.8,0);
        \draw[->, thick] ($(N.east)+(-0.25,-0.5)$) -- +(0.8,0);
        \draw[->, thick] ($(N.east)+(-0.25,0)$) -- +(0.8,0);
        \node[virtual, minimum height=15mm, minimum width=10mm] (S) at (1.25,0.5) {};
        \draw[<-, thick] ($(S.east)+(0,0.5)$) -- +(0,0.75);
        \draw[<-, thick] ($(S.east)+(-0.5,0.5)$) -- node[left,yshift=1mm] {${\rm E}$} +(0,0.75);
        \node[parameter, minimum width = 10mm] (E) at (1.5,2) {$e_i$};
        \node[virtual] (placeholder) at (0,-0.75) {};
    \end{mathtikz} = \begin{mathtikz}
        \node[parameter, minimum height=15mm] (x) at (0,0.5) {$v_x$};
        \node[virtual, minimum height=15mm, minimum width=10mm] (S) at (1.25,0.5) {$N_1'$};
        \draw[dashed, thick] (S.north west) -- (S.south west) -- (S.south east) -- ($(S.north east)+(0.5,0)$) -- (S.north west);
        \node[virtual, minimum height=15mm, minimum width=10mm] (T) at (3.25,0.5) {$N_2$};
        \draw[dashed, thick] (T.north west) -- ($(T.south west)-(0.5,0)$) -- (T.south east) -- (T.north east) -- (T.north west);
        \draw[->, thick] ($(x.east)+(0,0.5)$) -- ($(S.west)+(0.25,0.5)$);
        \draw[->, thick] ($(x.east)+(0,-0.5)$) -- ($(S.west)+(0.25,-0.5)$);
        \draw[->, thick] ($(T.east)+(-0.25,0.5)$) -- node[above,xshift=1mm] {$\rm P$} +(0.8,0);
        \draw[->, thick] ($(T.east)+(-0.25,-0.5)$) -- +(0.8,0);
        \draw[->, thick] ($(T.east)+(-0.25,0)$) -- +(0.8,0);
        \draw[->, thick] ($(S.east)+(0.125,0.5)$) -- ($(T.west)+(0.125,0.25)$);
        \draw[->, thick] ($(S.east)+(-0.125,-0.25)$) -- ($(T.west)+(-0.125,-0.5)$);
        \draw[<-, thick] ($(S.east)+(0,0.5)$) -- +(0,0.75);
        \draw[<-, thick] ($(S.east)+(-0.5,0.5)$) -- node[left,yshift=1mm] {${\rm E}$} +(0,0.75);
        \node[parameter, minimum width = 10mm] (E) at (1.5,2) {$e_i$};
        \coordinate (start) at ($(S.north east)!0.75!(T.north west)$);
        \coordinate (end) at ($(S.south east)!0.25!(T.south west)$);
        \draw[dashed, thick] ($(start)!-0.2!(end)$) -- ($(start)!1.2!(end)$) node[below] (label) {$\map{H}_C$};
    \end{mathtikz}
    $$
\caption{\label{fig:cut3}The network generating $\ket{\phi_{x,i}}$ and the cut of $N'$. Both $\ket{v_x}$ and $\ket{e_i}$ are regarded as parameters. $N_1'$ is obtained from $N_1$ by reversing all edges corresponding to the environment.}
\end{figure}

Let $d_{\rm E} = \dim\map{H}_{\rm E}$, and take the computational basis $\{\ket{e_i}\}_{i=1}^{d_{\rm E}}$ of $\map{H}_{\rm E}$. Being the computational basis means $\ket{e_i} = \ket{\overline{e_i}}$. Define $\ket{\phi_{x,i}} := (I_{\rm P}\otimes \bra{e_i})\ket{\Psi_x}  = (I_{\rm P}\otimes \bra{e_i})N_* \ket{v_x} \in \spc H_{\rm P}$, so that $\ket{\Psi_x}=\sum_{i=1}^{d_{\rm E}}\ket{\phi_{x,i}}\ket{e_i}$. Now we consider the compression for the (unnormalised) states $\{\ket{\phi_{x,i}}\}_{x\in\set{X},i=1,\dots,d_{\rm E}}$. In fact, this set of states is generated by the same network $N$ by reversing the edges for the environment. This results in a network $N'$, compatible with the template $\Temp'$. This is shown in Figure \ref{fig:cut3}.

We regard the cut $C$ for network $N$ also as a cut for $N'$. Let $\ket{\psi_{x,i}} = N_{1*}'\ket{v_x}\ket{e_i}$, and then $\ket{\phi_{x,i}} = N_{2*}\ket{\psi_{x,i}}$.

Applying Theorem \ref{thm:tncompression} for $\{\ket{\phi_{x,i}}\}$, there exists a partial isometry $V: \spc H_{\rm P} \to \spc{H}_C$ such that
\begin{align}
    V^\dag V \ket{\phi_{x,i}} = \ket{\phi_{x,i}}, ~\forall x\in X, i=1,\dots, d_{\rm E}
\end{align}

This $V$ is what we want. For all $x\in\set{X}$,
\begin{align}
    (V^\dag V \otimes I_{\rm E})\ket{\Psi_x} & = (V^\dag V \otimes I_{\rm E}) \sum_{i=1}^{d_{\rm E}}\ket{\phi_{x,i}}\ket{e_i}\\
    & = \sum_{i=1}^{d_{\rm E}} (V^\dag V\ket{\phi_{x,i}})\ket{e_i}\\
    & = \sum_{i=1}^{d_{\rm E}} \ket{\phi_{x,i}}\ket{e_i}\\
    & = \ket{\Psi_x}
\end{align}

\qed

\section{Proof of Lemma \ref{lem:MPSspan}} \label{app:basis}

\begin{proof}

For concreteness, we first prove the lemma  for MPSs with open boundary conditions, and then show how to generalise the proof to arbitrary boundary conditions.

We encode the coefficients into a vector $\ket{c}:= \sum_{k=1}^t c_{k}\ket{k}$. We assume $\ket{c}$ is normalised such that the linear combination $|\Psi\>  = \sum_k  c_k \,  |\Sigma_k\>$ has unit length.  Since $\{\ket{\Sigma_k}\}$ are MPSs, they have the following form:
\begin{align}
    \ket{\Sigma_k} =
    \begin{mathtikz}
        \node[tensor] (L) at (-1,0) {$L_k$};
        \node[tensor] (A1) at (0,0) {$A^{[1]}_k$} edge [->, thick] (L) edge [->, thick] (0,0.75);
        \node[tensor] (A2) at (1,0) {$A^{[2]}_k$} edge [->, thick] (A1) edge [->, thick] (1,0.75);
        \node (dots) at (2,0) {$\cdots$} edge [->, thick] (A2);
        \node[tensor] (An) at (3,0) {$A^{[n]}_k$} edge [->, thick] (dots) edge [->, thick] (3,0.75);
        \node[tensor] (R) at (4,0) {$R_k$} edge [->, thick] (An);
    \end{mathtikz}
\end{align}
To represent their linear combination, we define tensors $\Lambda$, $B^{[1]},\dots,B^{[n]}$, and $\Rho$ such that
\begin{align}
    \begin{mathtikz}
        \draw[use as bounding box, white] (-0.25,-0.75) rectangle (0.75,0.75);
        \node[tensor] (Lambda) at (0,0) {$\Lambda$} edge [<-, thick] (0.75,0);
        \node[virtual] (i) at (0,-1) {$k$};
        \draw[->, thick] (i) -- (Lambda);
    \end{mathtikz} =
    \begin{mathtikz}
        \node[tensor] (Lambda) at (0,0) {$L_k$} edge [<-, thick] (0.75,0);
    \end{mathtikz}, \quad
    \begin{mathtikz}
        \draw[use as bounding box, white] (-0.75,-0.75) rectangle (0.75,0.75);
        \node[tensor] (A) at (0,0) {$B^{[i]}$} edge [<-, thick] (0.75,0) edge [->, thick] (-0.75,0) edge [->, thick] (0,0.75);
        \node[virtual] (i) at (0,-1) {$k$};
        \draw[->, thick] (i) -- (A);
    \end{mathtikz} =
    \begin{mathtikz}
        \draw[use as bounding box, white] (-0.75,-0.75) rectangle (0.75,0.75);
        \node[tensor] (A) at (0,0) {$A^{[i]}_k$} edge [<-, thick] (0.75,0) edge [->, thick] (-0.75,0) edge [->, thick] (0,0.75);
    \end{mathtikz}, \quad
    \begin{mathtikz}
        \draw[use as bounding box, white] (-0.75,-0.75) rectangle (0.25,0.75);
        \node[tensor] (Lambda) at (0,0) {$\Rho$} edge [->, thick] (-0.75,0);
        \node[virtual] (i) at (0,-1) {$k$};
        \draw[->, thick] (i) -- (Lambda);
    \end{mathtikz} =
    \begin{mathtikz}
        \node[tensor] (Lambda) at (0,0) {$R_k$} edge [->, thick] (-0.75,0);
    \end{mathtikz}
\end{align}
for every $k\in \{1,\dots,t\}$ and $i\in \{1,\dots,n\}$. Then $\ket{\Sigma}$ can be represented as:
\begin{align}\label{eq:Sigma}
    \ket{\Psi} =
    \begin{mathtikz}
        \node[tensor] (alpha) at (-1.75,-0.75) {$c$};
        \node[tensor] (A0) at (-1,0) {$\Lambda$};
        \node[tensor] (A1) at (0,0) {$B^{[1]}$} edge [->, thick] (A0) edge [->, thick] (0,0.75);
        \node[tensor] (A2) at (1,0) {$B^{[2]}$} edge [->, thick] (A1) edge [->, thick] (1,0.75);
        \node (dots) at (2,0) {$\cdots$} edge [->, thick] (A2);
        \node[tensor] (An) at (3,0) {$B^{[n]}$} edge [->, thick] (dots) edge [->, thick] (3,0.75);
        \node[tensor] (R) at (4,0) {$\Rho$} edge [->, thick] (An);
        \draw[->,thick] (alpha) -| (R);
        \foreach \i in {0,1,2,n}
        {
            \draw[->,thick] (alpha-|A\i) -- (A\i);
            \fill (alpha-|A\i) circle [radius=0.05];
        }
        \draw[densely dotted] (-0.5,-1) -- (-0.5,0.5);
        \draw[densely dotted] (3.5,-1) -- (3.5,0.5);
    \end{mathtikz}
\end{align}
where  $\begin{mathtikz}\label{eq:connector}
        \fill (0,0) circle [radius=0.05];
        \draw[->, thick] (-0.5,0) -- (0.5,0);
        \draw[->, thick] (0,0) -- (0,0.5);
    \end{mathtikz}$ is the tensor defined by
\begin{align}
\begin{mathtikz}\label{eq:connector}
        \fill (0,0) circle [radius=0.05];
        \draw[->, thick] (-0.5,0) -- (0.5,0);
        \draw[->, thick] (0,0) -- (0,0.5);
    \end{mathtikz} := \sum_{k=1}^t \ket{k}\ket{k}\bra{k} \,.
\end{align}
Equation (\ref{eq:Sigma}) shows that $\ket{\Psi}$ is an MPS: the left boundary condition is to the left of the first dotted line, the right boundary condition is to the right of the second dotted line, and the tensor on each physical system is $B^{[i]}$ connected with a T-intersection (\ref{eq:connector}). There are two edges connecting consecutive physical systems, one has dimension $d_{\rm c}$, and the other one has dimension $t$. Therefore the bond dimension of $\ket{\Psi}$ is $t d_{\rm c}$.

For MPSs with  general  boundary conditions, one just needs to replace the boundary conditions by a suitable  tensor connecting $A^{[1]}_k$ and $A^{[n]}_k$ ($B^{[1]}$ and $B^{[n]}$), and the rest of the proof is identical.

To conclude the proof, we invoke the fact  that all MPSs with polynomial-size bond dimension can be prepared in polynomial time on a quantum computer \cite{cirac2009renormalization,eisert2013entanglement}.
\end{proof}

\section{Proof of Lemma \ref{prop:gap}}\label{app:gap}

The channel $\map R$ can be expressed as
\begin{align}
\map R  =  \frac 1 {r} \sum_{l=0}^{r-1}  \map R_l \, ,\qquad     \map R_l  (\rho)    : =  \frac 1 6  \,  \sum_{\alpha  \in  \{x,y,z\}}  \sum_{s\in  \{ +, -\}}  \,    U_{l,\alpha  ,  s}  \rho     U_{l,\alpha  ,  s}^\dag   \, \qquad       U_{l,\alpha  ,  \pm}   :  =  I_{\rm in}  - 2   |\Psi_{l, \alpha , \pm    }  \>\<\Psi_{l,\alpha,  \pm}|   \, .
\end{align}
Let $\spc S_l := \Span\{  |l\> ,  |l\oplus1\>\}$ and $\spc S_l^\perp$ be its orthogonal complement. For each $l$, both $\spc S_l$ and $\spc S_l^\perp$ are invariant subspaces of channel $\map R_l$. For $\rho$ with support only in $\spc S_l$, one can explicitly determine $\map R_l$ as
\begin{align}
\map R_l (\rho)   =     \,  \frac 43      \frac {  P_l}2 \Tr[\rho]   - \frac 13  \rho \, , ~ \forall \rho, \Supp(\rho)\subseteq \spc S_l \,,
\end{align}
where $P_l$ is the projector on $\spc S_l$.
For $\rho$ with support only in $\spc S_l^\perp$, the channel $\map R_l$ is just the identity. For a general $\rho$, one can decompose it as $\rho = P_l \rho P_l + P_l^\perp \rho P_l^\perp + \rho'$, where $P_l^\perp $ is  the projector on $\spc S_l^\perp$, and $\rho':=P_l \rho P_l^\perp + P_l^\perp \rho P_l$ contains the off-block-diagonal terms. We further observe that $\map R_l(\rho')=0$, because for any off-block-diagonal element $\sigma = \ket{j}\bra{k}$ (or $\sigma = \ket{k}\bra{j}$) with $j\in\{l,l+1\}$ and $k\notin\{l,l+1\}$, we have $U_{l,\alpha, +}  \sigma  U_{l,\alpha, +}^\dag  + U_{l,\alpha, -}  \sigma  U_{l,\alpha, -}^\dag=0, \forall \alpha\in\{x,y,z\}$, and thus $\map R_l(\sigma)=0$. We conclude that $\map R_l(\rho) = \map R_l( P_l  \rho   P_l )  +   \map R_l(   P_l^\perp \rho  P_l^\perp )$, and therefore for general $\rho \in S(\spc{H}_{\rm in})$,
\begin{align}
\map R_l (\rho)   =     \,  \frac 43      \frac {  P_l}2 \Tr[P_l \rho]   - \frac 13  P_l  \rho   P_l   +      P_l^\perp \rho  P_l^\perp  \, .
\end{align}

To find the eigenvalues of the channel $\map R$, we use the double-ket notation $|A\kk  :  =  \sum_{j,k}  \,  A_{jk}  \,  |j\>\otimes |k\>$, representing linear operators on $\spc H_{\rm in}$ as vectors in the tensor product space $\spc H_{\rm in}\otimes \spc H_{\rm in}$.    Using this notation, the eigenvalue equation $\map R (A)  =  \lambda \, A$ becomes     $   \check R  |A\kk  =  \lambda\,  |A\kk$, with
  \begin{align}
  \check R  :=  \frac 1 {r} \, \sum_{l=0}^{r-1} \check R_l  \,  ,  \qquad  \check R_l     :=    \frac 43       \frac { | P_l\kk  \bb   P_l|}2          -\frac 13    P_l  \otimes   P_l    +  P_l^\perp \otimes P_l^\perp  \, .
  \end{align}
Averaging over $l$, we finally obtain
\begin{align}
\check R   =  \check R_1 + \check R_2 + \check R_3
\end{align}
where
\begin{align}
\check R_1 := &~ \frac 8{3 r}   \,  \sum_{l=0}^{r-1}     \, \ketbra{l}\otimes\ketbra{l} + \frac2{3 r} \,  \sum_{l=0}^{r-1} ( \ket{l}\bra{l\oplus 1}\otimes\ket{l}\bra{l\oplus 1}+\ket{l\oplus 1}\bra{l}\otimes\ket{l\oplus 1}\bra{l} )\\
\check R_2 := &~ \frac  2{3 r}   \,  \sum_{l=0}^{r-1}    \,   |l\>\<l|\otimes  |l\oplus1\>\<l\oplus 1|  + \frac  2{3 r}   \,  \sum_{l=0}^{r-1}    \,  |l\oplus1\>\<l\oplus 1| \otimes |l\>\<l|\\
\check R_3 := &~  \left( 1  -  \frac 4 {r}\right) \,  \left(  I_{\rm in}\otimes I_{\rm in}\right) \, .
\end{align}

$\check R_3$ is proportional to the identity and does not contribute to the spectral gap. From $\check R_1 \check R_2 = \check R_2 \check R_1 = 0$, the supports of $\check R_1$ and $\check R_2$ are orthogonal subspaces, so we can consider the eigendecompositions of $\check R_1$ and $\check R_2$ separately. The spectral gap would then equals to the difference between the largest and second largest among the union of eigenvalues of $\check R_1$ and $\check R_2$. Notice that the support of $\check R_1$ is in the subspace $\Span\{\ket{l}\otimes\ket{l}\}_{l=0}^{r-1}$. Under the basis $\{\ket{l}\otimes\ket{l}\}_{l=0}^{r-1}$, $\check R_1$ is a Toeplitz matrix whose eigendecomposition has a simple form \cite{gray2006toeplitz}:

\begin{align}\label{eq:Reigs}
\check R_1 =  \frac 4r   \,  \sum_{k=0}^{r-1}     \, \frac{  2  +  \cos \frac {2\pi k}{r} }{3}  \,  |\Phi_k\>\<\Phi_k|
\end{align}

where each eigenvector $|\Phi_k\>$ is the Fourier vector defined as $|\Phi_k\>   : =  \sum_l  \,  e^{2\pi  i  kl/  d_{\rm in}}\,  |l\>\otimes |l\>  /\sqrt{d_{\rm in}}$.
Among the eigenvalues in Equation (\ref{eq:Reigs}), the largest eigenvalue is $\frac{  4 }{r}$, with eigenvector $\ket{\Phi_0}$. The second largest is $\frac 4r  \frac{  2  +  \cos \frac {2\pi k}{r} }{3} $ with eigenvectors $\ket{\Phi_1}$ and $\ket{\Phi_{r-1}}$. Now we turn to $\check R_2$, and observe that its only eigenvalue is $\frac  2{3 r}$, which is smaller than the second largest eigenvalue of $\check R_1$. We therefore conclude that the spectral gap equals to the difference between the two largest eigenvalues of $\check R_1$, which is
 \begin{align}
 \gamma_{\map R}   =    \frac{ 4  }{r} - \frac 4r   \left ( \frac{  2  +  \cos \frac {2\pi}{3} }{r}  \right)   =   \frac{  8\left(    \sin \frac {\pi }{r}  \right)^2 }{3r}  \, .
  \end{align}
 \qed

\section{Proof of Proposition \ref{prop:efficientlyprep}}\label{app:efficientlyprep}

The key of the proof is to show that Conditions \ref{cond:spol}-\ref{cond:seig} guarantee that every linear combination $\sum_k c_k |\Sigma_k\>$   with efficiently computable coefficients $\{c_k\}$  is efficiently preparable. Once this is done, we can simply construct an orthonormal basis from the states $\{|\Sigma_k\>\}$, and use the fiducial states in Equation (\ref{vectors}).

Any vector  $|\Psi\>  \in  \spc H_{\rm in}$ can be decomposed as  $|\Psi\>  =  \sum_k  \, c_k  \,  |\Sigma_k\>$, with  $c_k  =   \<  \Sigma_k  | F^{-1} | \Psi\>  $  and $F  =  \sum_k  |\Sigma_k\>\<\Sigma_k|$  \cite{casazza1998frames}.  Now, let ${\rm A}$ be an auxiliary  quantum system of dimension $s$, and consider the state
\begin{align}
|\psi\>  =    \frac{ \sum_k     c_k ~      |k\>}{  \sqrt{  \sum_k  |c_k|^2 }}    \, .
\end{align}
For given coefficients $\{c_k\}$, the state $|\psi\>$ is efficiently preparable, because it is a state in a Hilbert space of polynomial dimension. 

Now, consider  a coherent control mechanism that prepares the state $|\Sigma_k\>$ by a coherent process $\ket{k} \mapsto \ket{k}\otimes\ket{\Sigma_k}$, where $\{|k\>\}$ is a basis of an auxiliary system. 
 Setting the auxiliary system to $\ket{\psi}$, the resulting state is
\begin{align}
|\Gamma\>  =    \frac{ \sum_k     c_k ~    |k\> \otimes   |\Sigma_k\>}{  \sqrt{  \sum_k  |c_k|^2 }}    \, ,
\end{align}

Finally, projecting the auxiliary system on the vector $\sum_k  |k\> /\sqrt s $, one obtains the state $|\Psi\>$.
The probability that the projection takes place is  $p_\Psi   =  (  s   \sum_k  |c_k|^2  )^{-1} =   (s   \<  \Psi|  F^{-1}  |\Psi\>)^{-1}  \ge 1/(s \lambda_{\min})$, where $\lambda_{\rm min}$ is the minimum non-zero eigenvalue of $F$.   Note that the eigenvalues of $F$   are the same as the eigenvalues of the Gram matrix  $G  =  \<\Sigma_k|\Sigma_l\>$. This is the case because because one has $F  =  X X^\dag$ and $G  =  X^\dag X$, with $X   = \sum_k   |\Sigma_k\>\<k|$.
Since $\lambda_{\min}$ is assumed to be at least inverse polynomial,  the probability is guaranteed to be at least inverse polynomial.  This means that a polynomial number of repetitions of the above procedure are sufficient to generate the state $|\Psi\>$ with probability close to 1.  \qed

\end{document}